\documentclass{mn2e}
\usepackage{epsfig}
\usepackage{epsf}

\title[LBV diversity]{Luminous Blue Variable eruptions and related
  transients: Diversity of progenitors and outburst properties}

\author[N.\ Smith et al.]{Nathan Smith$^{1,2}$\thanks{Email:
    nathans@as.arizona.edu}, Weidong Li$^2$, Jeffrey M.\
  Silverman$^2$, Mohan Ganeshalingam$^2$, \newauthor \& Alexei V.\
  Filippenko$^2$ \\ $^1$Steward Observatory, University of Arizona,
  933 North Cherry Avenue, Tucson, AZ 85721, USA \\ $^2$Astronomy
  Department, University of California, Berkeley, CA 94720-3411, USA}

\begin{document}
\date{Accepted 0000, Received 0000, in original form 0000}
\pagerange{\pageref{firstpage}--\pageref{lastpage}} \pubyear{2009}
\def\arcdeg{\degr}
\maketitle
\label{firstpage}

\begin{abstract}

  We present new light curves and spectra for a number of
  extragalactic optical transients or ``supernova (SN) impostors''
  related to giant eruptions of luminous blue variables (LBVs), and we
  provide a comparative discussion of LBV-like giant eruptions known
  to date.  New data include photometry and spectroscopy of
  SNe~1999bw, 2000ch, 2001ac, 2002bu, 2006bv, and 2010dn.  SN~2010dn
  is a carbon copy of SN~2008S and NGC~300-OT, whereas SN~2002bu shows
  spectral evolution from a normal LBV at early times to a twin of
  these cooler transients at late times.  SN~2008S, NGC300-OT, and
  SN~2010dn appear to be special cases of a broader eruptive
  phenomenon where the progenitor star was enshrouded by dust, perhaps
  from a previous unseen eruptive episode.  Evidence suggests that
  their progenitors have initial masses in the range 10--20
  $M_{\odot}$, extending the range of masses susceptible to violent
  eruptive phenomenon below the canoncial LBV mass range.  Examining
  the full sample, SN impostors are characterized by strong
  photometric variability on a range of timescales from a day to
  decades, potentially suffering multiple eruptions of the same
  source.  The upper end of the luminosity distribution overlaps with
  the least luminous core-collapse SNe, but in most cases a
  distinction can be made based on spectra.  The low end of the
  luminosity distribution is far less well defined, and a distinction
  between LBV giant eruptions, S Doradus phases of LBVs, novae, and
  possible eruptions of intermediate-mass stars is not entirely clear.
  We discuss observational clues concerning stellar winds or shocks as
  the relevant mass-loss mechanism, and we evaluate possible ideas for
  the physical mechanism(s) of outbursts, but {\it there is still a
    great need for theoretical work on this problem}.  Although known
  examples of these eruptions are sufficient to illustrate their
  remarkably wide diversity in peak absolute magnitude, duration,
  progenitor stars, outburst spectra, and other observable properties,
  their statistical distribution is an area that will benefit greatly
  from upcoming transient surveys.  Based on the distribution of these
  eruptive properties, we propose that the prototypical object
  SN~1961V was not a member of this class of impostors after all, but
  was instead a true core-collapse SN~IIn that was preceded by a giant
  LBV eruption.
 
\end{abstract}


\section{INTRODUCTION}

This paper investigates observations of transient phenomena known
variously as luminous blue variable (LBV) eruptions, supernova (SN)
impostors, or other optical transients usually associated with massive
stars.  These are thought to be nonterminal eruptions or explosions
(i.e., not core-collapse) related to the extreme brightening events
observed in LBVs such as $\eta$~Carinae, although the physical
mechanism of the outbursts is not yet known.  The naming convention is
rather haphazard, with some earning official SN designations --- only
to be recognized later as ``impostors'' --- while others deemed
unworthy are demoted to generic optical transients at the time of
discovery.\footnote{While none of these names are ideal, we
  tentatively prefer ``LBV-like eruptions'', since it is based on an
  observationally established class of objects, while we remain
  cognizant of the possibility that LBV-like outbursts might also
  occur in cool (i.e., non-blue) stars like red supergiants, or stars
  that are not necessarily the most massive stars.  This paper
  attempts to provide a comparative study of the light curves and
  spectra for known examples of this class.  One must also be careful
  to distinguish between ``LBVs'' --- which refers to a particular
  class of variable stars, not all of which have been observed to
  suffer a giant eruption like $\eta$ Car --- and ``LBV-like
  eruptions'', which refers to the temporary brightening event that
  resembles the giant eruptions observed in LBVs like $\eta$ Car
  (i.e., not all LBVs have documented giant eruptions).}  When
designated as a SN, their spectra are classified as Type~IIn due to
the strong and narrow H~{\sc i} emission lines that arise from their
relatively slow winds or ejecta, typically moving at $\la$1000 km
s$^{-1}$.

Only two of these events have been witnessed in our own
galaxy,\footnote{If V838~Mon is a similar type of event, then it would
  be the third example in our galaxy.} both as historical naked-eye
transients: P~Cygni erupted in 1600~A.D., and $\eta$~Carinae suffered
its so-called Great Eruption in the mid 19th century.  While small in
number, these nearby events have had an enormous influence on our
understanding of the phenomenon, since their physical parameters are
relatively well constrained and we can verify that the stars survived
the eruptive events.  They are the only two outbursts where we can
directly measure the total mass ejected; analysis of their spatially
resolved circumstellar shells implies more than 10 $M_{\odot}$ in the
case of $\eta$ Carinae (Smith et al.\ 2003b) and only about 0.1
$M_{\odot}$ for P Cygni (Smith \& Hartigan 2006).  The radiated and
kinetic energy in these events also differed by more than two orders
of magnitude, so from just these two events we can already see a wide
diversity among the eruptions of LBVs, which is a major theme in this
paper.

Additional examples from nearby external galaxies are also known.
SN~1954J was the eruption of the bright blue irregular variable V12 in
NGC~2403 (Tammann \& Sandage 1968; Smith et al.\ 2001; Van Dyk et al.\
2005), and the famously weird object SN~1961V was originally
categorized as a Type V event (Zwicky 1964), but was later thought to
be an extreme version of a non-terminal eruption {\it a la} $\eta$ Car
(Goodrich et al.\ 1989; Filippenko et al.\ 1995).  Together with
P~Cygni and $\eta$ Car, these four historical LBV giant eruptions have
come to represent the class of SN impostors (Van Dyk 2005; Humphreys
et al.\ 1999).\footnote{As we argue in this paper, however, SN~1961V
  may be a true core-collapse SN~IIn event.  One day before submission
  of this paper, we learned that Kochanek et al.\ (private comm.)
  simultaneously reached a similar conclusion about SN~1961V based on
  the lack of an expected mid-IR counterpart.}  The eclipsing binary
HD~5980 (the most luminous star in the SMC) and V1 in NGC~2366 both
suffered eruptions in the mid 1990s, and over a dozen more examples
have been discovered in the past decade in the course of various SN
searches.  A list of these events is provided later in the paper.

LBVs are thought to be massive stars that are unstable because they
have reached a point in their evolution where they are dangerously
close to the classical Eddington limit, partly due to core evolution
and partly to mass loss in preceding phases (see e.g., Smith \& Conti
2008).  It was suggested long ago that cool temperatures in the
stellar envelopes may lead to an opacity-modified Eddington limit that
may play a role in initiating the outbursts (Lamers \& Fitzpatrick
1988; Appenzeller 1986), but further progress on the physical
mechanism causing LBV eruptions has been slow to enter the refereed
literature.  Most theoretical work on LBV eruptions so far has
focussed on the physics of driving powerful winds in quasi-steady
state when a star exceeds the Eddington limit (e.g., Shaviv 2000;
Owocki et al.\ 2004; Owocki 2005; Owocki \& van Marle 2007; van Marle
et al.\ 2008, 2009).  With the high mass-loss rates required for LBV
eruptions, the material must be optically thick and therefore
continuum driven or hydrodynamically launched, rather than line driven
(Smith \& Owocki 2006; van Marle et al.\ 2008).  For this reason,
these super-Eddington continuum-driven winds are of interest as a
potential mode of mass loss at low metallicity (Smith \& Owocki 2006).
Although the underlying mechanism behind the increased luminosity
remains unknown, the massive shells seen around many LBVs with nebular
masses of a few to 20 $M_{\odot}$, combined with the fact that these
episodes appear to recur, argue that the episodic ejection of the H
envelope in LBV eruptions is a dominant mode of mass loss for massive
stars (Smith \& Owocki 2006).

The traditional explanation for LBV eruption light curves in
historical examples (e.g., Humphreys et al. 1999; Humphreys \&
Davidson 1994) has been that a massive star increases its bolometric
luminosity output and then reaches or exceeds the classical Eddington
limit; this initiates catastrophic mass loss.  Dust condensation in
the ejected shell eventually obscures the star and causes the object
to fade at visual wavelengths.  Humphreys et al.\ (1999) suggested
that these ``giant eruptions'' are different from the more typical ``S
Doradus variability'' exhibited by LBVs in that their bolometric
luminosity increases, whereas the visual brightening in a normal S
Doradus episode is thought to be caused by a change in bolometric
correction at constant luminosity.  The traditional view has been that
LBVs should be relatively cool in their bright phases, exhibiting an F
supergiant-like spectrum (Humphreys \& Davidson 1994).  Modern
observations are revealing that these and other characterizations of
LBV eruptions, which are based on few examples, are not necessarily
true for the class, and so our understanding of these events is still
developing as we discover additional examples.


A qualitative shift in interpreting LBV giant eruptions came with the
recent recogition that strong shock waves may also play a role in some
of the outbursts.  This became apparent following the discovery of
very fast ejecta surrounding $\eta$~Carinae (Smith 2008), but it had
been suspected earlier based on the rough equipartition in the kinetic
and radiated energy budgets of its 19th century giant eruption (Smith
et al.\ 2003b).  Smith (2008) suggested that we may expect to see
X-rays or radio emission from some LBV eruptions, and that this
evidence for a shock would not necessarily implicate a core-collapse
event.  Since then, Dessart et al.\ (2010) have explored weak
explosions as a possible mechanism for some SN impostor events, and
additional observational evidence for an explosive component in LBV
eruptions is accumulating.  In particular, Smith et al.\ (2010a)
proposed that the fast ($\sim$5000 km s$^{-1}$) ejecta seen in
absorption in SN~2009ip may result from an explosion similar to that
inferred for $\eta$ Car, and that shock excitation may be important in
explaining some of the diversity among spectral properties of LBV
eruptions.  Preliminary reports of a high X-ray luminosity in the very
recent LBV eruption SN~2010da (Immler et al.\ 2010) may also suggest
the influence of a shock, but this new object is still being studied
at the time of writing (see below).  The influence of both shocks and
super-Eddington winds on observations of LBV eruptions was discussed
in detail by Smith et al.\ (2010a).  Although shocks may play a role
in a few cases, strong super-Eddington winds must operate in many of
the LBV eruptions.


Another key development in our interpretation of these eruptions is
that their progenitors may be substantially more diverse than
previously recognized.  SN~2008S and the 2008 optical transient in
NGC~300 (N300-OT hereafter) were similar in their observed properties
to other SN impostors, but Prieto and collaborators (Prieto 2008;
Prieto et al.\ 2008; Thompson et al.\ 2009) discovered that their
progenitors were faint and heavily obscured.  While only upper limits
were available for visual wavelengths, archival {\it Spitzer} data
suggested IR luminosities of $\la$10$^{4.9}$ $L_{\odot}$ before the
eruptions.  If the progenitor stars were cool, their observed IR
luminosities could be consistent with initial masses as low as 8--10
$M_{\odot}$, suggesting the intriguing possibility that these
eruptions might be associated with weak electron capture SNe in
extreme AGB stars (Thompson et al.\ 2009; Botticella et al.\ 2009), or
that they may be associated with obscured OH/IR stars (see also Khan
et al.\ 2010a).  On the other hand, if they were heavily obscured
supergiant stars, their IR luminosities would imply initial masses of
10--20 $M_{\odot}$ (Smith et al.\ 2009a, 2010a; Bond et al.\ 2009;
Berger et al.\ 2010).  Smith et al.\ (2010a) discussed this debate in
detail, showing that the IR luminosity of N300-OT, for example, was
quite similar to the progenitor luminosity of V12/SN~1954J.  For the
nearby case of N300-OT, at least, studies of the surrounding stellar
population favor an initial mass of 12-25 $M_{\odot}$ (Gogarten et
al.\ 2009), apparently ruling out the low-mass option.  In any case,
the progenitors of SN~2008S and N300-OT were probably less massive
than classical LBVs, which were thought to extend down to initial
masses of only 20--25 $M_{\odot}$ (Smith et al.\ 2004).

Initial masses below $\sim$20 $M_{\odot}$ for some of these events
have rather profound implications for the larger class of SN impostors
and LBV eruptions, because stars of this mass are not expected to
approach or exceed the Eddington limit during the normal course of
their post-main sequence evolution.  Together with evidence for
explosive shock waves described above, this seems to favor a
deep-seated energy injection, rather than a runaway near-Eddington
instability in the outer envelope.  Furthermore, if being dangerously
near the Eddington limit is {\it not} a necessary precondition for
these eruptions after all, then the same (or a related) mechanism that
drives giant eruptions of luminous stars like $\eta$ Car might also
operate in lower mass stars as well, perhaps even below 8 $M_{\odot}$.
In this context, relieved of the notion that LBV-like eruptions are
exclusive to the most massive stars, it is prudent to explore the
diversity in this class of non-terminal stellar eruptions.  As the
astronomical community embarks upon an era of more intensive transient
studies, more examples will hopefully illuminate and quantify the
statistical distribution across this diverse range of properties.

In this paper we collect examples of LBV giant eruptions known to
date, examining their light curves, spectra, and several derived
properties.  In \S 2 we present some unpublished data on previous SN
impostors as well as some recent examples.  In \S 3 we compile a list
of known events and present their light curves and spectra, and we
provide a detailed comparative discussion of their various
observational properties.  In \S 4 we discuss the diversity of the
sample and its implications for the physics behind these eruptions.
We also briefly discuss overlap with transients that may be related
but have not been considered as LBV eruptions so far, and discuss
which objects should belong in the class.

\begin{table}\begin{center}\begin{minipage}{3.1in}
\caption{New photometry for SN~1999bw}\scriptsize
\begin{tabular}{@{}lcccc}\hline\hline
JD          &$B$ mag &$V$ mag &$R$ mag  &$I$ mag \\ 
\hline
2451289.69  &19.27$\pm$0.14 &18.45$\pm$0.06	&17.99$\pm$0.07	&17.60$\pm$0.10 \\
2451291.70  &...            &18.36$\pm$0.05	&17.98$\pm$0.08	&17.67$\pm$0.06 \\
2451292.69  &...            &18.39$\pm$0.14	&17.87$\pm$0.12	&17.69$\pm$0.13 \\
2451295.72  &...            &18.37$\pm$0.10	&...            &17.83$\pm$0.30 \\
2451298.72  &...            &18.33$\pm$0.08	&17.82$\pm$0.08	&17.72$\pm$0.11 \\
\hline
\end{tabular}\label{tab:phot99bw}
\end{minipage}\end{center}
\end{table}

\begin{table}\begin{center}\begin{minipage}{3.1in}
\caption{New photometry for SN~2001ac}\scriptsize
\begin{tabular}{@{}lcccc}\hline\hline
JD          &$B$ mag        &$V$ mag            &$R$ mag        &$I$ mag \\ 
\hline
2451981.79  &18.77$\pm$0.11 &18.53$\pm$0.08	&18.53$\pm$0.08	&17.88$\pm$0.07 \\
2451982.76  &18.70$\pm$0.11 &18.71$\pm$0.07	&18.71$\pm$0.07	&18.15$\pm$0.11 \\
2451986.81  &19.09$\pm$0.12 &18.91$\pm$0.09	&18.91$\pm$0.09	&18.21$\pm$0.15 \\
2451990.79  &18.78$\pm$0.33 &18.61$\pm$0.30	&18.61$\pm$0.30 &...  \\
2451994.81  &19.68$\pm$0.24 &19.24$\pm$0.12	&19.24$\pm$0.12	&18.60$\pm$0.16 \\
2451998.78  &19.81$\pm$0.24 &19.13$\pm$0.11	&19.13$\pm$0.11	&18.54$\pm$0.13 \\
2452009.74  &19.90$\pm$0.23 &19.80$\pm$0.29	&19.80$\pm$0.29	&18.97$\pm$0.27 \\
2452013.76  &20.59$\pm$0.36 &19.91$\pm$0.27	&19.91$\pm$0.27	&19.16$\pm$0.28 \\
\hline
\end{tabular}\label{tab:phot01ac}
\end{minipage}\end{center}
\end{table}

\begin{table}\begin{center}\begin{minipage}{3.1in}
\caption{New photometry for SN~2002bu}\scriptsize
\begin{tabular}{@{}lcccc}\hline\hline
JD          &$B$ mag &$V$ mag  &$R$ mag  &$I$ mag \\ 
\hline
2452363.88  &16.20$\pm$0.03	 &15.60$\pm$0.02	&15.13$\pm$0.02	&...            \\
2452364.84  &15.98$\pm$0.02	 &15.46$\pm$0.02	&15.03$\pm$0.02	&14.70$\pm$0.02 \\
2452365.77  &15.84$\pm$0.02	 &15.33$\pm$0.02	&14.99$\pm$0.02	&14.65$\pm$0.02 \\
2452366.83  &15.74$\pm$0.02	 &15.26$\pm$0.02	&14.93$\pm$0.02	&14.62$\pm$0.02 \\
2452368.88  &15.67$\pm$0.02	 &15.23$\pm$0.02	&14.92$\pm$0.02	&14.65$\pm$0.02 \\
2452372.83  &15.71$\pm$0.02	 &15.34$\pm$0.02	&15.02$\pm$0.02	&14.80$\pm$0.02 \\
2452375.82  &15.84$\pm$0.02	 &15.42$\pm$0.02	&15.13$\pm$0.02	&14.86$\pm$0.02 \\
2452377.81  &15.89$\pm$0.02      &15.47$\pm$0.02	&15.20$\pm$0.02	&14.92$\pm$0.02 \\
2452380.76  &16.01$\pm$0.02      &15.57$\pm$0.02	&15.26$\pm$0.02	&14.97$\pm$0.02 \\
2452383.82  &16.14$\pm$0.02	 &15.63$\pm$0.02	&15.34$\pm$0.02	&15.04$\pm$0.02 \\
2452386.77  &16.24$\pm$0.02	 &15.71$\pm$0.02	&15.42$\pm$0.02	&15.10$\pm$0.02 \\
2452389.77  &16.36$\pm$0.03	 &15.74$\pm$0.03	&15.46$\pm$0.04	&15.10$\pm$0.02 \\
2452392.82  &16.30$\pm$0.05	 &15.77$\pm$0.02	&15.41$\pm$0.03	&15.05$\pm$0.02 \\
2452396.76  &16.50$\pm$0.02	 &15.80$\pm$0.02	&15.45$\pm$0.02	&15.06$\pm$0.02 \\
2452399.73  &16.52$\pm$0.02      &15.84$\pm$0.02	&15.48$\pm$0.02	&15.07$\pm$0.02 \\
2452403.72  &16.61$\pm$0.03	 &15.88$\pm$0.02	&15.48$\pm$0.02	&15.07$\pm$0.02 \\
2452411.70  &16.80$\pm$0.02	 &16.03$\pm$0.02	&15.59$\pm$0.02	&15.19$\pm$0.02 \\
2452419.70  &17.06$\pm$0.04	 &16.26$\pm$0.02	&15.74$\pm$0.02	&15.30$\pm$0.02 \\
2452423.69  &17.17$\pm$0.02	 &16.34$\pm$0.02	&15.85$\pm$0.02	&15.38$\pm$0.02 \\
2452427.70  &17.38$\pm$0.02	 &16.53$\pm$0.02	&16.03$\pm$0.02	&15.55$\pm$0.02 \\
2452431.74  &...                 &16.71$\pm$0.03	&16.17$\pm$0.02	&15.66$\pm$0.03 \\
2452438.69  &17.84$\pm$0.04	 &17.05$\pm$0.03	&16.40$\pm$0.03	&15.93$\pm$0.02 \\
2452445.70  &18.22$\pm$0.04	 &17.45$\pm$0.05	&16.81$\pm$0.03	&16.24$\pm$0.03 \\
2452452.70  &18.60$\pm$0.09	 &17.90$\pm$0.05	&17.15$\pm$0.05	&16.63$\pm$0.04 \\
\hline
\end{tabular}\label{tab:phot02bu}
\end{minipage}\end{center}
\end{table}

\begin{table}\begin{center}\begin{minipage}{2.0in}
\caption{Unfiltered KAIT photometry for SN~2006bv}\scriptsize
\begin{tabular}{@{}lcc}\hline\hline
JD          &mag &err \\ 
\hline
2453852.80  &18.51   &0.15 \\
2453864.80  &18.14   &0.07 \\
2453887.74  &18.70   &0.07 \\
\hline
\end{tabular}\label{tab:phot06bv}
\end{minipage}\end{center}
\end{table}

\begin{table}\begin{center}\begin{minipage}{2.0in}
   \caption{Additional unfiltered KAIT photometry for U2773-OT}\scriptsize
\begin{tabular}{@{}lcc}\hline\hline
JD          &mag &err \\ 
\hline
2455181.5 &17.64   &0.03 \\
2455184.5 &17.57   &0.03 \\
2455188.5 &17.57   &0.03 \\
2455191.5 &17.60   &0.03 \\
2455199.5 &17.53   &0.03 \\
2455202.5 &17.58   &0.03 \\
2455205.5 &17.56   &0.05 \\
2455211.5 &17.48   &0.05 \\
2455227.5 &17.50   &0.03 \\
2455238.5 &17.42   &0.04 \\
2455241.5 &17.41   &0.03 \\
2455244.5 &17.49   &0.03 \\
2455256.5 &17.46   &0.03 \\
2455266.5 &17.41   &0.03 \\
2455269.5 &17.37   &0.03 \\
2455272.5 &17.43   &0.05 \\
\hline
\end{tabular}\label{tab:photU2773}
\end{minipage}\end{center}
\end{table}

\begin{figure}\begin{center}
\includegraphics[width=3.1in]{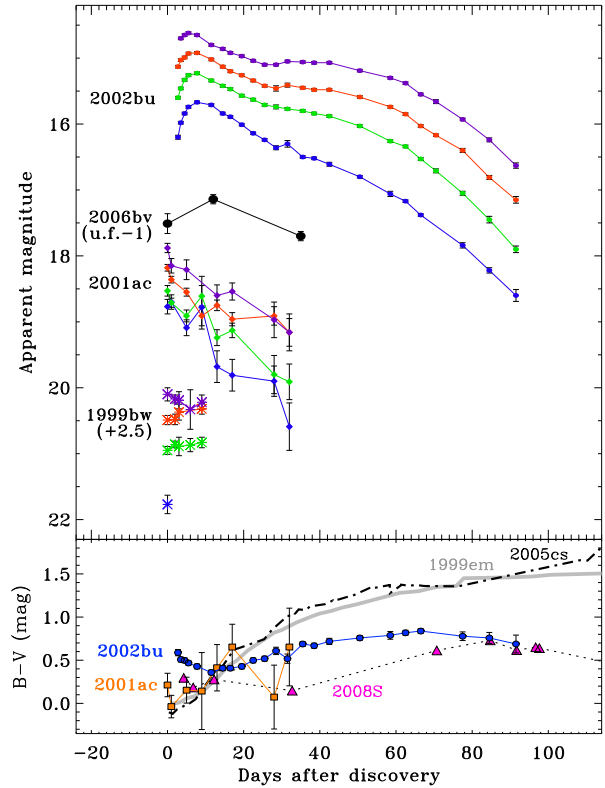}
\end{center}
\caption{{\it Top:} Apparent magnitude light curves for SN impostor
  photometry from KAIT, reported here for the first time (see
  Tables~\ref{tab:phot99bw}, \ref{tab:phot01ac}, \ref{tab:phot02bu},
  and \ref{tab:phot06bv}).  For SN~2002bu (filled dots) and SN~2001ac
  (filled diamonds), apparent $BVRI$ magnitudes obtained with KAIT are
  shown as observed; $B$, $V$, $R$, and $I$ are plotted as blue,
  green, red, and purple, respectively.  For clarity of display, the
  $BVRI$ photometry for SN~1999bw (asterisks) is shown offset by +2.5
  mag.  Unfiltered photometry for SN~2006bv (black dots) is offset by
  $-$1 mag.  {\it Bottom:} $B-V$ color curves of SN~2002bu (blue
  circles) and 2001ac (orange squares), compared to color curves of
  the SN impostor SN~2008S (magenta triangles; Smith et al.\ 2009a),
  the normal SN~II-P 1999em (gray; Leonard et al.\ 2002), and the
  faint SN~II-P 2005cs (black dot-dashed; Pastorello et al.\ 2009).}
\label{fig:phot}
\end{figure}

\section{New Observations}

For new observational material on SN impostors, our data were
collected as part of the Lick Observatory Supernova Search
(LOSS). Most of our photometry comes from the Katzman Automatic
Imaging Telescope (KAIT; Filippenko et al.\ 2001; Filippenko 2003),
while our new spectra listed below were obtained using the Kast
spectrograph (Miller \& Stone 1993) on the Shane 3m Reflector at Lick
Observatory, or at the 10m Keck Observatory using the Low Resolution
Imaging Spectrograph (LRIS; Oke et al.\ 1995) or the Deep Imaging
Multi-Object Spectrograph (DEIMOS; Faber et al.\ 2003). Details about
the new data are given in subsequent sections.

\begin{table}\begin{center}\begin{minipage}{3.1in}
\caption{New spectroscopy of SN impostors}\scriptsize
\begin{tabular}{@{}llrlccc}\hline\hline
Transient  &Obs.\ Date   &Day &Tel./Inst.  &$\delta\lambda$ (\AA) \\ 
\hline
SN~1999bw  &1999 Apr 24  &4   &Lick/Kast   &4300-7000    \\
SN~2000ch  &2000 May 31  &28  &Lick/Kast   &4250-6950    \\
SN~2000ch  &2004 Apr 26  &1456&Keck/LRIS   &3300-9400    \\
SN~2001ac  &2001 Mar 21  &9   &Lick/Kast   &3300-7830    \\
SN~2001ac  &2001 Mar 29  &17  &Keck/LRIS   &4350-6860    \\
SN~2002bu  &2002 Apr 08  &11  &Lick/Kast   &3300-10400   \\
SN~2002bu  &2002 Apr 20  &23  &Lick/Kast   &3300-10400   \\
SN~2002bu  &2002 May 07  &40  &Lick/Kast   &3300-10400   \\
SN~2002bu  &2002 Jun 08  &72  &Lick/Kast   &3100-10400   \\
SN~2002bu  &2002 Jun 17  &81  &Lick/Kast   &3100-10400   \\
SN~2010dn  &2010 Jun 08  &9   &Lick/Kast   &3430-10260   \\
SN~2010dn  &2010 Jun 11  &12  &Keck/DEIMOS &6101-7410    \\
SN~2010dn  &2010 Jun 18  &19  &Lick/Kast   &3510-9920    \\
\hline
\end{tabular}\label{tab:spec}
\end{minipage}\end{center}
\end{table}

\begin{figure*}\begin{center}
\includegraphics[width=4.8in]{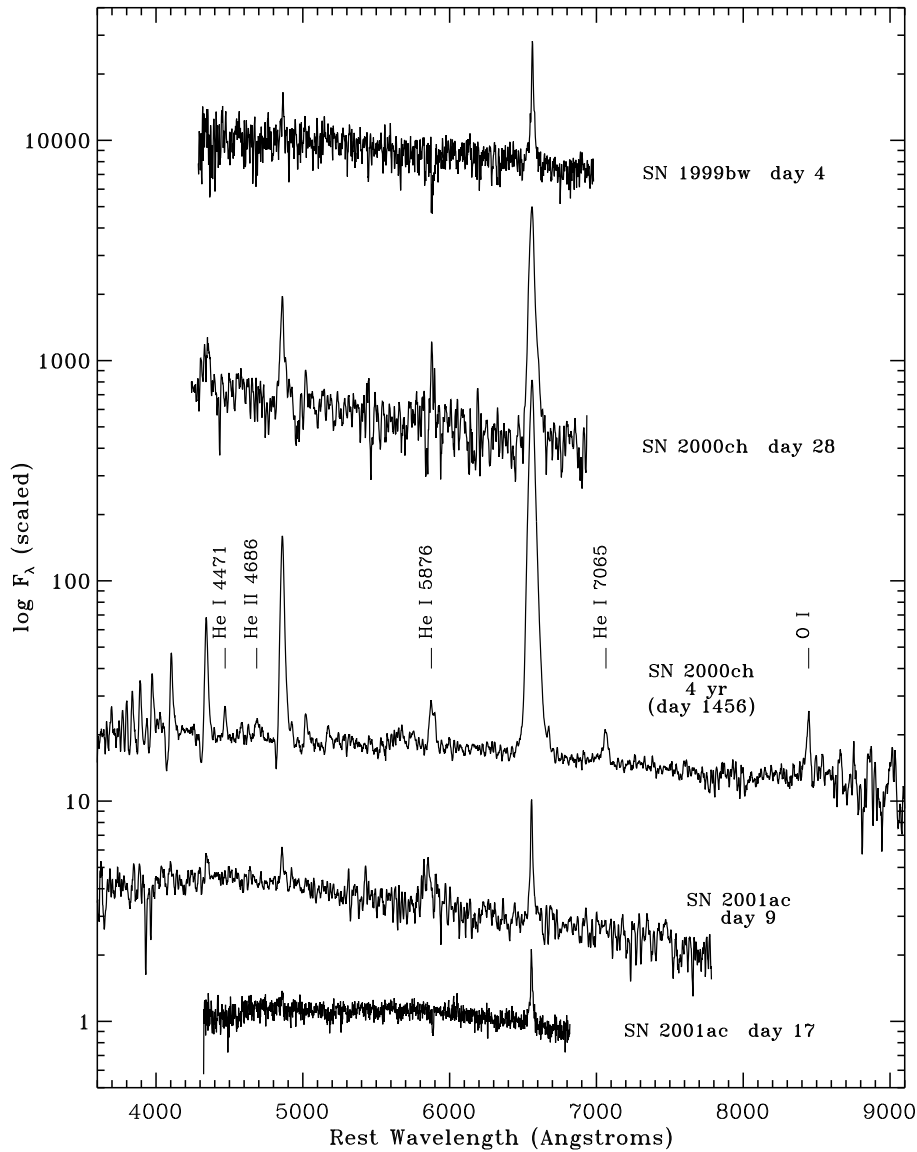}
\end{center}
\caption{Spectra of SN impostors obtained with the Lick 3m reflector
  (see Table~\ref{tab:spec}).  The day 28 spectrum of SN~2000ch was
  published previously by Wagner et al.\ (2004), but the others are
  previously unpublished.  SN~1999bw on day 4 is a fairly noisy
  spectrum dominated by Balmer lines.  The late-time spectrum of
  SN~2000ch obtained about 4 yr after discovery shows interesting
  changes from the earlier spectrum and covers a wider wavelength
  range.  The two spectra of SN~2001ac on days 9 and 17 show
  interesting evolution of the spectrum over a short time, where the
  broad He~{\sc i} $\lambda$5876 line disappears and the Balmer lines
  fade.}
\label{fig:spec99bw}
\end{figure*}

\subsection{New Photometry}

Optical photometry of the SN impostors were obtained with
KAIT. Several objects were followed in multiple passbands ($BVRI$)
soon after discovery. For several other objects, no dedicated followup
campaign was initiated, but their host galaxies were monitored without
using a filter during the course of our SN search, so we have
unfiltered data of the eruptions as a byproduct.  For the objects with
multi-color $BVRI$ photometry, we obtained calibrations of the fields
by observing them together with several Landolt (1992) star fields at
various airmasses in photometric nights. Deep template images of the
fields after the objects have faded beyond detection have also been
obtained. These template images and calibrations are then used in the
KAIT photometry pipeline (Ganeshalingam et al.\ 2010) to perform image
subtraction on the image data and calibration to the standard
photometry system.

For the objects with only unfiltered data, we treat the images as
taken with the $R$ band (Li et al.\ 2003).  Template images are
constructed for each field by choosing the best monitoring data and
then stacking them.  For photometric calibration, we use the red
magnitudes for the stars in the SN fields in the USNO~B1 catalog
(Monet et al.\ 2003).  Although the accuracy of this calibration is
only $\sim$0.2$-$0.3 mag for an individual star, there are usually
more than 10 stars available in each field, so the uncertainty due to
calibration is $< 0.1$ mag. The data are then reduced in a similar
fashion as the KAIT photometry pipeline.  The final photometry of the
objects are listed in Tables 1$-$5 and the apparent light curves are
shown in Figure~\ref{fig:phot}.

{\it SN 1999bw:} Unfortunately, SN~1999bw was not extensively observed by
KAIT, and the luminosity appears relatively constant over the $\sim$10
days when it was observed.  The apparent $B-V$ color at the time of
discovery is $\sim$0.8 mag, suggesting either that the eruption was
redder than a normal LBV, or that it suffered significant
circumstellar reddening.

{\it SN 2001ac:} Our KAIT $BVRI$ photometry of SN~2001ac covers about one
month after discovery, and seems consistent with a relatively fast and
steady decline (within uncertainty), fading by $\sim$1--1.8 mag (in
various filters) in 30 days.  This suggests that it may have been discovered
after the time of peak luminosity.  The apparent $B-V$ color evolves
only mildly during this time, from $\la$0.2 to $\sim$0.6 mag toward
the end of the observed epoch.  The increasing relative strength of
the $R$-band compared to the others at late times may result from the
very strong H$\alpha$ seen in the spectrum (see below).

{\it SN 2002bu:} The bright eruption of SN~2002bu was well observed by
KAIT.  We started observing it photometrically about 5 days before
maximum light, and followed it for almost 100 d thereafter when it had
faded by about 3 mag.  The light curve of SN~2002bu shows an initial
10--20 day rounded peak, followed by a ``hump'' (i.e., almost a
plateau) with a subsequent slower rate of decline; qualitatively, this
decline with a change in decay rate resembles that of SN~1997bs (Van
Dyk et al.\ 2000).  The apparent color reddens with time, from $B-V$
$\approx$ 0.45 mag at peak, to $\sim$0.8 mag at late times
(Figure~\ref{fig:phot}), similar to the color evolution of SN~2008S
(Smith et al.\ 2009a).  The color evolution is substantially different
from a normal SN~II-P, never getting as red as a SN~II-P and
apparently becoming slightly blue again as the object fades.

\begin{figure*}\begin{center}
\includegraphics[width=5.1in]{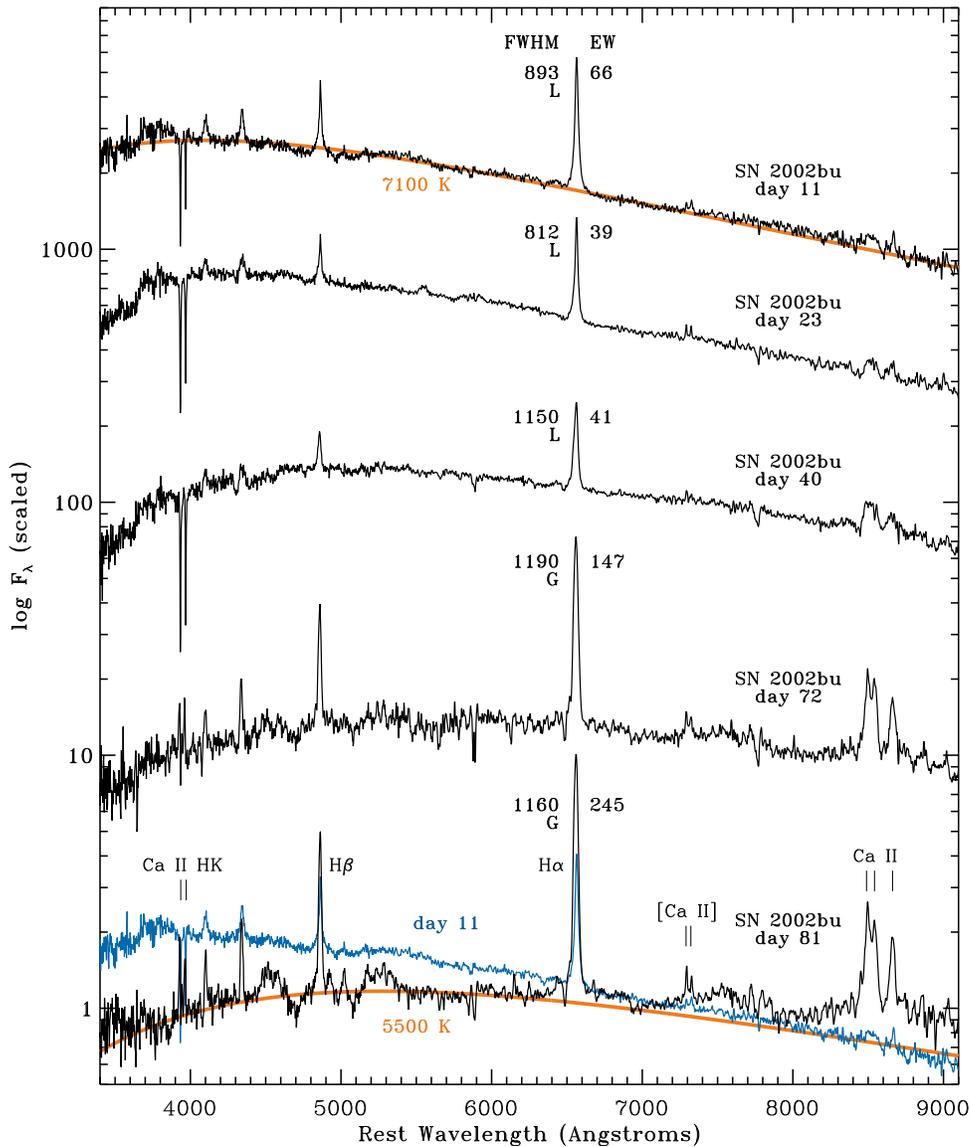}
\end{center}
\caption{Previously unpublished Lick/3m spectra of SN~2002bu on days
  11, 23, 40, 72, and 81.  The continuum shape gets redder with time,
  and the Balmer emission line equivalent widths get stronger as the
  continuum fades.  The FWHM (plus either ``G'' for Gaussian or ``L''
  for Lorentzian) and equivalent width of H$\alpha$ are listed aside
  the emission line for each epoch.  The spectrum transitions from a
  ``hot'' LBV at early times to a ``cool'' LBV at late times, with the
  red [Ca~{\sc ii}] doublet and the IR Ca~{\sc ii} triplet
  strengthening, while Ca~{\sc ii} H and K go from absorption to
  emission.  The blue tracing at the bottom is the day 11 spectrum
  plotted over the last day 81 spectrum to emphasize the changes in
  continuum shape and line intensities.  The orange curves show
  blackbodies. All epochs have been dereddened by the same value of
  $E(B-V)$ = 0.012 mag (i.e. correcting for Galactic reddening, but
  not any additional reddening that may be local, so the blackbody
  temperatures shown are lower limits).}
\label{fig:spec02bu}
\end{figure*}

{\it SN 2003gm:} We obtained only two unfiltered KAIT measurements of
SN~2003gm, including the discovery and one image 6 days later.  Both
were 17.0$\pm$0.1 mag; the limited light curve is not shown.

{\it SN 2006bv:} We obtained three unfiltered measurements of
SN~2006bv with KAIT, as listed in Table~\ref{tab:phot06bv} and shown
in Figure~\ref{fig:phot}, where it has been shifted by $-$1 mag for
clarity of display.  The peak occurred a few to 20 days after
discovery.  Unfortunately, no late-time measurements are available,
and we were not able to secure spectra of the eruption.

{\it U2773-OT:} We presented KAIT unfiltered photometry of this 2009
transient in UGC~2773 (U2773-OT hereafter) in Smith et al.\ (2010a),
but the transient has remained bright and has even continued its slow
rise in the year since then.  Table~\ref{tab:photU2773} gives
additional unfiltered KAIT photometry for this source, continuing
after the last data point in the previous paper.  See Smith et al.\
(2010a) for further details.


\subsection{Previously Unpublished Spectra}

All spectra were reduced using standard techniques (e.g., Foley et
al.\ 2003).  Routine CCD processing and spectrum extraction were
completed with the Image Reduction and Analysis Facility (IRAF), and
the data were extracted with the optimal algorithm of Horne (1986).
We obtained the wavelength scale from low-order polynomial fits to
calibration-lamp spectra.  Small wavelength shifts were then applied
to the data after cross-correlating a template sky to the night-sky
lines that were extracted with the SN.  Using our own reduction
routines, we fit spectrophotometric standard-star spectra to the data
in order to flux calibrate our spectra and to remove telluric lines
(Wade \& Horne 1988; Matheson et al.\ 2000). Most observations were
aligned along the parallactic angle to reduce differential light
losses (Filippenko 1982). Information regarding both our photometric
and spectroscopic data (such as observing conditions, instrument,
reducer, etc.) was obtained from our SN database (SNDB).  The SNDB
uses the popular open-source software stack known as LAMP: the Linux
operating system, the Apache webserver, the MySQL relational database
management system, and the PHP server-side scripting language (for
further details, see Silverman et al.\ 2010).

{\it SN 1999bw:} We were only able to obtain one spectrum of SN~1999bw
shortly after discovery on day 4, and unfortunately the spectrum is
rather noisy.  It shows the strong narrow Balmer emission lines
characteristic of LBVs.  The H$\alpha$ line has a Lorentzian shape
with a FWHM $\approx$ 630 km s$^{-1}$, but the broad wings extend to
roughly $\pm$3000 km s$^{-1}$ in our data.  This may be due to
electron scattering, but may also suggest that some of the mass is
moving rather fast, similar to SN~2009ip (Smith et al.\ 2010a) and
$\eta$ Car (Smith 2008).  No P~Cygni absorption is seen at this low
resolution in H$\alpha$.  Aside from H$\beta$, no other emission
features are seen in this wavelength range, but Na~{\sc i} D
absorption is present.

{\it SN 2000ch:} The light curve and spectra of SN~2000ch were already
discussed in detail by Wagner et al.\ (2004).  However, we obtained an
additional high signal-to-noise, late-time spectrum after that paper
was published.  The new spectrum of SN~2000ch in
Figure~\ref{fig:spec99bw} was obtained roughly 4 yr after discovery,
on 2004 April 26 using the Lick 3m reflector.  Even at this late time,
the spectrum still shows relatively broad (FWHM $\approx$ 1,500 km
s$^{-1}$) strong Balmer emission lines as well as prominent triplet
He~{\sc i} lines and even He~{\sc ii} $\lambda$4686.  O~{\sc i}
$\lambda$8446 is also seen.  The Balmer lines have strengthened
relative to the continuum, with about twice the equivalent width
compared to day 28.  In the higher quality spectrum we can now see
clear P~Cyg absorption features in the higher Balmer lines.  This 2004
spectrum is now quite valuable, because Pastorello et al.\ (2010) just
recently reported the discovery of multiple subsequent eruptions of
the same star that produced SN~2000ch, but much later in 2008 and
2009.  According to their photometry, our new spectrum showing a very
strong H$\alpha$ line with an emission equivalent width of 461 \AA\
was obtained at relative quiescence about halfway between the 2000 and
2008 eruptions.  It suggests that the wind speed during quiescence is
similar to that during the eruptive states.  Most of the same spectral
features (i.e. He~{\sc i} emission lines, etc), are also seen in
spectra of the subsequent outbursts.

{\it SN 2001ac:} The visual spectrum of SN~2001ac has a blue continuum
and narrow Balmer emission lines typical of LBVs.  The spectra are
rather noisy, so we cannot comment on many details.  One interesting
aspect is that over a relatively short time period of about a week,
between days 9 and 17, the prominent and broad emission feature at
around 5800 \AA \ disappears.  This could be a blueshifted emission
line of He~{\sc i} $\lambda$5876 from some hot and fast ejecta seen at
early times, but this is speculative with such a noisy spectrum.
After this broad emission fades, the spectrum closely resembles that
of SN~1999bw. Like many LBVs, SN~2001ac's H$\alpha$ line shows a
composite profile, with a narrow core that can be approximated with a
Gaussian FWHM $\approx$ 287 km s$^{-1}$ on day 9, but also with a
broader base that can be fit with a Gaussian with FWHM $\approx$ 1505
km s$^{-1}$ and extending to roughly $\pm$1500 km s$^{-1}$ at the
continuum level.  The emission equivalent width on day 9 is 46 \AA.
The day 17 H$\alpha$ profile is very similar, although Balmer lines
weaken and H$\beta$ disappears with time.  By comparison with previous
events, this probably indicates a relatively slow outflow speed of
around 290 km s$^{-1}$ for the bulk of the ejecta or wind, whereas the
broader base may be due to electron scattering wings.  It is of course
difficult to rule out the presence of unseen fast material, however.

\begin{figure}\begin{center}
\includegraphics[width=2.9in]{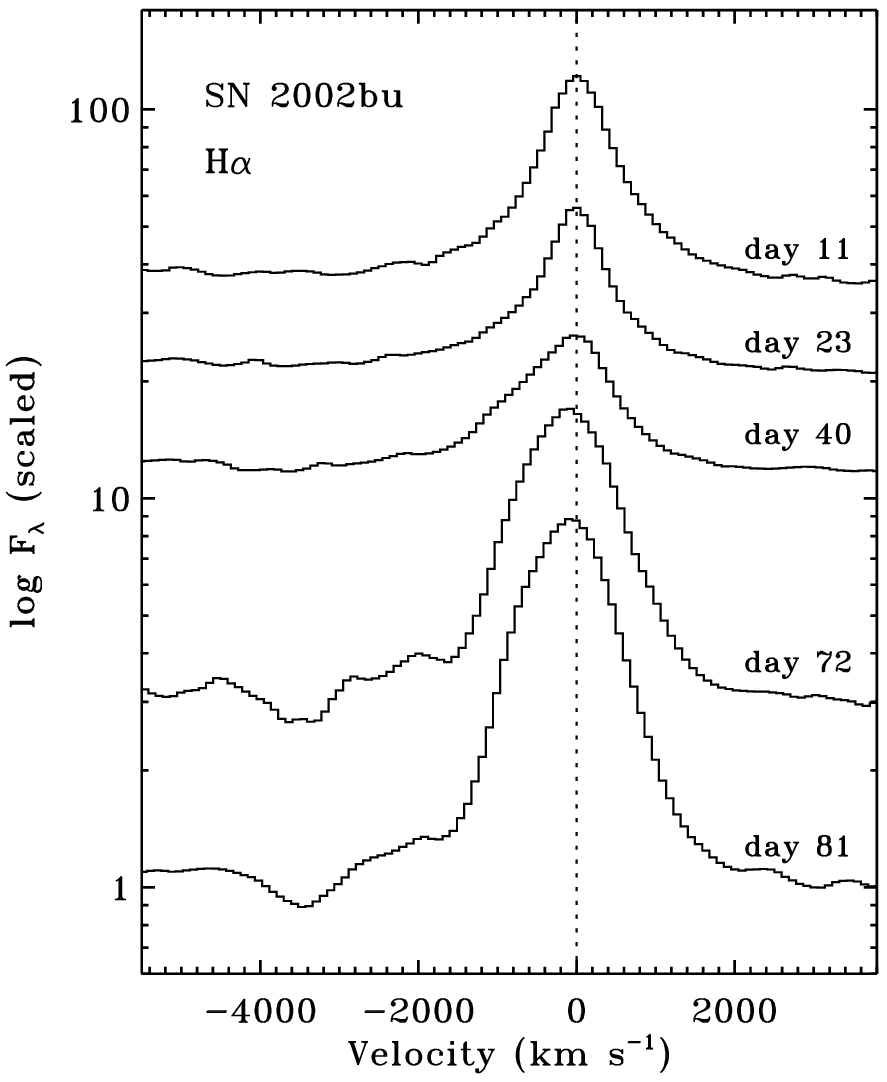}
\end{center}
\caption{Kast/Lick-3m spectra of the H$\alpha$ line velocity profile
  in SN~2002bu on days 11, 23, 40, 72, and 81.}
\label{fig:vel02bu}
\end{figure}

{\it SN 2002bu:} Our spectral coverage of SN~2002bu is much better
than the previous cases, with 5 epochs over the first $\sim$100 days
when we also have photometry.  Figure~\ref{fig:spec02bu} shows all 5
spectra, where we repeat our first spectrum (day 11) in blue at the
bottom for direct comparison with the late time (day 81) spectrum.

The most remarkable aspect of SN~2002bu's spectrum is its evolution
over time.  As the transient fades during the first $\sim$80 days, the
continuum gets substantially redder, while emission lines from the
[Ca~{\sc ii}] doublet and the Ca~{\sc ii} IR triplet strengthen
relative to the continuum.  Ca~{\sc ii} H and K transition from strong
absorption features at early times to narrow emission features at late
times, and a more complex absorption spectrum is evident in the last
two epochs on days 72 and 81.  Additionally, the H$\alpha$ line
profile (Figure~\ref{fig:vel02bu}) shows a change from a Lorentzian
profile for the first three epochs, similar to SN~2009ip (Smith et
al.\ 2010a), to a more Gaussian profile with an asymmetric shape.  The
red wing of H$\alpha$ appears to weaken at late times, perhaps
indicating the blueshift of lines that results when new dust formation
blocks emission from receding parts of the ejecta or CSM interaction
region as seen in some SNe~IIn (see, e.g., Smith et al.\ 2009b).
Given the dusty shells resolved around Galactic LBVs like $\eta$ Car,
dust formation in an eruptive event would not be surprising, although
direct evidence for it has been scant so far.  In the last spectrum on
day 81, the continuum cannot be fit with a single blackbody.  At
$\lambda < 7000$ \AA \ it appears well fit by a 5500 K blackbody, but
in this case the excess emission at longer wavelengths implies an IR
excess, perhaps due to hot dust emission.  The IR excess could be
caused by newly formed dust or an IR echo (or both; Fox et al.\ 2010;
Smith et al.\ 2009b), but the formation of new dust is consistent with
the H$\alpha$ line profile evolution.

Overall, the observed spectral changes signify a transition from a
spectrum that at early times resembles hot SN impostors with smooth
continua and strong Balmer lines like SN~1997bs and SN~2009ip, to one
that at late times looks cooler and develops the strong [Ca~{\sc ii}]
lines seen in SN~2008S and NGC~300-OT.  Smith et al.\ (2010a)
discussed the dichotomy of these ``hot'' and ``cool'' spectra in
various LBVs, but here {\it in SN~2002bu we see them both in the same
  object over time}.  A transition such as this could be quite common
among LBVs, since so far, few SN impostors have good spectral coverage
as the objects fade during a major eruption.  For example, there is
only one spectrum at day 2 available for SN~1997bs (Van Dyk et al.\
2000), to which the spectra of SN impostors are often compared.  This
provides yet another link in observed properties between LBVs and the
unusual transients with obscured progenitors, SN~2008S and NGC~300-OT
(see Smith et al.\ 2010a for further discussion of this link).  A more
inclusive comparison of the spectra of several LBVs is provided later
in the paper.

\begin{figure*}\begin{center}
\includegraphics[width=5.2in]{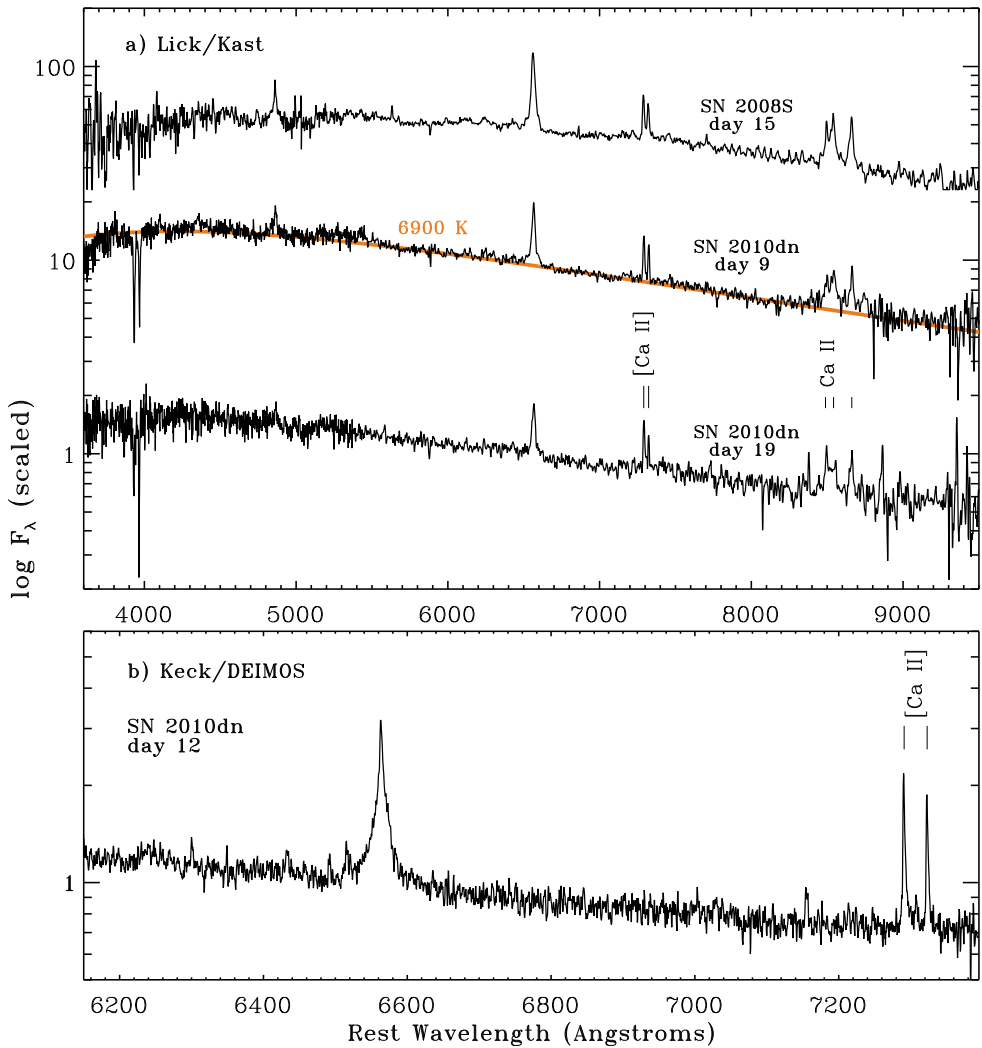}
\end{center}
\caption{(top) Lick/3m spectra of SN~2010dn obtained on days 9 and 19,
  compared to a spectrum of SN~2008S from Smith et al.\
  (2009a). (bottom) Keck/DEIMOS spectrum on day 12.  The bright lines
  are H$\alpha$ and the [Ca~{\sc ii}] doublet, while the fainter
  narrow lines are moslty Fe~{\sc ii}.}
\label{fig:spec2010dn}
\end{figure*}

The detailed evolution of the H$\alpha$ line in SN~2002bu is
interesting (Figure~\ref{fig:vel02bu}).  From day 11 to 23 the line
gets slightly narrower, and weakens relative to the continuum, showing
a Lorentzian profile at both epochs.  The day 40 profile is
transitional; it has about the same relative strength as day 23, but
is now somewhat broader and slightly asymmetric, with a devloping hump
on its blue side.  By the last two epochs, substantial changes are
apparent; the line now has a more Gaussian profile shape and is
broader with a FWHM of roughly 1200 km s$^{-1}$, and it seems to be
more asymmetric or blueshifted.  It is intriguing that this
progressive blueshift might be evidence for dust formation, as noted
above.  H$\alpha$ also seems to have developed blueshfted P~Cygni
absorption features at $-$3500 km s$^{-1}$ in the last two epochs,
similar to SN~2009ip (Smith et al.\ 2010a), although these might also
be due to some other absorption feature.  H$\beta$ also seems to
develop stronger broad blueshifted absorption, but it is at a
different velocity.  Higher resolution spectra of this transient would
have been quite valuable.

{\it SN 2010dn:} We present early-time spectra of SN~2010dn in
Figure~\ref{fig:spec2010dn}.  Moderate-resolution spectra obtained
with the Kast spectrograph at Lick observatory on days 9 and 19 after
discovery are shown in Figure~\ref{fig:spec2010dn}a, where they are
compared to the very similar spectrum of SN~2008S from Smith et al.\
(2009a) that was obtained with the same instrument.  Both objects show
strong narrow [Ca~{\sc ii}] and Ca~{\sc ii} emission, in addition to
the Balmer emission lines.  The overall continuum shape and the weak
spectral features in the blue are also remarkably similar in both
objects.  In fact, spectroscopically, SN~2010dn is a near twin of both
SN~2008S and N300-OT.  A 6900~K blackbody function is shown in orange
for comparison to the blue continuum shape of SN~2010dn on day 9.  On
days 9 and 19, we measure H$\alpha$ emission equivalent widths of 31.5
and 26.7 \AA \ ($\pm$2 \AA), respectively, in the Lick spectra.

\begin{figure}\begin{center}
\includegraphics[width=2.9in]{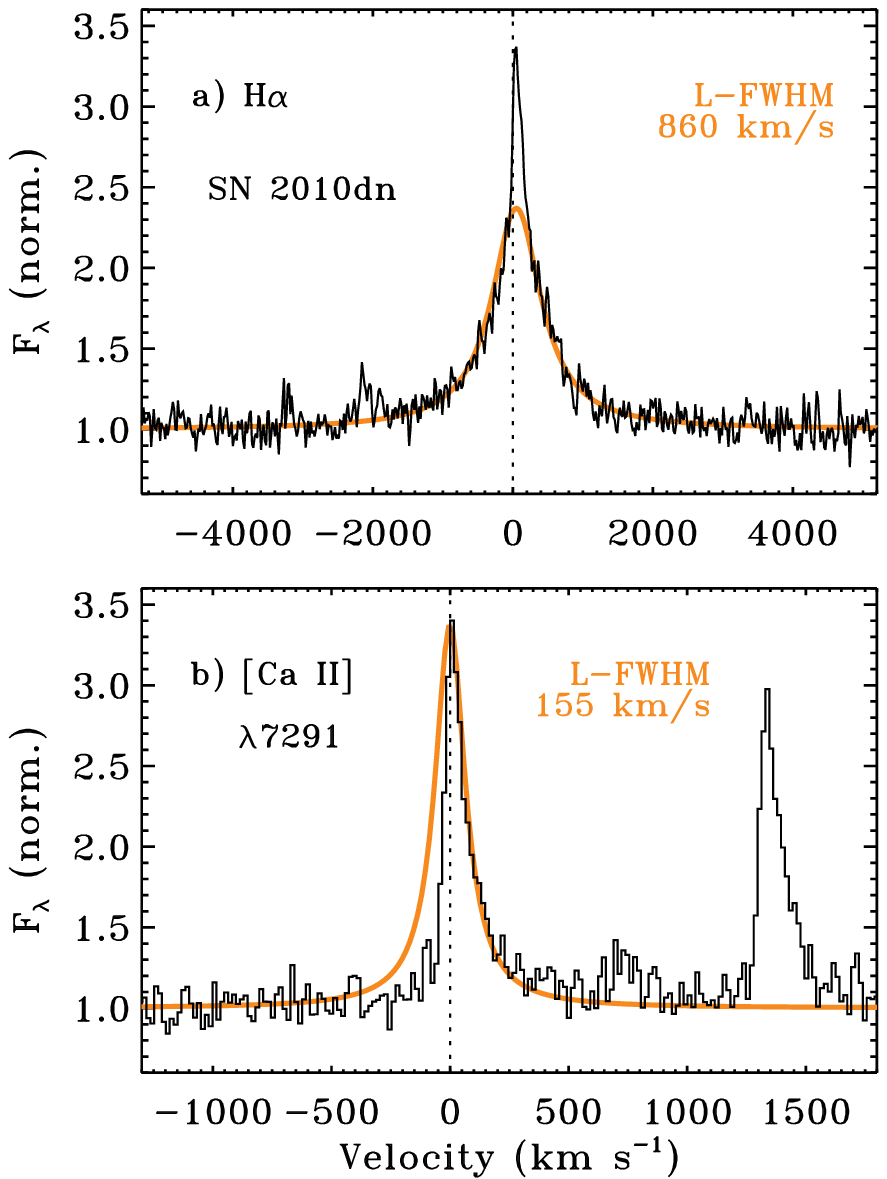}
\end{center}
\caption{Velocity profiles (a) of H$\alpha$ and (b) of [Ca~{\sc ii}]
  $\lambda$7291 from the day 12 Keck/DEIMOS spectrum of SN~2010dn.
  [Ca~{\sc ii}] $\lambda$7323 can also be seen at the right.
  Symmetric Lorentzian profiles with FWHM = 860 and 155 km s$^{-1}$,
  resepctively, are shown for comparison in orange.}
\label{fig:vel2010dn}
\end{figure}

We also obtained a high-resolution spectrum of SN~2010dn on day 12
after discovery using the DEIMOS spectrograph at Keck, shown in
Figure~\ref{fig:spec2010dn}b.  This was limited in wavelength coverage
to the red spectrum showing H$\alpha$ and the [Ca~{\sc ii}] doublet.
Velocity profiles of H$\alpha$ and [Ca~{\sc ii}] $\lambda$7291 from
the DEIMOS spectrum are shown in Figure~\ref{fig:vel2010dn}.  Most of
the H$\alpha$ flux can be accounted for with a broad Lorentzian
profile with FWHM $\approx$ 860 km s$^{-1}$, as shown in
Figure~\ref{fig:vel2010dn}a, but there is also excess emission from a
narrow component on top of this profile.  Qualitatively, the mostly
Lorentzian profile with a small contribution from very narrow emission
closely resembles that of the Type IIn SNe 1998S and 2006gy at early
times (Chugai 2001; Smith et al.\ 2010b), which was thought to be
indicative of diffusion of radiation through an opaque circumstellar
envelope or slow wind.  In the day 12 spectrum taken with DEIMOS, we
measure an H$\alpha$ equivalent width of 38.6 \AA \ ($\pm$2 \AA).

Superposed on this intermediate-width Lorentzian profile is a much
narrower H$\alpha$ line.  The narrow component of H$\alpha$ has the
same profile as the narrow emission seen in the pair of [Ca~{\sc ii}]
lines, shown in Figure~\ref{fig:vel2010dn}b; the [Ca~{\sc ii}] lines
show the narrow profile better because they are free from the
underlying broad profile.  These narrow components have FWHM values of
roughly 110--120 km s$^{-1}$, but they are asymmetric with a very
steep drop on the blue side of the line.  The red wing has a
Lorentzian shape that would imply FWHM = 155 km s$^{-1}$ if it were
symmetric, so perhaps this is a better indicator of the expansion
speed of the circumstellar gas emitting these narrow components.
These narrow [Ca~{\sc ii}] profiles are qualitatively identical to
those of the same lines in N300-OT, which also showed asymmetric
profiles with a Lorentzian red wing and a steep cutoff on the blue
side, with very similar widths of 140--190 km s$^{-1}$ at early times
(Berger et al.\ 2009).  Similar profiles were seen in other lines such
as [O~{\sc i}] $\lambda\lambda$6300,6364 in N300-OT as well (Berger et
al.\ 2009), and we see the same profile in the narrow component of
H$\alpha$ in SN~2010dn, so we infer that the shape is not the result
of some peculiar excitation/ionization effect unique to Ca~{\sc ii}.

So far, N300-OT and SN~2010dn are the only SN impostors with
comparable high-resolution spectra available for these [Ca~{\sc ii}]
lines, so their nearly identical asymmetric profiles are rather
intriguing.  The shape of these asymmetric {\it forbidden} lines has
not been explained, but suggests either an intrinsically aymmetric
distribution of emitting gas oriented the same way in both objects, or
dust obscuration of the blue wing with a particular geometry.

\begin{table*}\begin{center}\begin{minipage}{5.2in}
\caption{A List of SN impostors considered here}\scriptsize
\begin{tabular}{@{}lllclll}\hline\hline
Transient &Host Gal.  &Date    &mag (peak) &R.A./Dec. (J2000)   &Discoverer  &Refs.$^{a}$ \\ 
\hline
P Cygni    &MW        &1600-55     &2.8   &20 17 47.20 +38 01 58.5  &Blaeu              &[1]     \\
$\eta$ Car &MW        &1837-60     &-1.0  &10 45 03.59 --59 41 04.3  &Herschel           &[1]     \\
SN~1954J   &NGC~2403  &1954        &16.5  &07 36 55.36 +65 37 52.1  &Tammann, Sandage   &[2]     \\
SN~1961V   &NGC~1058  &1961        &12.5  &02 43 36.42 +37 20 43.6  &Wild               &[3,4]   \\
HD~5980    &SMC       &1993-94     &8.8   &00 59 26.57 --72 09 53.9  &Barb\'{a} \& Niemala &[5]   \\
V1         &NGC~2366  &1994-?      &17.4  &07 28 43.37 +69 11 23.9  &Drissen et al.     &[6]     \\
SN~1997bs  &NGC 3627  &1997 04 15  &17.0  &11 20 14.25 +12 58 19.6  &LOSS               &[7]       \\
SN~1999bw  &NGC 3198  &1999 04 20  &17.8  &10 19 46.81 +45 31 35.0  &LOSS               &this work \\
SN~2000ch  &NGC 3432  &2000 05 03  &17.4  &10 52 41.40 +36 40 08.5  &LOSS               &[8]       \\
SN~2001ac  &NGC 3504  &2001 03 12  &18.2  &11 03 15.37 +27 58 29.5  &LOSS               &this work \\
SN~2002bu  &NGC 4242  &2002 03 28  &15.5  &12 17 37.18 +45 38 47.4  &Puckett, Gauthier  &this work \\
SN~2002kg  &NGC 2403  &2002 10 26  &19.0  &07 37 01.83 +65 34 29.3  &LOSS               &[9,10]    \\
SN~2003gm  &NGC 5334  &2003 07 06  &17.0  &13 52 51.72 --01 06 39.2  &LOSS               &[9]       \\
2005-OT    &NGC 4656  &2005 03 21  &18.0  &12 43 45.84 +32 06 15.0  &Rich               &...       \\
SN~2006bv  &UGC 7848  &2006 04 28  &17.8  &12 41 01.55 +63 31 11.6  &Sehgal, Gagliano, Puckett &this work \\
SN~2006fp  &UGC 12182 &2006 09 17  &17.7  &22 45 41.13 +73 09 47.8  &Puckett, Gagliano  &...     \\
SN~2007sv  &UGC 5979  &2007 12 20  &17.4  &10 52 40.05 +67 59 14.2  &Duszanowicz        &...     \\
SN~2008S   &NGC 6946  &2008 02 01  &17.6  &20 34 45.35 +60 05 57.8  &Arbour             &[11]    \\
2008-OT    &NGC 300   &2008 05 14  &16.2  &00 54 34.16 --37 38 28.6  &Monard             &[12,13] \\
SN~2009ip  &NGC 7259  &2009 08 26  &17.9  &22 23 08.26 --28 56 52.4  &Maza, Pignata et al.&[14]   \\
2009-OT    &UGC 2773  &2009 08 18  &18.0  &03 32 07.24 +47 47 39.6  &Boles              &[14]    \\
2010da     &NGC 300   &2010 05 23  &16.0  &00 55 04.86 --37 41 43.7  &Monard             &[15,16]    \\
2010dn     &NGC 3184  &2010 05 31  &17.1  &10 18 19.89 +41 26 28.8  &Itagaki            &this work   \\
\hline
\end{tabular}
\label{tab:list1}
$^{a}$Primary references for sources of light curves and early
spectral analysis: [1] Smith \& Frew (2010); [2] Tammann \& Sandage
(1968); [3] Zwicky (1964); [4] Bertola (1963,1965); [5] Jones \&
Sterken 1997; [6] Petit et al.\ (2006); [7] Van Dyk et al.\ (2002);
[8] Wagner et al.\ (2004); [9] Maund et al.\ (2006); [10] Van Dyk et
al.\ (2006); [11] Smith et al.\ (2009a); [12] Bond et al.\ (2009);
[13] Berger et al.\ (2009); [14] Smith et al.\ (2010a); [15] Bond
(2010); [16] Chornock et al.\ (2010). The comment ``this work'' refers
to new data published in the present paper for the first time.
\end{minipage}\end{center}
\end{table*}

\section{COMPARATIVE RESULTS}

\subsection{Comments on Individual Events}

Here we briefly list relevant observational material for suspected
members of the class of SN impostors or giant LBV eruptions, collected
from the literature for the purposes of this discussion (see
Table~\ref{tab:list1}).  When published analyses exist, we refer to
those papers and adopt the same assumptions except where noted.  For
new observational material, our data were collected as part of the
Lick Observatory Supernova Search (LOSS), as noted above.  In most
cases below, we adopt distance moduli from the NASA Extragalactic
database\footnote{{\tt http://nedwww.ipac.caltech.edu/}} and we take
line-of-sight Galactic extinction values of $E(B-V)$ from Schlegel et
al.\ (1998).  Two of the transients listed below were discovered
recently.  We list them here and provide some initial details for
completeness, but cannot yet comment on their late-time behavior since
they are still being studied.

\begin{figure*}\begin{center}
\includegraphics[width=6.3in]{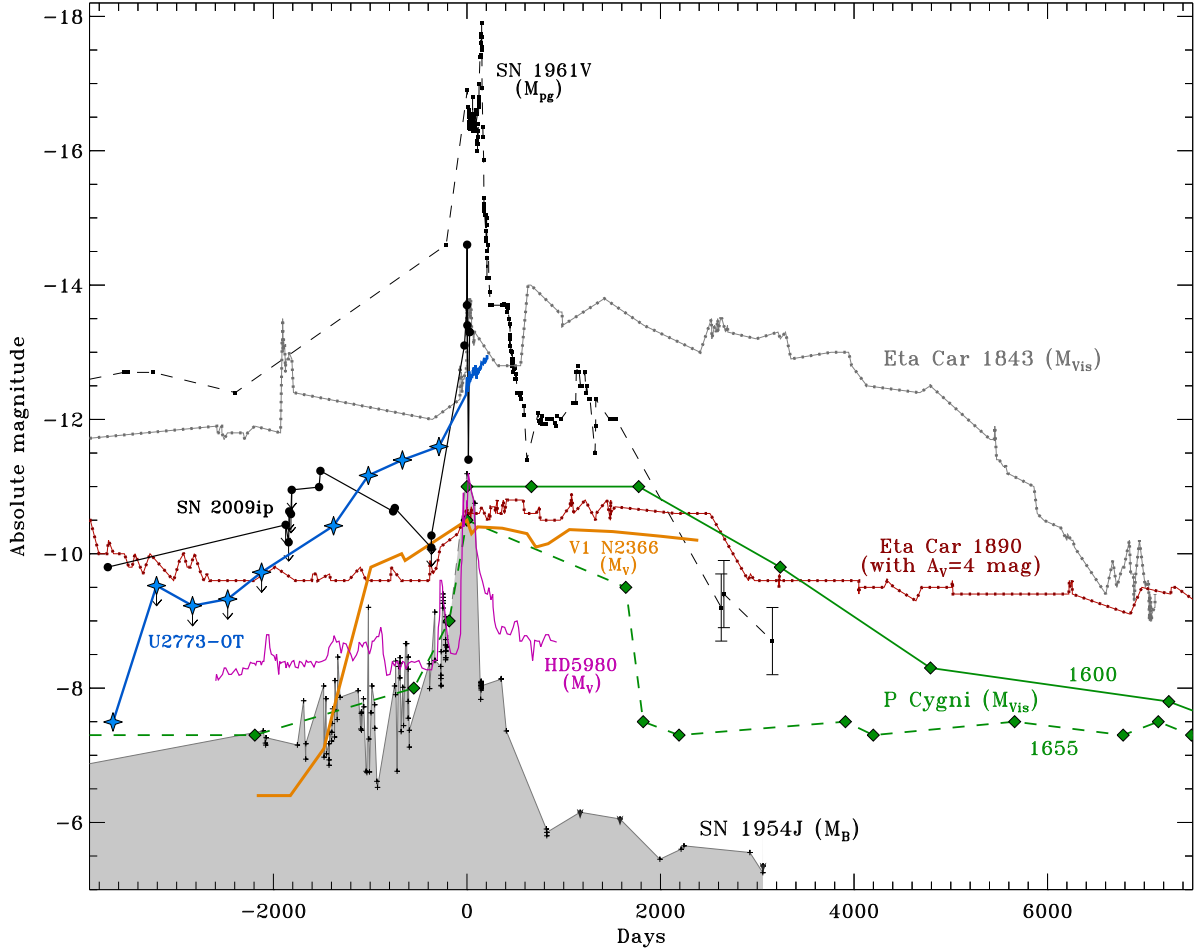}
\end{center}
\caption{Long timescale LBV light curves, adapted from Smith et al.\
  (2010a).  Absolute-magnitude light curves of LBV eruptions for cases
  where information is available over long (i.e. decade) timescales,
  including observations before the main eruptions.  We show the
  historical 19th century Great Eruption of $\eta$~Carinae from Smith
  \& Frew (2010; grey) as well as the 1890 outburst corrected for 4
  mag of visual extinction (brown).  We also show P Cygni's pair of
  eruptions in 1600 (green diamonds; solid) and 1655 A.D.\ (green
  diamonds; dashed) (see Smith \& Frew 2010 and references therein).
  The eruption of SN~1954J (V12 in NGC~2403; Tammann \& Sandage 1968)
  is shown as a gray shaded plot.  The absolute magnitude of SN~1961V
  corrected for $A_B$=0.26 mag is shown with small filled squares,
  compiled from photometry in Zwicky (1964), Bertola (1963,1965), and
  Bertola \& Arp (1970).  The thick orange curve is the LBV eruption
  of V1 in NGC~2366 (Drissen et al.\ 2001; Petit et al.\ 2006),
  although shifted to an arbitrary date.  The magenta curve is for the
  eruption of HD~5890 in the SMC during 1993-1994 (from Jones \&
  Sterken 1997).  We also show the decade-long pre-eruption light
  curves from the recent transients SN~2009ip (black filled circles)
  and U2773-OT (blue stars) from Smith et al.\ (2010a).  Unfiltered
  visual magnitudes are shown for $\eta$ Car and P~Cyg, $B$ magnitude
  for SN~1954J, photographic (approximately $B$-band) for SN~1961V,
  $V$ magnitudes for V1 and HD~5980, and unfiltered (approximately
  $R$-band) for SN~2009ip and U2773-OT.  Although these are different
  filters, our multi-band photometry of SN~2002bu shows that the $V$
  and $R$-band lightcurves are almost identical in shape.}
\label{fig:lc}
\end{figure*}

{\it P Cygni:} Although famous for its namesake line-profile shape,
P~Cygni is also notable as the {\it first} LBV, and for being only the
third variable star discovered --- after Tycho's SN and Mira.  (It was
of course not referred to as an LBV at the time, but was called a
nova.)  Despite the excitement it generated at the time, there are
only sparse observations of its 1600-1665 A.D.\ eruption, with only a
handful of surviving reports during the main light curve peak that
lasted $\sim$10 yr (see Figure~\ref{fig:lc}; from Smith \& Frew 2010,
in prep.).  These historical observations are valuable for recording
the long timescale variability of P Cygni, but the sparse sampling
suggests that if there had been short timescale variation as exhibited
by many other LBV eruptions around peak, then it could easily have
been missed.  Thus, a brief unobserved peak could have been more
luminous than the several-year sustained peak of the eruption, which
had an absolute magnitude of roughly --11 mag (according to Lamers \&
de Groot 1992).  Most of these observations were made with the newly
invented telescope, and a typical value for the uncertainty of these
observations is $\pm$0.2--0.3 mag, although the quality of
observations may vary considerbly from one epoch to another (see Smith
\& Frew 2010).  These are visual (i.e., unfiltered) observations,
converted approximately to modern $V$-band based on the likely color
of LBV outbursts, although this is uncertain and depends on reddening
and the strength of H$\alpha$.  The spectrum during outburst was not
recorded, of course.  P Cygni also suffered a second major outburst in
1655 (dashed in Figure~\ref{fig:lc}) that reached a peak almost as
luminous as the first, with an absolute magnitude of roughly
--10.5. After this second outburst, P~Cygni faded and remained faint
for several decades, but then brightened suddenly around 1700.  It has
been relatively tame and brightening very slowly since then.  From
modern observations of its shell nebula, we can infer that the
dominant expansion speed of the 1600 A.D. eruption was about 136 km
s$^{-1}$ (Smith \& Hartigan 2006).

\begin{figure*}\begin{center}
\includegraphics[width=6.6in]{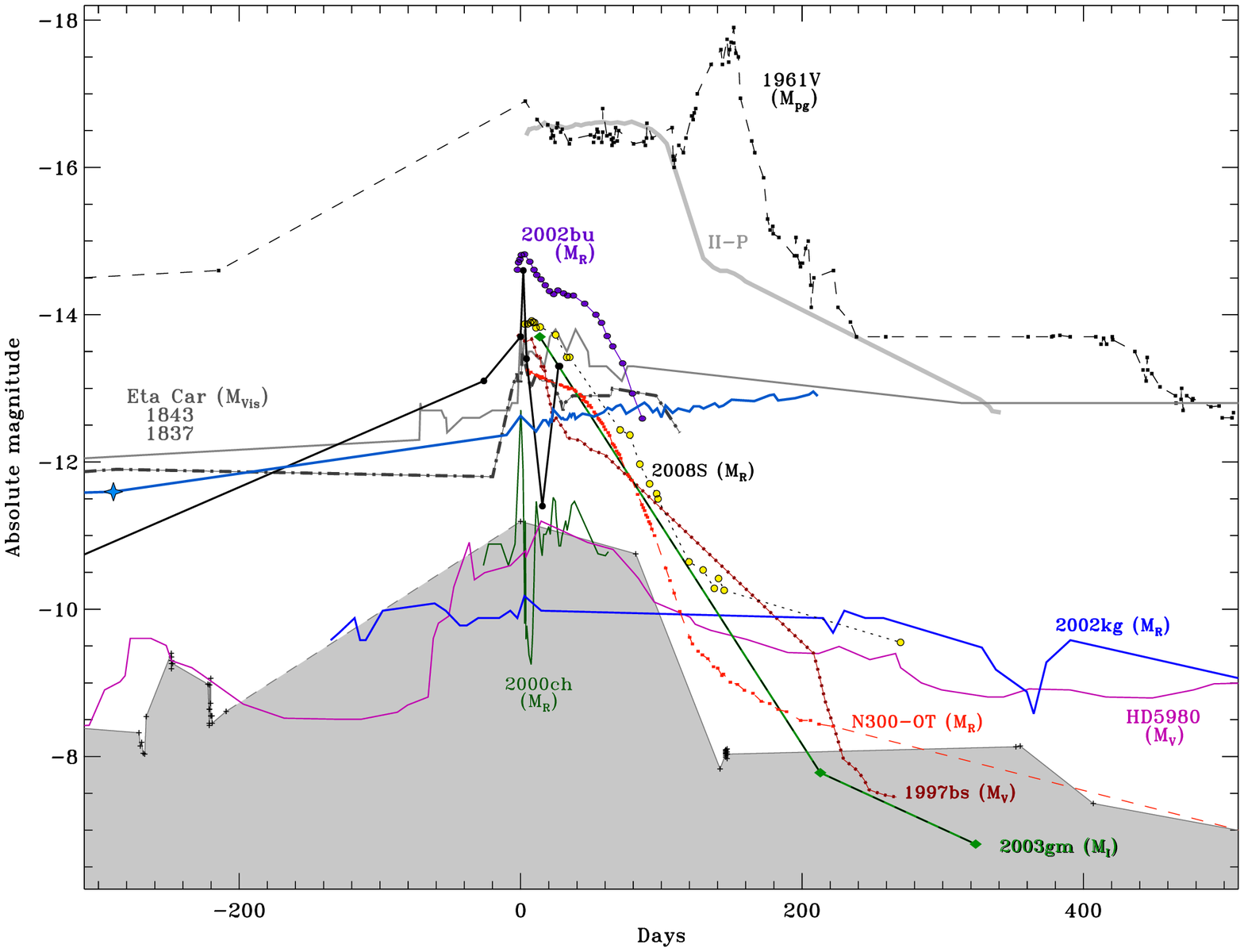}
\end{center}
\caption{Similar to Figure~\ref{fig:lc}, but zooming in on the time
  around peak brightness for several eruptive transients.  In addition
  to the light curves repeated from Figure~\ref{fig:lc}, we add the
  $V$-band light curve of SN~1997bs (Van Dyk et al.\ 2000; brown
  dotted), the $R$-band light curves of SN~2000ch (Wagner et al.\
  2004; green), SN~2002bu (this work; purple dots), SN~2002kg (Van Dyk
  et al.\ 2006; blue), SN~2008S (Smith et al.\ 2009a; yellow dots),
  and NGC~300-OT (Bond et al.\ 2009; red squares and dashed line), and
  the $I$-band light curve of SN~2003gm (Maund et al.\ 2006;
  green/black dashed line).  For comparison with a normal Type II-P
  event, we also show the $R$-band light curve of SN~1999em (Leonard
  et al.\ 2002; thick gray curve).  We show both the 1837 (solid gray)
  and 1843 (dot-dashed darker gray) precursor eruptions of $\eta$~Car
  (Smith \& Frew 2010).}
\label{fig:lc2}
\end{figure*}

{\it Eta Carinae:} The complex light curve of $\eta$ Car has a long
history of discussion that will not be repeated here (see Frew 2004).
A very recent study by Smith \& Frew (2010) recovered many new
historical observations from the 19th century and uncovered some
mistakes in earlier works going back to Herschel's original reports.
The new light curve of Smith \& Frew (2010) looks substantially
different in detail from previously published and often reproduced
light curves of $\eta$~Car (e.g., Innes 1903; see Frew 2004 for a
thorough discussion of the historical data), and the Smith \& Frew
light curve is used here (Figures~\ref{fig:lc} and \ref{fig:lc2}).
The star suffered two shorter-duration bursts in 1838 ($-$13.5 mag)
and 1843 ($-$13.8 mag), which preceded a final rise at the end of 1844
($-$14.0 mag), from which the star declined slowly for more than 10 yr
afterward. Again, we do not know what the spectrum looked like during
the 1840s eruption, but reports of its red or ``ruddy'' color probably
indicate strong H$\alpha$ emission.  Following another smaller
eruption in $\sim$1890, the star is apparently still slowly recovering
from the upheaval of its 19th century eruptions (e.g., Smith et al.\
2003a; Davidson et al.\ 2005).  The 1890 eruption was probably much
more luminous than it looked, since the star is thought to have been
buried in $\sim$4 mag of visual extinction at that time (Humphreys et
al.\ 1999).  The light curve of the 1890 event is also shown in
Figure~\ref{fig:lc}, with a correction for this extinction applied.
Based on the kinematics of the bipolar Homunculus nebula, the polar
expansion speed for the bulk of the matter ejected in the major
eruption was 650 km s$^{-1}$, dropping to values as low as 40 km
s$^{-1}$ at the pinched equatorial waist (Smith 2006).  However, deep
spectroscopy of the surroundings outside the Homunculus reveal that
the 19th century eruption also ejected a small amount of extremely
fast material moving at $\sim$5000 km s$^{-1}$, probably requiring a
strong shock wave during the event (Smith 2008).  A smaller bipolar
nebula called the Little Homunculus is growing inside the larger one
(Ishibashi et al.\ 2003), and its kinematics suggest that it was
ejected in the smaller 1890 event (Smith 2005).  Both the kinematics
of the Little Homunculus and historical spectra obtained during the
1890 event suggest an ejection speed of around 200 km s$^{-1}$ (Smith
2005; Whitney 1952; Walborn \& Liller 1977).

{\it SN~1954J/V12:} This LBV outburst in NGC~2403 was well observed
photometrically by Tammann \& Sandage (1968), although no spectra of
the outburst are available.  The massive star was clearly an irregular
blue variable star (V12) for a decade before the peak of its giant
eruption in 1954, and the star apparently survived the event as a
faint reddened star (Smith et al.\ 2001; Van Dyk et al.\ 2005).  A
spectrum obtained by Van Dyk et al.\ (2005) in November 2002 revealed
a narrow H$\alpha$ profile suggesting an expansion speed of $\sim$700
km s$^{-1}$, although the expansion speed during the peak of the
eruption is not known since spectra during the event are not available
(in the case of $\eta$ Car, however, it is reassuring that the
present-day wind speed is similar to that of the Homunculus nebula;
Smith 2006).  As with the case of P~Cygni, a brief peak in the light
curve with a brighter maximum might have been missed due to a
relatively long gap in the observations just before the recorded peak
(Tammann \& Sandage 1968).

{\it SN~1961V:} Of all the ``SN impostors'', SN~1961V is one of the
most controversial, due to its very high luminosity ($M_{\rm pg}$ at
peak was almost $-$18 mag) that blurs any clear distinction between
real core-collapse SNe~IIn and LBV-like eruptions, if it is indeed an
LBV.  Whether or not the surviving star is detected is key, but this
question has advocates on both sides (Van Dyk et al.\ 2002; Chu et
al.\ 2004; Filippenko et al.\ 1995; Goodrich et al.\ 1989).  Because
this source is so controversial and so much brighter than the rest of
the SN impostors, we feel that it needs special consideration and we
discuss it in more detail in \S 4.4.  The light curve shown in
Figure~\ref{fig:lc2} is compiled from Zwicky (1964), Bertola (1963,
1965), and Bertola \& Arp (1970).  Zwicky (1964) estimated ejection
speeds of 3700 km s$^{-1}$ from the width of H$\alpha$ in spectra
obtained during the main eruption.

{\it HD~5980:} This remarkable WR+LBV eclipsing binary is the most
luminous star in the Small Magellanic Cloud (SMC).  It had already
been an interesting object of study for decades as a massive eclipsing
binary of two WN stars (one is actually a WNH star; see Smith \& Conti
2008), where one star had a somewhat variable spectral type changing
from WN to O7.  It then surprised astronomers when it suffered a giant
LBV eruption around 1993-1994, at which time the primary star
brightened and changed its spectral type from a WN star to a H-rich B
supergiant (Bateson, Gilmore, \& Jones 1994; Barb\'{a} \& Niemala
1994; Barb\'{a} et al.\ 1995).  A substantial literature has built up
about this star, and Koenigsberger (2004) has provided a recent review
of the spectral and photometric properties of the binary and its
outburst.  In Figure~\ref{fig:lc} we use a smoothed version of the
visual light curve (i.e. ignoring measurements during eclipses)
adapted from Jones \& Sterken (1997), which is shown here for
reference.  After two main peaks during the main outburst that lasted
almost one year, the star has apparently taken about a decade to
settle back to its pre-outburst state; a recent study by Koenigsberger
et al.\ (2010) finds variability over several decades in historical
data from the mid-20th century.  It is noteworthy that the B1.5~Ia+
spectal type of the erupting star in 1994 implies a significantly
hotter temperature than the canonical $\sim$8,000 K F-type supergiant
expected in LBV eruptions.  This turns out to be the case for a number
of SN impostors, including V1 (Drissen et al.\ 2001), SN~2000ch, and
SN~2009ip (Smith et al.\ 2010a); we will return to these ``hot'' LBV
eruptions later.  The wind speed of the erupting component of the
binary system was estimated as 600 km s$^{-1}$ (Koenigsberger et al.\
1998).

{\it V1 in NGC~2366:} This source has been discussed in detail in a
series of papers by Drissen and collaborators (Drissen et al.\ 1997,
2001; Petit, Drissen, \& Crowther 2006).  It is located in the giant
starbursting H~{\sc ii} region NGC~2363 within the dwarf irregular
galaxy NGC~2366.  Its eruption began when the star brightened rapidly
in 1994, and it seems to have stayed at near-maximum ever since (see
Figure~\ref{fig:lc}).  Interestingly, while the visual magnitude
remained roughly constant at $M_V$ of about -10.2 mag during this
time, the UV flux actually brightened and the temperature indicated by
spectral analysis increased (Petit et al.\ 2006).  As with HD~5980,
this once again contradicts the traditional view of pseudo
photospheres in LBV eruptions having an F-type supergiant spectrum at
maximum light.  V1 has been subject to detailed modeling of its
optical and UV spectrum, which can be matched by a stellar wind with
$\dot{M} \approx 5 \times 10^{-4} M_{\odot}$ yr$^{-1}$ and an average
wind terminal speed of 300 km s$^{-1}$ (Petit et al.\ 2006).  Unlike
some other LBV eruptions, the spectrum is not consistent with that of
an explosion, but is consistent with a strong supergiant wind.  This
provides compelling evidence that in some cases, LBV eruptions are
indeed wind driven while in others they seem to be partly explosive.


{\it SN~1997bs:} This SN of questionable integrity in M66 was the
first ``SN'' discovered by the LOSS, but subsequent analysis revealed
that it was most likely not a genuine core-collapse event (Van Dyk et
al.\ 2000).  The $V$ and $R$-band light curves and optical spectra we
use here were discussed extensively by Van Dyk et al.\ (2000), shown
in Figure~\ref{fig:lc2}.  The progenitor was identified in
pre-discovery images as a luminous star with $M_V \approx -8.1$ mag,
and Van Dyk et al.\ (2000) conjectured that the star may have
survived.  Li et al.\ (2002) did not detect the star in late-time
follow-up images obtained with {\it HST}, suggesting that the star may
have disappeared or perhaps that it may have been deeply enshrouded in
dust.  Van Dyk et al.\ measured a FWHM of 765 km s$^{-1}$ from
H$\alpha$ in spectra obtained on day 2 after discovery, and noted
Lorentzian line wings extending to $\pm$3,000 km s$^{-1}$.

{\it SN~1999bw:} We discovered SN~1999bw in NGC~3198 on KAIT images
taken as part of the LOSS on 1999 Apr.\ 20.2 UT (Li et al.\ 1999), and
the first spectrum revealed that it was similar to SN~1997bs
(Filippenko et al.\ 1999).  As noted earlier, H$\alpha$ in our day 4
spectrum can be approximated by a Lorentzian profile with a FWHM of
$\sim$630 km s$^{-1}$, with a broader base with FWZI of roughly 3000
km s$^{-1}$. Sugerman \& Meixner (2004) reported the detection of
SN~1999bw in archival infrared data obtained in 2004 with the {\it
  Spitzer Space Telescope}, but no detailed follow-up study has been
performed.  A single spectrum of SN~1999bw was published by Matheson
(2005).  We obtained limited $BVRI$ and unfiltered photometry of
SN~1999bw with KAIT, as listed in Table~\ref{tab:phot99bw}.  We
obtained a spectrum at Lick during the initial peak, listed in
Table~\ref{tab:spec} and shown in Figure~\ref{fig:spec99bw}.  We adopt
$m-M = 30.42$ mag for the host galaxy NGC~3198, and $E(B-V)$ = 0.012.
This suggests that SN~1999bw had a peak absolute $R$ magnitude of
about $-$12.65, intermediate between those of P~Cygni and
$\eta$~Carinae.

{\it SN~2000ch (LBV1 in NGC 3432):} Observations of SN~2000ch were
discussed in detail by Wagner et al.\ (2004; see also Van Dyk 2005),
showing that it had a spectrum similar to that of SN~1997bs with a
smooth continuum and bright Balmer emission lines.  Its light curve
was quite different, however, with a sharp rise and dip over a
timescale of $\sim$5 days, superposed on a relatively constant
plateau.  The peak absolute $R$ magnitude was $-$12.8, and the plateau
at roughly $-$10.6 mag may have been either the quiescent state of a
very luminous star or a prolonged S~Dor-like eruption.  The $R$ light
curve from Wagner et al.\ (2004) is shown in Figure~\ref{fig:lc2}.  We
obtained additional spectra at Lick, as listed in Table~\ref{tab:spec}
and shown in Figure~\ref{fig:spec99bw}.  The spectrum during outburst
(day 28) has a strong H$\alpha$ line with a Lorentzian profile shape
and a FWHM of 1400 km s$^{-1}$.  The line is somewhat asymmetric, with
a blue wing extending to $-$2500 km s$^{-1}$ and the red wing reaching
$+$4300 km s$^{-1}$.  In our late-time spectrum from 2004 discussed
above, the H$\alpha$ line width was similar with FWHM $\approx$ 1500
km s$^{-1}$.

As this paper was in the final stages of preparation, Pastorello et
al.\ (2010) presented additional data showing that the same LBV star
that erupted as SN~2000ch also suffered three additional eruptions in
2008 and 2009.  Perhaps this object should be called ``LBV1'' in
NGC~3432.  Like the 2000 transient, these later eruptions were erratic
and fast variations with peak $R$-band absolute magnitudes of $-$12.1
to $-$12.7.  LBV1 showed rapid dips and recovery on timescales of a
few days, similar to SN~2009ip, but repeatedly over several years.
These recurring eruptions are not reproduced in Figure~\ref{fig:lc2},
but this interesting object is discussed thoroughly by Pastorello et
al.\ (2010).  Pastorello et al.\ (2010) drew comparisons between LBV1
and the binary HD~5980, suggesting similar binary encounters as a
possible mechnism, among other, behind the rapid and erratic
variability.  The erratic variability with multiple fast peaks and
dips is qualitatively similar to the wild pre-1954 variability of
SN~1954J/V12.  If this comparison is appropriate, then the erratic
variability may signify a growing instability, and we should not be
surprised if LBV1 culminates with a more major eruption in the next
decade or so.  In any case, we should keep an eye on this star!

{\it SN~2001ac:} We discovered SN~2001ac on KAIT images taken on 2001
Mar.\ 12.4 and 13.3 UT (Beckmann \& Li 2001) as part of the LOSS.  A
spectrum obtained by Matheson et al.\ (2001) was similar to those of
SNe~1997bs and 1999bw, with a blue continuum and strong Balmer
emission lines.  No detailed analysis of SN~2001ac has been published;
we presented limited $BVRI$ and unfiltered KAIT photometry
(Table~\ref{tab:phot01ac}) and spectra near maximum light
(Table~\ref{tab:spec}, Figure~\ref{fig:spec99bw}).  We adopt $m-M$ =
32.22 mag and $E(B-V)$ = 0.027 mag for the host galaxy NGC~3504,
suggesting a peak absolute $R$ magnitude of roughly $-$14.1 mag,
comparable to that of $\eta$~Car.  As noted earlier, from our spectrum
on day 9, the H$\alpha$ line profile has a composite shape that can be
fit with a Gaussian core FWHM of 287 km s$^{-1}$ and broader Gaussian
wings with FWHM of 1505 km s$^{-1}$.

{\it SN~2002bu:} Located in NGC~4242, SN~2002bu was discovered by
Pucket \& Gauthier (2002) on 2002 Mar.\ 28.26 UT.  The progenitor was
not detected to limiting red magnitudes of 20.5-21.0 mag.  Preliminary
reports of the spectrum indicated that it resembled a Type~IIn or an
LBV, with strong and narrow Balmer emission lines and a flat continuum
(Ayani \& Kawabata 2002).  We obtained extensive $BVRI$ photometry
with KAIT (see Table~\ref{tab:phot02bu}) as well as a series of
spectra from Lick (Table~\ref{tab:spec}, Figure~\ref{fig:spec02bu}).
We adopt $m-M$ = 29.71 mag and $E(B-V)$ = 0.012 mag for NGC~4242.
This suggests a peak absolute $R$ magnitude of roughly $-$15, which
places it among the brightest examples of the known SN impostors.  The
multi-color light curves and spectral evolution are described in some
detail above.  From spectra on day 11, H$\alpha$ exhibits a Lorentzian
profile with FWHM of 893 km s$^{-1}$ and wings that extend to
$\pm$2500 km s$^{-1}$.  The temporal evolution of the spectrum was
discussed above.

{\it SN~2002kg/V37:} The progenitor of this transient in the nearby
spiral galaxy NGC~2403 (also host to V12/SN~1954J) was first
identified by Van Dyk (2005) as Variable 37 from Tammann \& Sandage
(1968), a bright blue irregular variable like the Hubble-Sandage
variables (i.e., a classical LBV).  Observations of the increased
brightness that was dubbed SN~2002kg have been discussed in detail by
Maund et al.\ (2006) and Van Dyk et al.\ (2006).  Its absolute peak
$V$ magnitude during outburst\footnote{There is some inconsistency in
  the literature about the peak absolute magnitude of SN~2002kg.  At
  different places in their paper, Maund et al.\ (2006) quote $M_V$
  values of --9, --9.6, and --10.4 mag.  Van Dyk et al.\ (2006) quote
  $M_V$ = --9.8 mag in their text.} was roughly $-$10 mag, making it
one of the faintest of the recognized SN impostors, and the total
brightening compared to its progenitor was only about 2 mag.  We show
the KAIT $R$-band light curve from Van Dyk et al.\ (2006) in
Figure~\ref{fig:lc2}. Although SN~2002kg is usually discussed with the
other SN impostors that are attributed to giant LBV eruptions like
$\eta$ Car, it seems plausible that SN~2002kg was not really a giant
LBV eruption, but rather, a normal S~Dor phase of a massive LBV star.
This is based on its relatively modest increase in brightness that may
be consistent with a change in bolometric {\it correction} only.  The
difference between these two is that a giant LBV eruption is defined
as an increase in bolometric {\it luminosity}, whereas it is thought
that the bolometric luminosity remains roughly constant in a normal
S~Dor phase (see Humphreys et al.\ 1999).  Van Dyk estimated $M_{bol}$
= --9.8 mag for the progenitor star, which is consistent with the
observed peak absolute visual magnitude.  The outburst SN~2002kg was
very similar in magnitude to the previous eruptions experienced by V37
around 1920 and 1930 (Tammann \& Sandage 1968).  Weis \& Bomans (2005)
also associated SN~2002kg with V37, and claimed that the surviving
star was detected again several years after the outburst.  Both Maund
et al.\ (2006) and Van Dyk et al.\ (2006) presented spectra of
SN~2002kg, showing strong narrow Balmer emission lines with a narrow
component having widths of 330--370 km s$^{-1}$ and broader wings with
widths of 1500--1900 km s$^{-1}$, consistent with electron scattering
wings.  P Cygni absorption features in Balmer lines also suggest
expansion speeds of roughly 350 km s$^{-1}$.  The spectra revealed
strong narrow [N~{\sc ii}] $\lambda\lambda$6548,6583 emission, similar
to other LBVs and probably indicating the presence of a massive
circumstellar nebula.

{\it SN~2003gm:} This transient source in NGC~5334 was also analyzed
in detail by Maund et al.\ (2006).  Unfortunately, photometric data
are sparse and the available spectra are rather noisy.  We show the
$I$-band light curve from Maund et al.\ (2006) in Figure~\ref{fig:lc2}
(green/black).  There are only 3 photometric $I$-band points, but as
noted by Maund et al.\ (2006), the absolute peak magnitude and the
decay rate appear very similar to SN~1997bs.  The peak $M_I$ was about
--13.7, and the $M_R$ and $M_V$ values are probably similar to within
$\pm$0.8 mag.  We obtained unfiltered KAIT photometry on days 0
(discovery) and day 6, both of which were 17.0$\pm$0.1 mag.  Early
time spectra of SN~2003gm are similar to SN~2000ch, with H$\alpha$
having a rather narrow width of only $\sim$131 km s$^{-1}$, but with
broader wings having a width of 1472 km s$^{-1}$ (Maund et al.\ 2006).
Maund et al.\ estimated the metallicity of the host galaxy as 0.7
$Z_{\odot}$.  Like SN~2002kg, the progenitor star was identified as a
luminous star in pre-explosion data, and Maund et al.\ (2006)
estimated $M_V$ $\approx$ --7.5 mag for the progenitor, indicating
that the star brightened by more than 5 magnitudes during its giant
LBV eruption.

{\it N4656-OT (2005):} This optical transient source in NGC~4656 was
discovered by Rich (2005) with an unfiltered magnitude of 18.0 mag on
2005 Mar.\ 21 and 22 UT (the transient was also visible in unfiltered
images taken a few days earlier at 18.5 and 18.3 mag; Yamaoka 2005).
Elias-Rosa et al.\ (2005) reported that a spectrum of this transient
had a blue continuum with strong narrow Balmer emission lines having
widths of 730 km s$^{-1}$, but with no broad base as seen in normal
SNe~IIn, suggesting that it was an LBV-like event similar to SN~1997bs
and not a SN.  The spectrum also showed narrow Ca~{\sc ii} H and K
absorption.  Unfortunately, we did not obtain additional data on this
transient, and no other comprehensive analysis has been published to
date.  Adopting m--M = 28.69 mag (5.47 Mpc; from Tully et al.\ 2009)
and $A_R$ = 0.035 (Schlegel et al.\ 1998), the peak absolute
unfiltered ($\sim$$R$-band) magnitude is $-$10.73 mag at the time of
discovery.  If indeed N4656-OT is a giant LBV eruption, this makes it
one of the less luminous examples, comparable to SN~1954J or HD~5980.
No information is available about its progenitor star.

{\it SN~2006bv:} Occurring in UGC~7848, SN~2006bv was discovered by
Sehgal et al.\ (2006) on 2006 Apr. 28.36 UT in unfiltered images with
a magnitude of 17.8.  A spectrum obtained 2 days later showed a smooth
blue continuum and very narrow Balmer emission lines with FWHM of 400
km s$^{-1}$ (Blondin et al.\ 2006).  Immler \& Pooley (2006) presented
optical and UV photometry obtained a few days later and an upper limit
to the X-ray flux, and concluded that it was fading rapidly.  We
secured some limited photometric measurements of SN~2006bv
(Table~\ref{tab:phot06bv}), but we were unable to obtain spectra (this
makes it difficult for us to confirm from an independent analysis that
this is indeed an LBV and not a faint SN).  We adopt $m-M$ = 33.0 mag
and $E(B-V)$ = 0.015 mag for UGC~7848, suggesting a peak absolute
unfiltered ($\sim$$R$) magnitude of roughly $-$15.2.  This places
SN~2006bv among the most luminous of the SN impostors, comparable to
SN~2002bu.

{\it SN~2006fp:} This object was discovered on 2006 Sep.\ 17 UT by
Puckett \& Gagliano (2006) with an unfiltered magnitude of 17.7, while
images obtained the following night had a slightly brighter magnitude
of 17.6.  The progenitor was not detected a year earlier to a limiting
magnitude of 19.6.  Spectra obtained by Blondin et al.\ (2006)
revealed a reddened continuum with strong Balmer emission lines with
very narrow widths of 300 km s$^{-1}$, but with a slightly broader
base of around 1000 km s$^{-1}$ (FWHM).  The spectrum was most similar
to previous SN impostors SN~1999bw and SN~2001ac, and Blondin et al.\
noted that the peak absolute magnitude corresponding to the discovery
magnitude was roughly --14 mag, similar to other luminous LBV giant
eruptions.  Since the object was reddened, however, the true absolute
magnitude at peak was more luminous.  No comprehensive study of this
object has been presented, and we did not secure additional photometry
or spectroscopy.  Adopting $m-M$ = 32.0 mag for UGC~12182, and
correcting also for a rather large line-of-sight Galactic extinction
$A_R$ = 1.168 mag (Schlegel et al.\ 1998) implies that the peak
unfiltered absolute magnitude (approximately $R$-band) was about
--15.47 mag.  This makes SN~2006fp among the most luminous giant LBV
eruptions at its peak.

{\it SN~2007sv:} Located in UGC~5979, SN~2007sv was discovered by
Duszanowicz (2007) on 2007 Dec 20.9 UT with an unfiltered magnitude of
17.4.  Low-resolution spectra obtained by Haratyunyan et al.\ (2007)
show a blue continuum and narrow (FWHM $<$ 1000 km s$^{-1}$) Balmer
emission lines. No comprehensive study of this object has been
presented, and we did not obtain additional photometry or
spectroscopy.  Adopting $m-M$ = 31.57 mag (20.6 Mpc) and a small
Galactic reddening of $A_R$ = 0.046, we find that the absolute
magnitude of SN~2007sv at discovery was roughly $-$14.2 mag,
comparable to several other luminous giant LBV eruptions.

{\it SN~2008S:} This optical transient has been discussed extensively
in the recent literature, with comprehensive photometric and
spectroscopic datasets published by Smith et al.\ (2009a) and
Botticella et al.\ (2009).  The progenitor was very faint at optical
wavelengths, but interestingly, was detected as a bright IR source in
pre-discovery archival {\it Spitzer} data (Prieto 2008; Prieto et al.\
2008; Thompson et al.\ 2009).  Its optical spectrum had bright
[Ca~{\sc ii}] and Ca~{\sc ii} emission lines and bright narrow Balmer
emission lines, with a spectrum very similar to the yellow hypergiant
IRC+10420 (Smith et al.\ 2009a).  Its peak $R$-band absolute magnitude
was $-$13.9 mag, and the light curve from Smith et al.\ (2009a) is
shown in Figure~\ref{fig:lc2}.  Smith et al.\ (2009a) noted expansion
speeds of about 1000 km s$^{-1}$ near the time of peak luminosity,
dropping to about 550 km s$^{-1}$ after a few months.

{\it N300-OT (2008):} This transient was a near twin of SN~2008S in
most ways, including the obscured nature of its progenitor (Prieto
2008; Prieto et al.\ 2008; Thompson et al.\ 2008).  Comprehensive
analyses of the optical photometry and spectra were presented by Bond
et al.\ (2009) and Berger et al.\ (2009).  Its peak $M_R$ was $-$13.3
mag.  Berger et al.\ inferred an expansion speed of 560 km s$^{-1}$
from the widths of emission lines, whereas Bond et al.\ (2009)
suggested a slower speed of only 75 km s$^{-1}$ based on the
separation of the double peaks in emission lines.  To be consistent
with expansion speeds inferred from line widths in other objects, we
adopt FWHM = 560 km s$^{-1}$ as the representative expansion speed in
the discussion below.

{\it SN~2009ip:} This SN impostor is the first in modern times to be
discovered with precursor LBV-like eruptive variability in the decade
leading up to its peak brightness, and its photometry and spectra were
first analyzed in detail by Smith et al.\ (2010a).  A subsequent
analysis of similar spectra by Foley et al.\ (2010) confirmed the
conclusions of Smith et al.\ (2010a).  The unfiltered light curve
presented by Smith et al.\ (2010a) is shown in both
Figures~\ref{fig:lc} and \ref{fig:lc2}.  Smith et al.\ (2010a)
inferred expansion speeds of roughly 600 km s$^{-1}$ from H$\alpha$
line widths, but also noted faster material at 3,000--5,000 km
s$^{-1}$ seen in broad P Cygni absorption features of lines like
He~{\sc i} $\lambda$5876.  This is the first time we have seen
conclusive evidence for a second component with such high speeds
reminiscent of the blast wave around $\eta$ Car (in most other LBVs,
the presence of broad emission wings can be explained by electron
scattering and does not necessarily implicate faster moving ejecta).
Days before submission of this paper, Drake et al.\ (2010) reported
another subsequent outburst of SN~2009ip, apparently satisfying the
expectation of Smith et al.\ (2010a) that ``we should not be surprised
if the eruption continues.''

{\it U2773-OT (2009):} As for SN~2009ip, Smith et al.\ (2010a)
discovered that U2773-OT was also an LBV that exhibited eruptive
variability in the decade leading up to its discovery.  Its spectra
were different, however, with narrower lines and a cooler spectrum,
more closely resembling that of a cool S~Dor phase.  A subsequent
analysis of similar spectra by Foley et al.\ (2010) supported these
conclusions.  We adopt the unfiltered light curve and optical spectra
presented by Smith et al.\ (2010a), although we also update the light
curve with new photometric observations obtained with KAIT (see
Table~\ref{tab:photU2773} and Figure~\ref{fig:lc2}).  Spectra of this
transient indicated expansion speeds of order 350 km s$^{-1}$ (Smith
et al.\ 2010a; Foley et al.\ 2010).

{\it SN~2010da:} This transient occured in NGC~300, following 2 yr
after the well-studied and obscured optical transient in the same
galaxy, and was also discovered by Monard (Monard 2010).  A
comprehensive study has not yet been published for this very recent
source, but based on initial reports (Khan et al. 2010b; Brown 2010;
Elias-Rosa et al.\ 2010; Chornock \& Berger 2010; Berger \& Chornock
2010; Immler et al.\ 2010; Bond 2010), this object appears consistent
with an LBV giant eruption or SN impostor similar to SN~1997bs in some
respects.  Its progenitor was relatively faint at visual wavelengths
but was apparently enshrouded in dust based on the bright IR source at
the same position (Khan et al.\ 2010a, 2010b).  Another preliminary
analysis by Laskar et al.\ (2010) found pre-eruption variability
analogous to U2773-OT and SN~2009ip (Smith et al.\ 2010a), but
detected in the near-IR.  At discovery, it had an apparent $R$
magnitude of roughly 16.0 (Monard 2010), suggesting an absolute $R$
magnitude of $-$13.5, similar to other LBV eruptions and almost
identical to the 2008 transient in NGC~300.  Further detailed study of
this object is currently underway by several groups.

{\it SN~2010dn:} This recent transient was discovered by K.\ Itagaki
(see Nakano 2010) in the nearby galaxy NGC~3184, and the initial
spectrum taken two days after discovery showed narrow Balmer emission
lines and a blue continuum similar to some LBVs, plus visible emission
of [Ca~{\sc ii}] and Ca~{\sc ii} (Vinko et al.\ 2010).  We adopt $m-M$
= 30.4 mag (a distance of roughly 12 Mpc) and $E(B-V)$ = 0.017 mag for
NGC~3184, although the presence of [Ca~{\sc ii}] emission suggests a
dusty circumstellar environment (Smith et al.\ 2009a, 2010a; Prieto et
al.\ 2008) so the true reddening may be higher.  On day 2, SN~2010dn
had an unfiltered magnitude of 17.1 (Nakano 2010), corresponding to a
peak absolute magnitude of roughly $-$13.3.  This is similar to
N300-OT.

We presented the first published spectra of SN~2010dn in
Figures~\ref{fig:spec2010dn} and \ref{fig:vel2010dn}, and we noted
that the overall character of the spectrum was almost identical to
those of SN~2008S and N300-OT, with strong narrow [Ca~{\sc ii}] lines,
intermediate-width emission from the Ca~{\sc ii} IR triplet and Balmer
lines, and a similar continuum suggesting a temperature around 7000 K.
In our high-resolution Keck/DEIMOS spectrum on day 11, the H$\alpha$
line had a FWHM of 900 km s$^{-1}$, which is similar to SN~2008S, but
it displayed a mostly Lorentzian line profile shape unlike its close
cousins.  Atop the Lorentzian H$\alpha$ profile, it also had a weak
additional narrow component, perhaps indicating that it has additional
dense slow CSM irradiated by the transient.  The asymmetric profiles
of the [Ca~{\sc ii}] lines are qualitatively identical to those of
N300-OT (Berger et al.\ 2009). According to Berger (2010),
non-detection of the progenitor in archival {\it HST} images obtained
$\sim$9 yr before discovery implies a $V$-band (F555W) absolute
magnitude fainter than $-$6.3 mag.  Based on the [Ca~{\sc ii}]
emission lines and associations with dusty environments (Smith et al.\
2009a, 2010; Prieto et al.\ 2008), however, it is possible that the
progenitor could potentially be intrinsically more luminous than this.
Study of this transient is still underway at the time of writing.

\subsection{Light Curve Morphology}

The light curves of SN impostors in Figures~\ref{fig:lc} and
\ref{fig:lc2} exhibit a wide variety in both peak luminosity,
duration, and light curve shape.  As we discuss later, light curve
behavior is not necessarily correlated with their spectral properties,
nor does the duration of the event seem to scale with the mass
ejected.  Some events show extremely complex and rapid rises and dips
in absolute magnitude, sometimes multiple times, whereas other events
exhibit a simple 10-yr plateau or single 100 d exponential
decay. There does not seem to be any simple way in which the light
curve properties scale, probably because there are a number of
different physical parameters that can vary in each system: the
progenitor mass and luminosity, the ejected mass and velocity,
different input radiative and kinetic energy budgets, possible binary
encounters, etc.  Without knowing the physical mechanism(s) at work,
it seems difficult to derive clearly meaningful information from the
light curves alone, especially without the benefit of photometric
observations in multiple filters.

\subsection{Color and Temperature Evolution}

Not much is known about the color evolution of SN impostors, and the
question is mired by the possible presence of severe CSM dust and
reddening. There is a relatively poor observational record of the UV
characteristics during SN impostor outbursts.

It is becoming clear, however, that some long-held paradigms are
certainly wrong.  The common wisdom has been that LBV eruptions should
always be seen at relatively cool temperatures of no cooler than
$\sim$7500 K with an F-supergiant like spectrum (e.g., Humphreys \&
Davidson 1994; Davidson 1987), and that they therefore have small or
zero bolometric correction at their peak. The reasoning behind this
expectation is that LBVs develop opaque winds during their eruptions,
with pseudo-photospheres that always tend toward these temperatures
because of the opacity in the wind (see Davidson 1987).

This picture does apply to the cool states of normal S~Doradus
episodes of LBVs (Humphreys \& Davidson 1994; Smith et al.\ 2004), but
it apparently is not always the case in giant eruptions.  Some giant
eruption events do inded fit the bill, of course, with apparent
temperatures of $\sim$7000~K and F supergiant-like spectra (e.g.,
U2773-OT, SN~2002bu, SN~2010dn, etc.).  Other SN impostors such as
SN~2009ip clearly do not fit this expectation, however.  Detailed UV
and optical spectroscopy of V1 in NGC~2366, for example (Petit et al.\
2006), showed a much hotter temperture during its eruption, and
revealed that the temperature and bolometric luminosity actually
increased with time while the optical photometry showed a plateau.
Also, SN~2009ip and other events showed hotter temperatures and a
spectrum unlike an F supergiant.

In this older view, one would expect the coolest temperatures to
coincide with peak luminosity when the pseudo photosphere is the
largest, and that the effective photosphere would either stay at
constant temperature or even get hotter as the eruption subsides,
causing the mass-loss rate and opacity to drop.  Instead, in some
cases where information is available, we see much warmer temperatures
of 12,000 K or more at peak, with the temperture then getting cooler
as the object fades.

Substantial redward evolution with time may be expected from an
explosion that suffers adiabatic cooling as the photosphere recedes
through ejecta, as in a Type II-P event.  Figure~\ref{fig:phot} shows
that the redward color evolution of SN impostors is different from a
normal SN II-P, never becoming as red.  The more subtle drop in the
characteristic emitting temperature with time is similar to a Type~IIn
SN powered by CSM interaction (e.g., Smith et al.\ 2010b), where the
apparent temperature drops with time because the blast wave
decelerates as it sweeps up large amounts of mass (van Marle et al.\
2010).  Dust formation in these events may also cause increased
extinction and reddening, and may therefore have a strong influence on
the apparent color with time.  Further investigation of the color
evolution of SN impostors, including both UV and IR observations, are
sorely needed.

\begin{figure}\begin{center}
\includegraphics[width=3.1in]{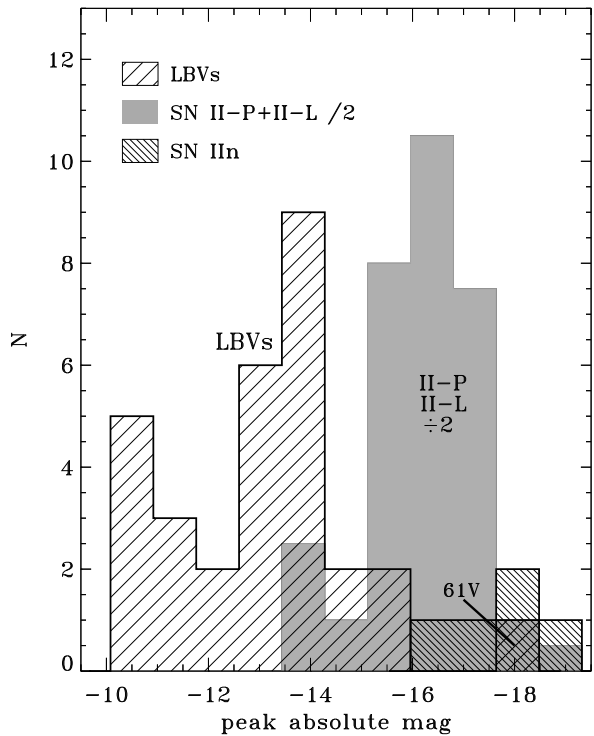}
\end{center}
\caption{Histogram of the peak absolute magnitudes (mostly $R$, but
  some are $V$; from Table~\ref{tab:list2}) for giant LBV
  eruptions. These are compared to the absolute $R$ magnitudes of the
  KAIT sample (Li et al.\ 2010) of 52 normal SNe II-P and 10 SNe~II-L
  (added together and divided by 2; shaded gray), and 5 SNe~IIn.
  Notice SN~1961V to the lower right as the only LBV overlapping with
  SNe~IIn.}
\label{fig:peak}
\end{figure}
\subsection{Peak Absolute Magnitudes}

A key parameter for any transient source is its peak absolute visual
magnitude.  In the case of LBV giant eruptions, this is critical for
evaluating the extent to which the star exceeded its own Eddington
limit.  Unfortunately, we do not have photometry in the same filters
for each source, so we must compare $R$ and unfiltered magnitudes
(generally about the same) to $V$-band or unfiltered visual magnitudes
for historical sources like $\eta$ Car an P Cygni.  For sources where
both $V$ and $R$ are available, the difference is typically not more
than a few tenths of a magnitude, and the qualitative shapes of the
light curves in different filters are similar (e.g., SN~2002bu;
Figure~\ref{fig:phot}).  Absolute peak magnitudes and the
corresponding filters are listed in Table~\ref{tab:list2}.

Figure~\ref{fig:peak} shows a histogram of peak absolute magnitudes
for SN impostors (hatched), compared to the distribution of peak $R$
magnitudes for normal SNe~II-P and II-L (shaded gray) and SNe~IIn
(narrow hatched), taken form the KAIT sample (Li et al.\ 2010).  LBV
eruptions clearly form a separate population that is distinct from
SNe.  While the SNe~II-P and II-L favor peak absolute magnitudes of
roughly $-$16.5$\pm$1.5, the LBVs are skewed to lower luminosity.

There appears to be one clear outlier among the LBVs in
Figure~\ref{fig:peak}, and that is the well-known event SN~1961V with
a peak absolute magnitude of $-$17.8. SN~1961V is an unusual case, and
later in this paper we consider the question of whether it is really
an LBV eruption or something else --- perhaps a genuine SN~IIn
resulting from core collapse, but preceded by an LBV eruption (see
below).  SN~1961V is well within the range observed for SNe~IIn
(Figure~\ref{fig:peak}).

Excluding SN~1961V, all the LBV eruptions span a range of absolute
magnitudes from about $-$10 to $-$16.  There is some overlap with the
low-luminosity tail of the SNe~II-P distribution; we note that objects
like SN~2005cs, SN~2001dc, and SN~1999br (see Pastorello et al.\
2007a) are included in this sample.  As noted by Smith et al.\
(2009a), the color evolution and spectral properties of these
low-luminosity SNe~II-P are quite distinct from SN impostors.  (The
highly reddened SN~2002hh is also included in the SN~II-P sample of
Figure~\ref{fig:peak}, and its true absolute magnitude is more
luminous than its apparent value because of local extinction.)

The distribution of LBV peak magnitudes in Figure~\ref{fig:peak} hints
that there may be two subgroups --- a more luminous class with peak
magnitudes clustered around $-$14$\pm$1.5, and a less luminous group
with peaks of $-$10 to $-$11.5 mag.  Since the SN impostors in this
sample were discovered with widely differing search parameters (it
includes historical objects in our galaxy as well as some discoveries
by amatueres and by systematic surveys like LOSS), we cannot test the
statistical significance of these two luminosity classes. SN impostors
are faint compared to true SNe, and it is thus likely that their
discoveries are highly imcomplete in most SN searches. This is
particularly true for the less luminous group ($-$10 to $-$11.5 mag),
so the true luminosity function of the impostors may look quite
different from what is displayed in Figure~\ref{fig:peak}.  For
example, in a volume-limited sample, there could be more $-$10 to
$-$11.5 mag impostors than the more luminous examples.  This is an
area where future discoveries of SN impostors will be highly
beneficial.  Is it really two groups, or is it a continuous
distribution of luminosities?  How low does the distribution of SN
impostor peak luminosities go?  For the purposes of discussion in this
paper, we tentativey refer to these as relatively low- and
high-luminosity events, while being mindful that there may not be a
true physical separation.

In any case, there is a practical problem with identifying eruptive
peak magnitudes of roughly $-$10 or fainter, which is that we enter
the territory of quiescent absolute bolometric magnitudes for very
luminous stars.  Namely, at the low end of this distribution,
incomplete observational data will make it difficult to reliably
distinguish genuine LBV giant eruptions (increase in bolometric
luminosity) from the more common S~Dor-type excursions.  Recall that
in an S Dor excursion, the star is supposed to brighten at visual
wavelengths because of a change in apparent temperature and radius,
but not necessarily luminosity, which in turn alters the bolometric
correction, so either UV data or detailed atmospheric models become
necessary.  However, recent studies of S Dor excursions may hint that
the traditional view of constant luminosity may need to be modified as
well (e.g., Groh et al.\ 2009), making the situation more murky.

At first glance, it may be tempting to naively group relatively low
and high luminosity eruptions into different subclasses, but this
would be too oversimplified and not necessarily helpful.  In several
cases we have well-studied LBVs where the star suffers multiple
eruptions, qualifying for the low- or high-luminosity category {\it in
  subsequent eruptions of the same star}.  Instead,
Figure~\ref{fig:peak} should be taken as a demonstration of the rather
wide diversity of the eruptive phenomenon in general.  A theory that
attempts to explain the mechanism of the eruptions should strive to
reproduce this diversity.

\begin{table*}\begin{center}\begin{minipage}{5.2in}
\caption{A collection of observed parameters for SN impostors}\scriptsize
\begin{tabular}{@{}lclccccc}\hline\hline
Transient  &M(peak) &$t_{1.5}$  &$V_{exp}$     &EW(H$\alpha$) &Multi-peak   &Sharp Dip  &M(prog.)  \\ 
           &(mag/filt.) &(days)&(km s$^{-1}$) &(\AA)        &(Y/N)        &(Y/N)      &(mag/filt.)    \\
\hline
P Cygni     &-11,-10.5/Vis.             &1800,3600         &136          &...  &Y &N &-9.7/Vis   \\
$\eta$ Car  &-13.5,-13.8,-14,-10.8/Vis. &110,200,4400,3000 &200,650,5000 &...  &Y &Y &-12/Vis   \\
SN~1954J    &-11.3/B   &100            &700           &...   &Y &Y &-7.5/B   \\
SN~1961V    &-17.8/pg  &$\sim$200      &3700          &...   &Y &N &-12.4/pg \\
HD~5980     &-11.1/V   &200            &600           &...   &Y &N &-8.1/V  \\
V1          &-10.5/V   &$>$2000        &300           &...   &N &N &...   \\
SN~1997bs   &-13.8/V   &45             &765           &119   &N &Y &-8.1/V  \\
SN~1999bw   &-12.65/R  &$>$10          &630           &53.7  &N &N &...   \\
SN~2000ch  &-12.8,-12.7,-12.3,-12.1/R &2,25,50,8 &1400 &461 &Y &Y &-10.6/R \\
SN~2001ac   &-14.0/R   &$\sim$50       &287           &46.4  &N &N &...   \\
SN~2002bu   &-14.97/R  &70             &893           &66    &N &N &...   \\
SN~2002kg   &-10.4/R   &365,$\sim$600  &350           &39    &Y &Y &-8.2  \\
SN~2003gm   &-14.4/I   &$\sim$65       &131           &...   &N &N &-7.5  \\
NGC~4656-OT &-10.7/u.f.(R)  &?         &730           &...   &N &N &...   \\
SN~2006bv   &-15.24/u.f.(R) &$>$40     &400           &...   &N &N &$>$-13.5 \\
SN~2006fp   &-15.6/u.f.(R)  &?         &300           &...   &N &N &$>$-13.6  \\
SN~2007sv   &-14.25/u.f.(R) &?         &$<$1000       &...   &N &N &$>$-11.7  \\
SN~2008S    &-13.9/R   &75             &1100          &53.7  &N &N &-6.5/IR   \\
NGC~300-OT  &-13.2/R   &80             &560           &...   &N &N &-7.5/IR   \\
SN~2009ip   &-14.5/u.f.(R)  &7,$>$40   &600,5000      &214   &Y &Y &-9.8/R    \\
UGC~2773-OT &-13?/u.f.(R)   &$>$400    &350           &33    &N &N &-7.8/R    \\
SN~2010da   &-13.5/R   &?              &660           &...   &? &? &...   \\
SN~2010dn   &-13.3/R   &?              &155,860       &...   &? &? &$>$-6.3/V  \\
\hline
\end{tabular}
\label{tab:list2}
\end{minipage}\end{center}
\end{table*}

\begin{figure}\begin{center}
\includegraphics[width=2.9in]{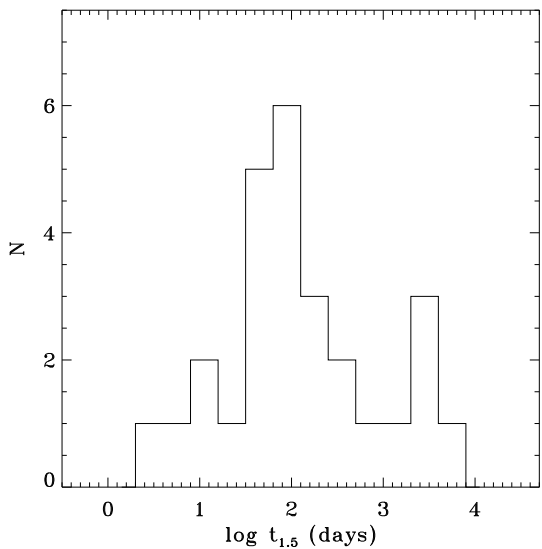}
\end{center}
\caption{Histogram of the logarithm of values of $t_{1.5}$ listed in
  Table~\ref{tab:list2}.}
\label{fig:t15}
\end{figure}
\subsection{Characteristic Rise and Fall Timescales}

The rise timescales are poorly constrained for most SN impostors,
because they are most often discovered around the time of peak
luminosity due to the smaller telescopes used for most transient
searches, while larger telescopes are used mainly for followup
observations.  Adequate information about fainter progenitors is
therefore rare, limited to cases with good archival data.  This
situation is, of course, improving with time as high-quality archives
become more populated with observations, and with transient searches
conducted with larger telescopes.

Discovery near peak implies a rather sudden onset for the brightening
of some SN impostors, and hence, a fast rise timescale comparable to
SNe.  We do see examples, however, of very slow rise times as well, as
in the case of U2773-OT, which rose steadily for at least 5 years, and
is in fact still rising.  This, as well as precursor eruptions and
variability seen in some objects like $\eta$ Car, V12/SN~1954J, and
SN~2009ip, points toward a significant ``preparatory'' phase in SN
impostors.  This is physically meaningful because it signals a growing
instability, rather than a sudden event.  This echoes the fact that
LBV eruptions are apparently sometimes a preparatory phase for the
eventual core-collapse SN, at least in the case of luminous SNe~IIn
(e.g., Smith et al.\ 2007, 2008, 2010b; Gal-Yam \& Leonard 2009; Foley
et al.\ 2007; Pastorello et al.\ 2007).


Timescales for SN impostors to fade from maximum are much better
characterized than their rise times.  In SNe, the rate of decline
provides information about the ejecta mass (i.e., the diffusion time)
and energy source (i.e., radioactive decay rates). In SN impostors,
the direct meaning of the decline rate is not immediately clear, but
characterizing the distribution of relative rates at which SN
impostors fade may eventually help distinguish between models.

For comparison among the sample of sources, we define a timescale,
$t_{1.5}$, as the time in days for a transient to fade by roughly 1.5
magnitudes from its peak.  This is either the time beginning at
discovery or peak visual luminosity, depending on the available
information.  Some cases require exceptions, such as the brief
precursor eruptions of $\eta$~Car in 1838 and 1843, when observations
are imcomplete and we are not sure if the source actually faded by a
full 1.5 mag.  In cases such as this, the value of $t_{1.5}$ is
approximate, and represents the time over which the star appeared to
be fading back to its quiescent level.  The resulting values of
$t_{1.5}$ are listed in Table~\ref{tab:list2}, and a historgram of the
values is plotted in Figure~\ref{fig:t15}.  Cases where more than one
value of $t_{1.5}$ is listed in Table~\ref{tab:list2} correspond to
more than one major outburst observed from the same source.  Since
this is not a complete sample with uniform coverage for each source,
the histogram in Figure~\ref{fig:t15} is meant to convey the range of
timescales observed, rather than a statistically significant
distribution.  Figure~\ref{fig:t15} does not show the typical
timescales for SNe~II-P, which is always close to 100 days.

SN impostors span a wide range of fading timescales, clearly peaked at
durations around $\sim$10$^2$ days.  There are also examples that fade
quickly in only a few days, and several cases that last for a decade.
An important point that distinguishes SN impostors from true SNe is
that the fading timescale does not necessarily tell us anything about
the amount of mass ejected.  Both $\eta$ Car and P Cygni had durations
of $\sim$10 yr for their major eruptions, but from measurements of
their nebulae we know that $\eta$~Car ejected more than 10 $M_{\odot}$
(Smith et al.\ 2003b), whereas P Cygni only ejected about 0.1
$M_{\odot}$ (Smith \& Hartigan 2006).  Furthermore, $\eta$~Car also
showed two brief events of $\sim$100 days duration, when it is
possible that much of the mass may have been ejected (Smith \& Frew
2010), and it had another decade-long outburst in the 1890s when only
0.1--0.2 $M_{\odot}$ was ejected (Smith 2005).  Unfortunately,
measuring the mass for extragalactic SN impostors, where circumstellar
nebulae are not resolved, is impossible without a detailed
understanding of the physical mechanism and the radiative transport
involved in the outbursts.  The sustained energy source of the
decade-long eruptions is therefore unclear, but it is probably not
caused by diffusion from an extremely large mass of ejecta.

The very fast declines correspond to obvious sharp dips in the light
curves, in many cases corresponding to drops in magnitude to the
quiescent progenitor's luminosity or even fainter.  Notable cases of
these sharp dips are SN~2009ip (Smith et al.\ 2010a), SN~2000ch (or
LBV1 in NGC~3432; Wagner et al.\ 2004; Pastorello et al.\ 2010),
SN~2002kg (Van Dyk et al.\ 2006; Maund et al.\ 2006).  The cause of
these is unclear, but Smith et al.\ (2010a) have hypothesized that
they may correspond to sudden ejections of massive shells of material,
which expand quickly and cool while remaining opaque.  After the shell
finally becomes optically thin, the emergent luminosity may return to
its previous level.  This is of course just a working hypothesis;
detailed radiative transfer calculations that include sudden massive
shell ejections would be valuable.

Several objects show evidence for multiple peaks or multiple separate
eruptions.  This is discussed further below, but it is also relevant
to mention here that repeated eruptions in the same source can often
have different timescales.  The delay time or dormant time between
these multiple outbursts may also be highly relevant, perhaps
indicating a recovery timescale or orbital timescale in the case of
binary encounters.  Due to incomplete archival information, it is of
course very difficult to constrain the possible occurance of previous
eruptions in objects where precursors have not been documented.

\begin{figure}\begin{center}
\includegraphics[width=2.9in]{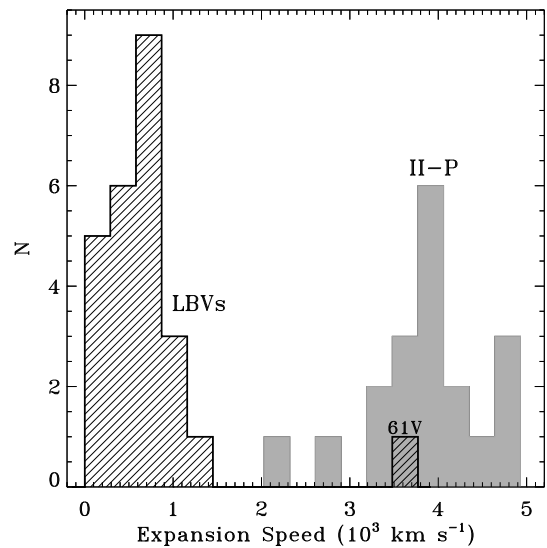}
\end{center}
\caption{Histogram of the expansion speeds for LBV eruptions, measured
  by the FWHM of H$\alpha$ in most cases when spectra are available
  during the eruption, or from the expansion speed of the resulting
  nebulae in the cases of P~Cygni and $\eta$~Car (see
  Table~\ref{tab:list2}).  Expansion speeds for the KAIT sample of
  SNe~II-P is shown for comparison, measured from Fe~{\sc ii} lines in
  the middle of the plateau (see Poznanski et al.\ 2009).  True
  core-collapse SNe~IIn apparently fill the gap between LBVs and SNe
  II-P, with typical line widths of 1000--4000 km s$^{-1}$.}
\label{fig:vexp}
\end{figure}

\subsection{Expansion Speeds and Line Profiles}

While the ejecta mass is difficult to estimate from observations of SN
impostors, their spectra are extremely valuable for providing direct
estimates of the expansion speeds during an eruption.  Aside from
studying the kinematics of spatially resolved circumstellar nebulae in
the nearest examples, which may have been decelerated through
interaction with pre-existing CSM, direct spectra of outbursts are the
only way to understand the kinematics of the ejection.  The speed of
ejection provides important clues to the nature of the star, because
in some scenarios one expects the expansion speed to be related to the
star's escape velocity (i.e., RSG stars have very slow winds of
$\sim$20 km s$^{-1}$, LBVs and blue supergiants have speeds of a few
hundred km s$^{-1}$, and compact WR stars typically have fast winds of
more than 1000 km s$^{-1}$).  The observed expansion speed and its
change with time through the outburst are also critical for
understanding the physics of the eruptive event (explosive blast wave
or sustained wind).  Understanding the distribution of SN impostor
expansion speeds is also relevant for understanding the pre-SN
evolution of SNe~IIn, where narrow lines from the pre-shock CSM can be
observed.

To assess the distribution of expansion speeds for our sample of SN
impostors, we generally took the FWHM value of the H$\alpha$ emission
line, either measured directly in our spectra or quoted from the
literature, as the primary indicator of the expansion speed for the
bulk of the material in the ejecta/wind of SN impostors.  However,
this was supplemented with other information.  For $\eta$~Car and P
Cygni, values of $V_{\rm exp}$ were inferred from the kinematics of
their expanding nebulae (Smith 2005, 2006, 2008; Smith \& Hartigan
2006), and even this is incomplete (e.g., we only listed the polar
expansion speed for the Homunculus nebula of 650 km s$^{-1}$, while a
latitude-dependent range of speeds is seen down to 40 km s$^{-1}$ at
the equator; Smith 2006).  Also, in cases such as SN~2009ip, fast
speeds of $\sim$5000 are seen in absorption only, in He~{\sc i} and
Balmer lines (Smith et al.\ 2010a; Foley et al.\ 2010).  The adopted
values of $V_{\rm exp}$ are listed in Table~\ref{tab:list2} and the
distribution is shown in Figure~\ref{fig:vexp}.

The dominant outflow speeds in SN impostors span a wide range from
around 100 km s$^{-1}$ up to somewhat more than 1000 km s$^{-1}$, with
a peak in the distribution around 600-800 km s$^{-1}$
(Figure~\ref{fig:vexp}).  Note that in Figure~\ref{fig:vexp} we are
aiming for the dominant outflow speed, so we did not include the fast
material moving at $\sim$5000 km s$^{-1}$ around $\eta$~Car or
SN~2009ip, because in both cases this fast material is thought to
constitute a very small fraction of the total mass (Smith 2008; Smith
et al.\ 2010a).  Contrast this with the case of SN~1961V, which showed
an H$\alpha$ FWHM of 3700 km s$^{-1}$ in spectra taken during the peak
of the eruption (Zwicky 1964).  The observed expansion speed in
H$\alpha$ is one of several ways in which SN~1961V is a clear outlier
among the SN impostors, and in fact, we argue later that SN~1961V is
not a SN impostor at all, but is instead a true core-collapse SN.
From Figure~\ref{fig:vexp}, one can see that the expansion speed of
SN~1961V is much closer to the range of speeds seen in SNe II-P than
to the SN impostors.  It is also worth noting that there is little
overlap between speeds in SNe II-P and SN impostors; the KAIT sample
of SNe~II-P in Figure~\ref{fig:vexp} from Poznanski et al.\ (2009)
includes faint and low-energy SNe~II-P such as SN~2005cs.  Speeds
observed in the intermediate components of H$\alpha$ in SNe~IIn occupy
the intermediate zone between SN impostors and SN~II-P, with typical
speeds of 1000--4000 km s$^{-1}$.

In addition to the FWHM values for H$\alpha$, SN impostors also show
remarkable diversity in the detailed shapes of their line profiles.
As described above, one can see lines dominated by a Lorenztian shape,
a Gaussian shape, or a combination with a narrow Gaussian core and
Lorentzian wings.  In several cases (e.g., SN~2002bu discussed
earlier) one sees a transition from a Lorentzian line shape at early
times to a Gaussian shape in the same object, which is also seen in
SNe~IIn (Smith et al.\ 2008, 2010b).  This diversity is not
understood, but is likely related to the changing optical depth of the
wind or ejecta, since electron scattering through dense material will
produce Lorentzian shapes.  Lorentzian profiles are more noticable in
SN impostors than in core-collapse SNe because the intrinsic line
width is narrower, making the scattering wings out to a few 10$^3$ km
s$^{-1}$ easier to see.\footnote{For this reason, SN observers may be
  unfamiliar with Lorentzian line wings, and consequently, some
  observers have fit individual broad components to the line wings and
  erroneously inferred the presence of very fast moving gas.}  Perhaps
detailed radiative transfer modeling of these evolving line shapes
will lead to an understanding of the density of the winds and ejecta,
and hence, the mass ejected in a given event.  SN impostors also show
a range of asymmetry in their line profiles, with some very symmetric
emission lines and some with strong blueshifted absorption (some rare
cases even show {\it redshifted} absorption at high resolution; Berger
et al.\ 2009).  This may depend on either different viewing geometry
from one object to the next or different optical depths and ionization
levels in the winds.

\begin{figure*}\begin{center}
\includegraphics[width=4.8in]{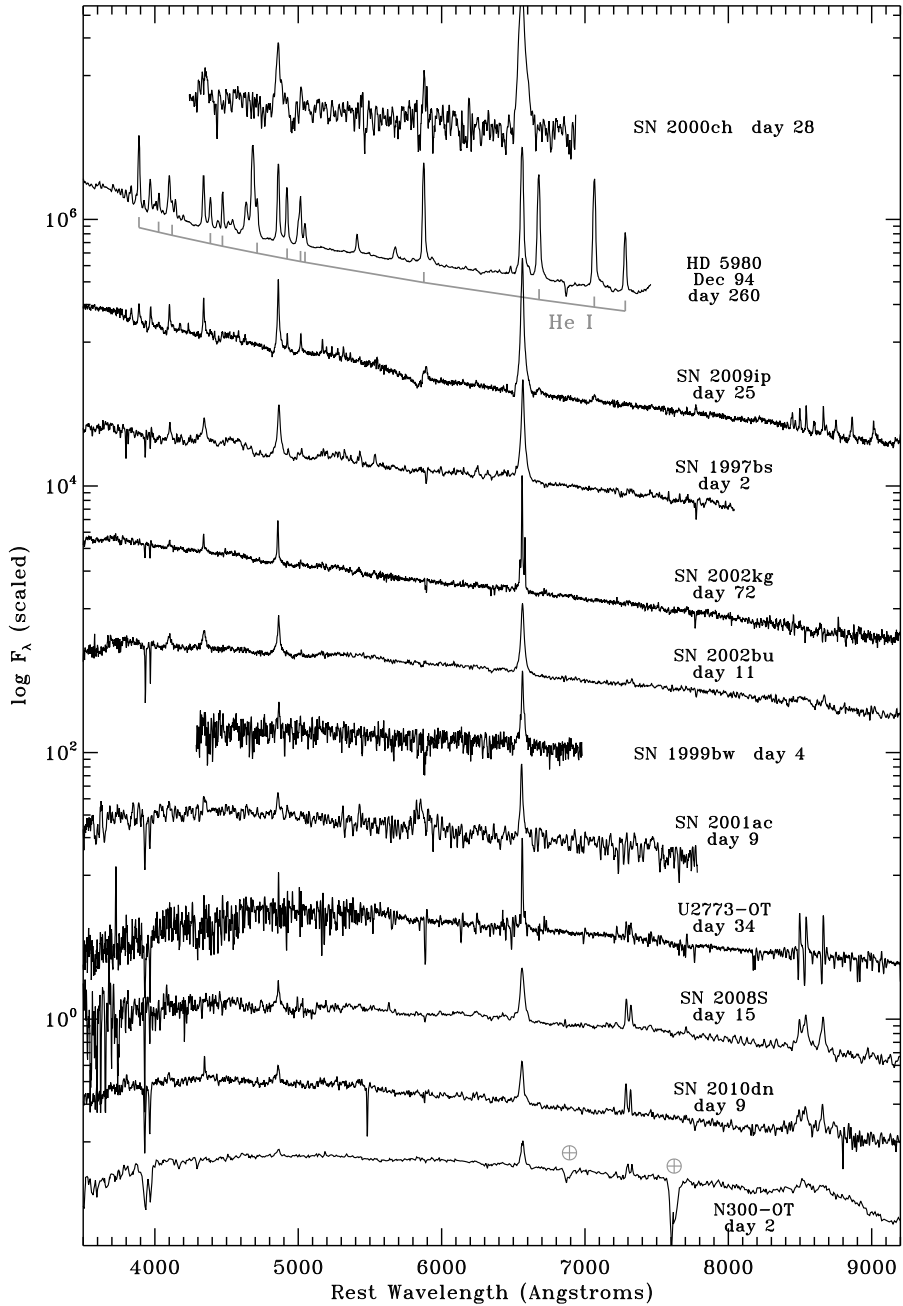}
\end{center}
\caption{A comparison of most of the known SN impostors near their
  time of maximum light.  The spectrum of SN~2008S is from Smith et
  al.\ (2009a), and the spectra of UGC~2773-OT and SN~2009ip are from
  Smith et al.\ (2010a).  NGC~300-OT is a low-resolution spectrum
  taken from Bond et al.\ (2009), which is not corrected for telluric
  absorption (major telluric bands are marked).  Most of the remaining
  spectra are from our own spectral database, although the spectra of
  SN~2002kg (Van Dyk et al.\ 2006) and SN~1997bs (Van Dyk et al.\
  2000) were previously published.  We also show the spectrum of
  HD~5980 a few months after the peak of its LBV eruption on Dec.\ 31
  1994; this is an {\it HST}/STIS spectrum from Koenigsberger et al.\
  (1998), and the locations of numerous He~{\sc i} lines are
  indicated.}
\label{fig:spec}
\end{figure*}

\subsection{Spectral Morphology}

Figure~\ref{fig:spec} displays a gallery of visual-wavelength spectra
of a number of SN impostors, collected from our database and from the
literature (see the caption for Figure~\ref{fig:spec} for references).
We have attempted to gather spectra as close to maximum light as
possible, but complete coverage of SN impositors is not always
available, and so we see a distrubition of times after peak or
discovery.  The spectrum can evolve with time, as illustrated in cases
such as SN~2002bu, so some caution is needed to interpret
Figure~\ref{fig:spec}.  Readers are referred to individual papers to
examine the spectral evolution of each object; this is too large a
topic to describe here.  Nevertheless, it is clear that we do {\it
  not} see consistent spectral evolution in all objects.  While some
transients like SN~2002bu evolve from characteristically ``hot'' to
``cool'' with time, there are other examples which are cool at early
times or examples that remain hot at late times.  Thus, the diversity
in spectral characteristics shown in Figure~\ref{fig:spec} is real and
is likely representative of the class.

All the SN impostors share the common property of strong Balmer line
emission with relatively narrow lines compared to SNe (this is, in
fact, one of the criteria used to classify them as a SN impostor, in
addition to their relatively faint absolute magnitudes).  Beyond that,
there seems to be a wide range of qualitative properties that can be
attributed to the characteristic temperature of the emitting
photospheres or pseudo-photospheres.  Smith et al.\ (2010a) have
discussed the dichotomy of relatively ``hot'' LBVs like SN~2009ip and
relatively ``cool'' objects like U2773-OT, while both are
characteristic of LBVs.  In Figure~\ref{fig:spec} we have attempted to
organize the spectra very roughly with the hotter objects on the top
half and the characteristically cooler objects toward the bottom.  The
``hot'' objects are characterized by smoother and steeper blue
continua, stronger and broader Balmer lines, relatively weak
absorption, and less complex spectra in general.  The ``cool'' objects
tend to have redder continua, weaker and narrower Balmer lines, strong
[Ca~{\sc ii}] and Ca~{\sc ii} emission, deeper P Cygni absorption
features, and in some cases stronger absorption spectra similar to
F-type supergiants (U2773-OT is the best case of this) or to yellow
hypergiants like IRC+10420 (see Smith et al.\ 2009a).  As noted
earlier, there are intermediate objects, and there are some that
transition from reatively hot to cool as time passes.  We do not see
any trend that the hotter objects are necessarily more luminous,
although they do tend to have stronger and somewhat broader Balmer
lines.  It is interesting that the narrow [Ca~{\sc ii}] emission that
was so remarkable in SN~2008S and N300-OT is actually present to
varying degrees in many of the SN impostors. Smith et al.\ (2010a)
have discussed the [Ca~{\sc ii}] and Ca~{\sc ii} lines in more detail,
while Smith et al.\ (2010a) and Prieto et al.\ (2009) have suggested
that these lines may be related to the presence of pre-existing
circumstellar dust.

Few of the objects are hot enough to exhibit He~{\sc i} emission
lines, which is generally quite weak if present, and often fades
quickly with time.  HD~5980 is peculiar in this sense, because it has
extremely strong He~{\sc i} and He~{\sc ii} emission lines (see
Figure~\ref{fig:spec}).  In this case, however, we know that the
eruptive star in the HD~5980 eclipsing binary system has a very
luminous and hot WR companion star, which may strongly influence the
observed spectrum during outburst.  SN~2009ip, SN~2000ch, and one
early epoch of SN~2001ac also show evidence for weaker He~{\sc i}
emission.

\begin{figure*}\begin{center}
\includegraphics[width=4.8in]{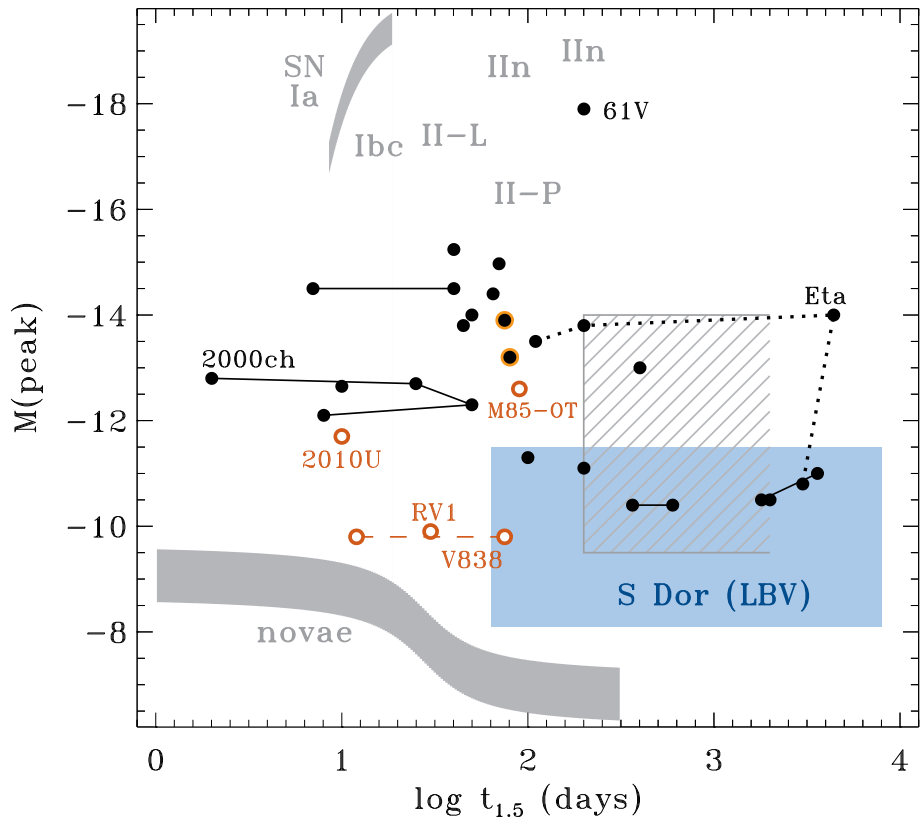}
\end{center}
\caption{Peak absolute magnitude as a function of the characteristic
  fading timescale, $t_{1.5}$, defined in \S 3.5.  SN impostors from
  Table~\ref{tab:list2} are shown as solid black points.  The two of
  these that are outlined in orange are the notable and often debated
  cases of SN~2008S and N300-OT, which do not stand out among the
  population of SN impostors.  SN impostor outbursts for which more
  than one eruption has been documented in the same source are
  connected by solid or dashed black lines.  For comparison in gray,
  we also show the peak luminosity vs.\ $t_{1.5}$ relations for for
  novae (adapted from Della Valle \& Livio 1995), SNe~Ia (adapted from
  Phillips et al.\ 1999), and very rough locations for SNe~II-P, II-L,
  and IIn on this plot (note that SNe~IIn occupy a large area above
  the plot as well, due to very luminous examples).  Locations of the
  intermediate-luminosity transients discussed in \S 4.5, which are
  reputed to be something different from SN impostors, are plotted in
  red (again, two peaks of the V838 Mon eruption are connected by a
  dashed line).  The hatched gray box shows the range of parameter
  space that has been attributed to LBVs in the past (e.g., Kulkarni
  et al.\ 2007), but this is clearly incomplete and only captures a
  few of the SN impostors.  The solid light blue box corresponds to
  normal S Doradus-type outbursts of LBVs (thought to exhibit no
  substantial increase of bolometric luminosity), which partly
  overlaps with giant eruptions of LBVs where the bolometric
  luminosity does increase.  Confidently distinguishing between the
  two cases requires knowledge of the star's quiescent luminosity.  We
  caution that the left side of the plot at moderate luminosities
  might be highly incomplete for LBVs, due to selection effects.}
\label{fig:compare}
\end{figure*}

\subsection{Correlations in observed properties?}

This sub-section could be made very brief by simply stating that there
are no obvious correlations among various observed properties of SN
impostors.  We do not, for example, see any trend between luminosity
and expansion speed, since both fast and realtively slow expansion speeds
are seen among both luminous and relatively faint SN impostors.
Although the description of spectral morphology is more qualitative,
we also see no trend that SN impostors with characteristically ``hot''
spectra are more luminous, or vice versa.  Instead, the main lesson
seems to be that LBVs/SN impostors are highly diverse, occupying a
range of parameters without obvious correlations.

As an example, consider the plot shown in Figure~\ref{fig:compare},
which relates the duration of a transient event to its peak
luminosity, as is commonly done for transient sources.  This is
adapted from a similar plot shown by Kulkarni et al.\ (2007) and
others, although here we have expressed the duration in terms of a
somewhat different quantity, $t_{1.5}$.  SNe~Ia and novae obey clear
relations described elsewhere, and core-collapse SNe tend to be fairly
localized (except for SNe~IIn).  However, the eruptions of LBVs or SN
impostors essentially fill the entire range of the so-called ``gap''
in Figure~\ref{fig:compare} between SNe and novae, covering timescales
from a day to decades, and ranging over two orders of magnitude in
peak luminosity.  As noted elsewhere in this paper, we even have cases
where the same star suffered multiple eruptions that appear in very
different places in Figure~\ref{fig:compare} (these cases are
connected by solid or dashed lines).  Although there is no obvious
``main sequence'' along which LBVs reside in Figure~\ref{fig:compare},
there does seem to be a concentration around $t_{1.5} \approx 50-100$
days and M(peak) $\simeq$ $-$14 mag.  This common location for SN
impostors includes the well-studied and often debated events SN~2008S
and N300-OT.  The transient M85-OT also appears to reside comfortably
among the most common types of SN impostors, and is not exceptional in
this regard.

As a starting point, then, it may be prudent for theory to focus on
possible physical mechanisms that can lead to $\la$100 day events with
peak luminosities of 2$\times$10$^7$ $L_{\odot}$, and to then explore
variations in physical parameters that extend this parameter space.
The wide observed diversity may be attributed to a huge range in
possible physical parameters, such as ejected mass, explosion energy,
progenitor mass and luminosity, Eddington factor, etc., which can
obviously affect quantities like the relevant thermal or diffusion
timescales, and the photospheric radius with time during a transient.
The few cases where we have detailed estimates of ejected mass and
energy already exhibit differences of orders of magnitude.

Some of the observed diversity may also depend strongly on previous
recent mass-loss history.  For example, the amount of local dust
extinction around the progenitor, and by extension the presence of
strong [Ca~{\sc ii}] emission lines, IR excess, or perhaps absorption
lines in the spectrum, may depend on how recently the star suffered a
previous outburst that created a dense and dusty CSM.  The same
progenitor star might look extremely different depending on how much
time has elapsed since the last eruption, and how much mass was
ejected in that event.  A hypothetical eruption of $\eta$~Carinae that
is identical to its 1890 event could look very different if it were to
occur a few hundred years from now when the Homunculus nebula has
largely dispersed. Thus, the observed diversity in spectra and color
of transient events may not necessarily be tied to diverse physical
properties of the outbursts themselves.  There is no clear reason to
expect the previous mass-loss history to be correlated with other
observed properties, of course.

To make matters worse, it is well established that the eruptive mass
loss of LBV eruptions can be strongly nonspherical, and so observed
properties may depend on viewing angle.  For example, the outflow
speed that one would derive from spectra of $\eta$~Carinae would
appear to be $\sim$650 km s$^{-1}$ if it were viewed from a latitude
near the pole, but one would infer a much slower outflow speed of
40--100 km s$^{-1}$ if an observer happened to be looking from a low
latitude projected along the equator (see Smith 2006).  This may well
play a role in some of the diversity in outflow speed of SN impostors
in Figure~\ref{fig:vexp}.  Similarly, the amount of line-of-sight
extinction toward a source may be very latitude dependent if it arises
in the local circumstellar environment.  If a progenitor star were
surrounded by a dusty torus such as those commonly thought to reside
around supergiant B[e] stars, for example, an observer situated near
the equator might deduce that the progenitor was completely obscured
and enshrouded.  (A cautionary note is that this same observer would
then {\it underestimate} the star's bolometric luminosity by a factor
of 5-10 if that estimate were based on the measured IR luminosity.)
We might expect something like 10\% of SN impostors to be viewed from
low latitudes, so perhaps a few cases of heavily obscured progenitors
is not so surprising.  Again, there is no expectation that viewing
angle will correlate with any other observed property, except perhaps
extinction.

\section{DISCUSSION}

\subsection{Progenitor Star Diversity}

One of the most important clues to the nature of SN impostors is the
initial mass and evolutionary stage of the progenitor star in its
quiescent state before the outburst.  Unfortunately, this information
is rarely available and hard to come by, and detection bias for
progenitors tends to favor cases where the progenitor star is
relatively luminous.  An added difficulty is that, in the absence of
eclipsing binaries like HD~5980, it is of course always difficult to
measure the star's mass, which depends on evolutionary models, assumed
reddening, uncertain bolometric corrections, assumed inclination and
geometry of obscuring material, etc.

It has been well-established that the instability we associate with
LBVs occupies a large range of initial mass, from the most massive
stars that may exist down to about 20-25 $M_{\odot}$ (Smith et al.\
2004).  It is therefore no surprise that several of the SN impostors
appear to have very luminous progenitor stars within this range (e.g.,
SN~2009ip, SN~1997bs, $\eta$ Car, P Cygni, HD~5980, etc.).  A few
examples seem to have progenitor stars around the lower bound of this
mass range at $\sim$20 $M_{\odot}$ (U2773-OT, V12/SN~1954J, V1 in
NGC~2366), while there is suggestive evidence that the $\sim$18--20
$M_{\odot}$ progenitor of SN~1987A may have experienced an LBV-like
episode in its pre-SN evolution (Smith 2007).

However, the recent discovery of the relatively faint obscured
progenitors of objects like SN~2008S and N300-OT was a surprise, and
seems to extended this range of initial masses well below 20
$M_{\odot}$.  Given the slope of the initial mass function, we may
expect to see more of these events in coming years, hopefully with
identifiable progenitor stars.  An extremely interesting open question
raised by SN~2008S and N300-OT is just how low in initial mass stars
may experience extreme LBV-like eruptions.  Does the eruptive
phenomenon extend even below the lower limit for core-collapse SNe at
$\sim$8 $M_{\odot}$, and if so, are sudden energetic bursts important
for the formation of some planetary nebulae?  Thompson et al.\ (2009)
have touched upon this issue, but more examples and better constraints
in the progenitor stars are needed.  This is discussed further below.

The recent recognition that eruptions similar to LBVs may occur in
moderately massive stars with initial masses below 20 $M_{\odot}$ has
rather profound consequences.  While the ultimate trigger and physical
mechanism for LBV giant eruptions remains unknown, it has generally
been accepted that the eruptive behavior is the consequence of these
stars approaching or exceeding the classical Eddington limit.  If the
progenitor stars of SN~2008S and N300-OT really did have initial
masses well below 20 $M_{\odot}$, this is surprising and informative,
since {\it stars in this mass range will never approach the classical
  Eddington limit in the normal course of their evolution}.  The most
massive stars with initial masses above 60 $M_{\odot}$ will naturally
and unavoidably be driven to a super-Eddington state in their
post-main-sequence evolution, while stars with initial masses of
25--40 $M_{\odot}$ may approach the Eddington limit in a post-RSG
phase, after they have shed significant mass and thereby raised their
L/M ratio.  Stars below 20 $M_{\odot}$, however, have relatively tame
luminosities and do not have mass-loss rates high enough to bring
their L/M ratios to such dangerous levels.  Thus, while more massive
stars can easily exceed the Eddington limit temporarily with small
adjustments in opacity or stellar structure, lower mass stars require
a substantial input of extra energy to bring them to the exceptional
peak luminosities observed and to successfully eject large amounts of
mass.  While $\eta$~Car was about 5 times the Eddington luminosity at
the peak of its giant eruption, the SN impostors SN~2008S and N300-OT
that reached a similar peak luminosity had Eddington factors of more
like 40--80 (see Smith et al.\ 2009; Bond et al.\ 2009).

\subsection{Pre-Outburst Variability (or not) and Multiple Eruptions}

Very few of the eruptive transients discussed here have information
about the pre-outburst progenitor star.  When this information is
available, though, it is extremely valuable.  While progenitor
detections are hard to come by, multi-epoch progenitor detections are
even more rare.

Nevertheless, based on improving archival data, the observational case
is building that several SN impostors experience a phase of growing
instability that can precede the most dramatic brightening (usually
associated with the time of discovery) by a few years or decades.  A
classic example of this is $\eta$ Carinae, which showed a slowly
increasing visual magnitude for a century before its mid-19th century
eruption, but then -- more remarkably -- showed several very brief
precursor brightening events before the main extended bright phase of
its eruption (see Smith \& Frew 2010).  V12/SN~1954J is another key
historical example, which showed very peculiar and erratic variabiltiy
for 5--10 yrs before its giant eruption.  More recent examples include
SN~2009ip and U2773-OT, which showed slow $\sim$5 yr episodes
preceding their eruptions (Smith et al.\ 2010a). SN~2000ch has shown
multiple recurrances of brief brightening episodes (Pastorello et al.\
2010), and SN~2009ip has now exhibited another eruption $\sim$1 yr
later (Drake et al.\ 2010).  HD~5980 exhibited some minor brightening
episodes before its major eruption, and of course P Cygni suffered a
second eruption 55 years after the beginning of its first major
eruption (Smith \& Frew 2010). At the very least, the presence of
multiple recurring outbursts is a strong indication that the stars
survive these events, and that the underlying physical mechanism is
not a terminal event such as a core collapse, an electron capture SN,
or a failed SN.  This is discussed more below.

A critical point is that if these stars can experience multiple
outbursts on relatively short timescales, then {\it there is no
  gaurantee that a given transient event is the first one experienced
  by that star.}  A given progenitor may be in a state where it is
still recovering from a previous recent burst, which may have been an
extremely disruptive event, while any dusty CSM surrounding that
progenitor may have been ejected in a very recent but undocumented
previous eruption.

Thus, one must be cautious in interpreting the significance of a given
progenitor's observed properties -- especially if it is based on a
single epoch or a brief range of time.  This is perhaps an area where
much longer time baselines from plate archives may be of substantial
benefit.  The luminosity one infers from an observation of a
progenitor is not necessarily the {\it quiescent} bolometric
luminosity of the star or the normal state of that star, and may
therefore cause erroneous estimates of that star's initial mass.
Furthermore, if the progenitor was heavily obscured, one must be
careful in making direct comparisons to classes of stars that are
always heavily enshrouded, like OH/IR stars or AGB stars, because the
obscuring dust may have a very different origin in a recent eruption.
Thus, initial masses and evolutionary states derived from progenitor
observations must be taken with a grain of salt.  Ideally, one would
like to combine information about the progenitor with estimates of the
ages of surrounding stars, as Gogarten et al.\ (2009) did for N300-OT.
More studies of this type may help advance the field significantly.

\subsection{Outburst Diversity:  Explosions or Winds?}

All massive stars in the local universe have considerable
radiation-driven stellar winds, and these winds become a dominant and
defining characteristic for evolved supergiants and hypergiants.
Namely, the strong emission lines that define WR stars, LBVs, blue
supergiants, and yellow hypergiants are caused by their extended and
often partially opaque stellar winds.  Given the similarity between
the spectra of SN impostors and those of Galactic LBVs and
hypergiants, it is natural to conclude that extreme winds are also the
key physical mechanism in SN impostors.  Detailed modeling of the
spectra for a few events, like V1 in NGC 2366 (Petit et al.\ 2006) has
demonstrated this.

However, some recent clues also suggest that hydrodynamic expulsion of
the stellar envelope may be at work in some eruptions.  This was
suspected based on the ratio of kinetic to radiated energy in
$\eta$~Carinae (Smith et al.\ 2003b), and the presence of an energetic
blast wave was later confirmed by the discovery of extremely fast
ejecta surrounding this star (Smith 2008).  Similar fast material has
now also been seen in SN~2009ip (Smith et al.\ 2010a; Foley et al.\
2010).  Thus, these strong blast waves imply that dynamic explosions
are an important ingredient in at least some LBV giant eruptions, in
addition to extremely strong winds.  While the total mass lost in an
event may be the same for an extreme but temporary wind as compared to
an explosion, the corresponding implication for the underlying
physical mechanism is quite important.  An explosion implies a severe
restructuring of the star on a dynamical timescale, requiring a deep
deposition of energy inside the star, and the radiative transient we
see is an after-effect.  In the case of a wind, the implication is
that the luminosity of the star has increased, and that increase in
luminosity causes mass to be lifted from the surface of the star in
quasi-steady-state.  Given the complex light curves of some SN
impostors, it is easy to imagine a hybrid situation where an initial
shock heats the envelope, and the resulting increase in radiative
luminosity drives a strong wind.  Furthermore, one can imagine that a
range of energy deposition could lead to a large diversity of observed
phenomena ranging from enhanced winds to explosions, as explored by
Dessart et al.\ (2009).  In other words, it seems possible that the
diversity in winds and explosive phenomena might be different
manifestations of the same basic energy deposition.

With the possibility of explosive mechanisms for the origin of SN
impostors also comes the possbility that their luminosity might be
enhanced or even dominated by interaction of the blast wave with dense
CSM, as in traditional SNe~IIn, but with lower-energy shock waves.
This hypothesis has not been explored much in models for the spectra
of SN impostors, but seems quite promising given the remarkable
spectral similarities between SN impostors and SNe~IIn (see Smith et
al.\ 2010a; Foley et al.\ 2010).

\begin{figure}\begin{center}
\includegraphics[width=3.0in]{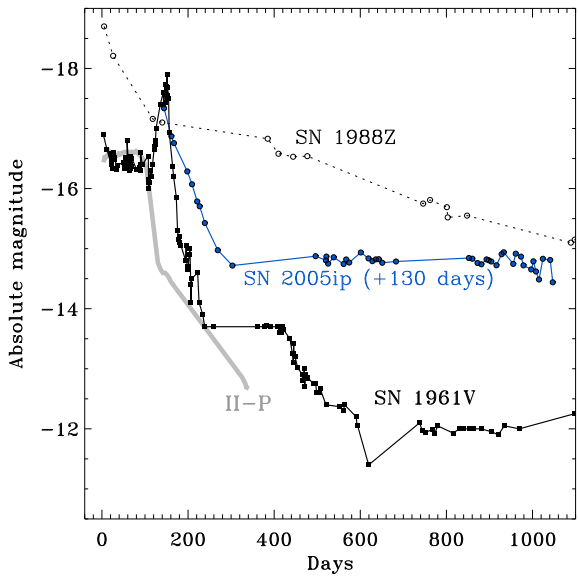}
\end{center}
\caption{Same as Figure~\ref{fig:lc2}, but comparing the light curve
  of SN~1961V to that of a normal SN~II-P and two examples of
  well-studied SNe~IIn.  The light curves of SN~1999em (II-P),
  SN~2005ip (IIn), and SN~1988Z (IIn) are the same as they appear in
  Figure~1 of Smith et al.\ (2009b), except that the light curve of
  SN~2005ip is shifted by +130 days for reasons discussed in the
  text.}
\label{fig:61v}
\end{figure}

\subsection{SN~1961V:  LBV mega-eruption or a true core-collapse SN~IIn?}

In light of the distribution of LBV eruption properties, a close
re-examination of SN~1961V is worthwhile, since it is the prototype of
Zwicky's original Type~V class of SNe, it has colored much
interpretation of the SN impostors because of some similarity to
$\eta$ Carinae, and it is a rare case where the progenitor star was
identified in the decades before the event.  Goodrich et al.\ (1989)
first made the case that SN~1961V was not actually a core-collapse SN,
but was instead an exaggerated $\eta$ Carinae-like outburst; this was
based on the detection of intermediate-width (2000 km s$^{-1}$)
H$\alpha$ emission at the expected position of the SN in a
ground-based spectrum taken 25 yr after the peak of the eruption.
Filippenko et al.\ (1995) tentatively identified a red source seen in
early {\it Hubble Space Telecope} ({\it HST}) images, suggesting that
this may be the dusty source predicted by Goodrich et al.  Later
imaging studies with the refurbished {\it HST} disagreed on which
source was coincident with SN~1961V (Van Dyk et al.\ 2002; Chu et al.\
2004).  Despite this disagreement about which star is the survivor,
SN~1961V is usually regarded as a prototype of the SN impostors
because it was well studied and was the original ``Type V'' supernova.
It is ironic, then, that our present comparison finds SN~1961V to be
an extreme outlier among the class of SN impostors in every measurable
way.  This begs the obvious question: {\it Was SN~1961V really an
  impostor?}

After considering the distribution of properties among LBVs and
SNe~IIn, the answer seems to be ``Probably not.''  The original
motivation for linking SN~1961V to $\eta$~Car was the relatively
narrow width of its emission lines compared to SNe~II-P, plus its slow
and unusual light curve evolution.  However, these and essentially all
of its observed properties are consistent with the class of true
SN~IIn, where the narrow lines and extra luminosity are thought to
arise from core-collapse explosions interacting with dense CSM.  The
Type~IIn class was not yet recognized at the time of the outburst
(leading to Zwicky's suggestion of a new Type V), but there has been
much progress in understanding the properties of SNe~IIn in recent
years.

Based on photographic spectra taken during the peak of the outburst,
Zwicky (1964) inferred an expansion speed of 3700 km s$^{-1}$ from the
width of H$\alpha$, while Branch and Greenstein (1971) estimated
$V_{\rm exp}$ $\simeq$ 2000 km s$^{-1}$ based mainly on calculated
fits to Fe~{\sc ii} and similar lines in a series of spectra taken at
different times during the event.  Goodrich et al.\ (1989) estimated
2100 km s$^{-1}$ from H$\alpha$ in the very late-time spectra.
Humphreys \& Davidson (1994) contended that such narrow lines meant
that SN~1961V was ``definitely'' not an ordinary SN and more closely
resembled $\eta$~Car.  However, the conjecture that SN~1961V was not a
true SN based on its narrow lines is not valid.  Most {\it bona-fide}
SNe~IIn have line widths of 1000--4000 km s$^{-1}$ (e.g., Chugai et
al.\ 2004; Smith et al.\ 2008, 2009b, 2010b; Filippenko 1997); even
some of the most luminous SNe known have lines as narrow as 1000 km
s$^{-1}$ (Smith et al.\ 2008, 2010b; Prieto et al.\ 2007).  Looking at
Figure~\ref{fig:vexp}, the expansion speed inferred from line widths
in SN~1961V is clearly more in-line with normal SNe than with the rest
of the LBV eruptions.

The light curve of SN~1961V -- while complex and quite unusual -- also
does not provide a very compelling case that it was not a true SN.
Figure~\ref{fig:61v} compares the light curve of SN~1961V to that of a
normal SN~II-P and to those of two well-studied SNe IIn: SN~1988Z and
SN~2005ip.\footnote{Here we find that the peak absolute magnitude was
  almost $-$18 mag.  Humphreys \& Davidson (1994) chose to adopt the
  closest of published distances to NGC~1058 ($m-M$=28.6 mag; 5.3
  Mpc), making the peak magnitude $-$16.4, which is still brighter
  than any other SN impostor and comparable to normal SNe II-P.  Most
  estimates, however, favor a larger distance and therefore a higher
  luminosity for SN~1961V.  The expanding photosphere method applied
  to SN~1969L gives $m-M$=30.13$-$30.25 mag for NGC~1058 (Schmidt et
  al.\ 1992; 1994), whereas the Hubble flow distance (assuming $H_0$ =
  73.0 km s$^{-1}$ Mpc$^{-1}$) gives $m-M$=29.77 mag.  We therefore
  adopt $m-M$=30.0 mag, and also correct the light curve for a
  Galactic extinction value of $A_B$=0.27 mag (the $B$-band extinction
  is probably most appropriate for the photographc magnitudes in the
  light curve), making the peak absolute magnitude roughly $-$17.8,
  far exceeding any other SN impostor.}  The long decay time for
SN~1961V is easily accounted for by continued CSM interaction at late
times; both SN~2005ip and SN~1988Z were more luminous for a longer
time.  SN~1961V is thus intermediate between these classic SNe~IIn and
a normal SN~II-P (i.e., at no time is it less luminous than a normal
SN II-P), making it a somewhat less extreme version of CSM interaction
than SN~2005ip.  The absolute magnitude of the brightest peak in
SN~1961V's light curve was almost identical to that of SN~2005ip.  The
rather stark interruptions in its decline (interpreted as late
``plateaus'' by Humphreys et al.\ 1999) also find clear precedent in
SN~2005ip, whose light curve declined rapidly until day 160 when it
abruptly hit a floor and remained at the same luminosity (or even rose
slightly) for years afterward (Smith et al.\ 2009b).

Still, the light curve of SN~1961V is admittedly a bit unusual
compared to most SNe~IIn.  The key property that makes it seem unique
is that it shows an initial luminous plateau at $-$16.5 mag for
$\sim$105 days, followed by a second more extreme peak reaching almost
$-$18 mag before declining rapidly thereafter.  This can potentially
be understood as a superposition of a normal SN~II-P plateau (like
SN~1999em) followed by a late-time addition of luminosity from
enhanced CSM interaction, as seen in a SN~IIn like SN~2005ip.  This
superposition is shown schematically in Figure~\ref{fig:61v} with the
light curve of SN~2005ip (Smith et al.\ 2009b) shifted by +130 days
for comparison.  The only requirement here, from the CSM-interaction
point of view, is that the CSM shell had an inner cavity of lower
density than the main shell, so that the time when the blast wave
struck the densest part of the CSM shell was delayed by 120-130 days.
A delayed turn-on of the CSM interaction luminosity is understandable
with a thin dense shell at a large radius (e.g., van Marle et al.\
2010).  A late turn-on even has clear observational precedent among
SNe~IIn: an extreme case is the recent SN~IIn~2008iy, which started
with a luminosity comparable to a SN II-P, but continued to rise
slowly for $\sim$400 days (Miller et al.\ 2010).  

With a SN blast wave expansion speed of $\sim$4000 km s$^{-1}$
(adopting Zwicky's estimated speed from spectra during the event), the
shock would have struck the shell after day $\sim$105 when the rise to
the bright peak began if the shell had a radius of $\ga$250 AU.  This
is entirely plausible given the observed shells around known LBVs and
the shells inferred around other SNe~IIn.  If that LBV shell had
initially been ejected at a few hundred km s$^{-1}$ (also typical of
LBVs) it would imply that the shell had been ejected within $\sim$5 yr
before the final SN explosion.  Indeed, $\la$1 yr prior to the
beginning of the main peak, SN~1961V was already in a precursor
outburst state with an absolute magnitude of $-$14.5 (see
Figure~\ref{fig:lc2}), which is quite similar to $\eta$~Car and other
LBV eruptions.  This provides a self-consistent picture, where
SN~1961V suffered a precursor LBV outburst that was followed within a
few years by a true core-collapse SN~IIn.

While somewhat complicated, this scenario fits in well with current
ideas about SNe IIn, and it is appealing because it no longer requires
the progenitor to have been an astoundingly massive $\ga$240
$M_{\odot}$ star that is substantially more massive than any known in
the local universe.  If the high pre-maximum luminosity is attributed
to an LBV-like outburst rather than the quiescent star, and if the
peak outburst was a genuine core-collapse SN explosion, then
conjectures that the progenitor was incredibly massive (up to 2000
$M_{\odot}$; Utrobin 1984, Goodrich et al.\ 1989; Humphreys \&
Davidson 1994) are clearly erroneous.  It also relieves the difficulty
of trying to account for the tremendous energy budget of SN~1961V with
a non-terminal event.

If this surmize is true, then it is the first definitive detection of
a precursor LBV outburst prior to a SN~IIn, further strengthening the
LBV/SN~IIn connection.  This builds upon earlier results of the
precursor LBV-like outburst before the unusual Type~Ibn event
SN~2006jc (Pastorello et al.\ 2007b; Foley et al.\ 2007), as well as
the detected progenitor of the SN~IIn 2005gl that was inferred to be
an LBV-like star (Gal-Yam \& Leonard 2009).  In fact, it remains
possible that the progenitor identified as a possible LBV by Gal-Yam
\& Leonard (2009) could have been in an eruptive state at the time the
pre-discovery archival data were obtained, although this is
uncertain. A decade before SN~1961V, the progenitor was observed at a
bolometric magnitude of about $-$12.4, similar to the quiescent
bolometric luminosity of $\eta$~Carinae.  Like Goodrich et al.\
(1989), we infer that this is likely to be the quiecent bolometric
luminosity of the progenitor star, making it comparable to the most
luminous stars known.  Since the progenitor resided in a giant H~{\sc
  ii} region similar to the Carina Nebula, this appears reasonable.

In this continued CSM-interaction context, the undulations in the late
time decay of SN~1961V are also easily explained by simply assuming
that the expanding blast wave overtook a series of additional shells
at larger radii, causing a small and temporary enhancement in the
luminosity.  SN~1988Z, SN~2005ip, and other SNe~IIn demonstrate that
this is achievable.  The luminosity of a SN~IIn can turn on or off at
any time, depending on the density of material it is running into.  If
the CSM environment was dusty, some of this late-time luminosity may
also be attributable to a reflected light echo (remember that the
SN~1961V historical light curve is in photographic magnitudes, which
favor blue wavelengths).  As noted earlier, putative detections of the
surviving star in recent times have been controversial, so proof of
the SN~IIn hypothesis remains elusive based on modern data.  Even if
the a source is detacted at the correct position, however, it may
still be fueled by weak CSM interaction at late times, or it may be
another star in the crowded star cluster.  For example, Li et al.\
(2002) detected the SN IIn 1995N many years after explosion, while
SN~1988Z still remains luminous.  The conjecture by Chu et al.\ (2004)
that the H$\alpha$ source identified in their data (object 7) ``cannot
be the SN or its remnant because of the absence of forbidden lines''
is incorrect if the late-time luminosity is powered by CSM interaction
rather than by the radioactive decay tail.  Stockdale et al.\ (2001)
detected a non-thermal radio continuum source at the position of
SN~1961V, and Chu et al.\ (2004) showed that this radio source is
coincident with the only strong H$\alpha$ emission-line star in the
cluster.  The radio and narrow H$\alpha$ emission are certainly
consistent, in principle, with the strong continued CSM interaction
that one may expect in the SN~IIn hypothesis.  Furthermore, the
presence of dust inferred from strong IR emission in the late time
data of SN~1961V is also consistent with the SN~IIn hypothesis, as a
strong IR excess from new dust formation and from an IR echo were both
seen in SN~2005ip (Smith et al.\ 2009b; Fox et al.\ 2010).

We conclude, therefore, that the peak of SN~1961V was probably not a
SN impostor after all, but a {\it bona-fide} SN~IIn caused by a
core-collapse event.  We suggest that the initial peak for the first
$\sim$105 d was akin the plateau of a normal SN~II-P (this does not
exclude the possibility of narrow lines from CSM interaction being
present at that time), while the so-called ``super outburst'' when
SN~1961V reached $M_{\rm pg} \approx -18$ mag and then faded rapidly
may have been powered by CSM interaction as in a SN~IIn like
SN~2005ip.  In this scenario, the essential difference between SN
1961V and a conventional SN~IIn is that the CSM interaction was {\it
  delayed}, probably because the CSM shell was at a large radius with
an interior cavity.  In future studies, readers are therefore advised
to disregard the fact that SN~1961V was included in figures in this
paper comparing the light curves and other properties of LBV
eruptions, at least to the extent that these plots are taken as
indicative of LBVs.

\begin{figure}\begin{center}
\includegraphics[width=3.0in]{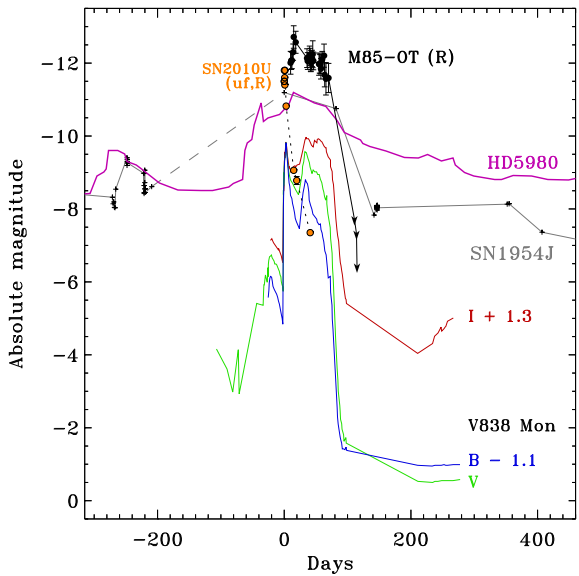}
\end{center}
\caption{Absolute magnitude light curves of the transients V838~Mon
  and M85-OT, along with the LBVs V12/SN~1954J and HD~5980 for
  comparison.  V12 and HD~5890 are the same as in previous figures in
  this paper.  The $B$, $V$, and $I$-band light curves (blue, green,
  and red, respectively) of V838~Mon are from Sparks et al.\ (2008),
  while the $R$-band light curve of M85-OT is from Kulkarni et al.\
  (2007).  The combined unfiltered and $R$-band light curve of
  SN~2010U is also shown as orange dots and a dotted black line, from
  Humphreys et al.\ (2010; and references therein).}
\label{fig:novae}
\end{figure}

\subsection{Other Intermediate-Luminosity Transients} 

While the distinction between LBV eruptions and true core-collapse SNe
may be more clear after considering the distribution of LBV eruption
properties, the bottom end of the LBV distribution remains nebulous.
Drawing a clear dividing line between true LBV giant eruptions and
``normal'' S~Doradus eruptions is not as easy as previously suggested
(e.g. Humphreys \& Davidson 1994), since the notion that S~Doradus
eruptions always occur at constant bolometric luminosity has not
withstood rigorous analysis (Groh et al.\ 2009), and the conjecture
that these eruptions should always have atmospheres with temperatures
around 8000~K is also incorrect.  For example, it is unclear if the
eruptions of HD~5980 and SN~2002kg do indeed qualify as giant LBV
eruptions, since it is not clear that they experienced a substantial
increase in bolometric luminosity, and the total amount of mass and
energy lost were not much in excess of their quiescent states.  Of
course, defining a transient as a giant LBV eruption or S~Dor outburst
at the lower end of SN impostor luminosities (see
Figure~\ref{fig:compare}) also requires reliable knowledge of the
progenitor's luminosity --- in order to decide of the bolometric
luminosity has indeed increased --- which is often not available.
(For the SN impostors with relatively high peak luminosities above
$-$13 mag, this is not a problem because no stars have a quiescent
luminosity this high, and so the bolometric luminosity must have
increased substantially.)


Furthermore, the bottom end of the luminosity distribution for SN
impostors also overlaps with transients that may not really be LBV
eruptions, and might not even be associated with massive stars.
Thompson et al.\ (2009) have proposed a new sub-class of transients
where the progenitor star was heavily obscured and had relatively low
bolometric luminosities, exhibiting [Ca~{\sc ii}] emission in addition
to narrow Balmer emission lines.  This was inspired largely by the
discovery and detailed observations of SN~2008S and N300-OT, discussed
extensively above.  However, whether these transients constitute an
entirely new class of outbursts, or if they instead represent an
extension of LBV-like eruptions to lower masses than previously
thought (i.e. below 20--25 $M_{\odot}$) is controversial (see Smith et
al.\ 2010a for a recent summary of the debate; see also Thompson et
al.\ 2009; Smith et al.\ 2009; Prieto et al.\ 2010).  The source of
the disagreement is that all of the properties attributed to this
putatively new class of objects are already observed among known LBVs.
Obscuring dust shells are certainly common among known LBVs, while we
have demonstrated here (Figure~\ref{fig:spec}) that the presence of
[Ca~{\sc ii}] emission is seen in many of the SN impostors to varying
degrees, although it had not been emphasized in discussions before
2008, and these lines are seen in hypergiants with very strong winds
like IRC+10420 (Smith et al.\ 2009).  

Recently, Prieto et al.\ (2009) presented a mid-IR spectrum of N300-OT
that contained an emission feature reminiscent of the polycyclic
aromatic hydrocrbon (PAH) features seen in some proto-planetary
nebulae, and they interpreted this as indicative that the progenitor
of N300-OT was a carbon-rich super-AGB star.  However, the presence of
PAH emission features does not necessarily constitute C-rich
chemistry.\footnote{PAH emission features are seen in proto-planetary
  nebulae and H~{\sc ii} regions mainly because there is sufficient
  near-UV radiation to excite them, not because the gas is highly
  carbon-enriched.}  Moreover, prominent PAH features have been seen
in the mid-IR spectra of known LBVs such as HD~168625 (e.g., Umana et
al.\ 2010), which has a luminosity corresponding to an initial mass of
about 25 $M_{\odot}$.  One cannot rely on the inference of
amorphous carbon grains as necessarily indicative of carbon-rich gas
chemistry either, since carbon grains form at much higher temperatures
than silicates, and the conditions for rapid dust formation in ejected
shells may be very different from the conditions in RSG/AGB winds.
The presence of PAH features in a SN impostor spectrum, or the
presence of carbon grains, is therefore not necessarily indicative of
a carbon-rich AGB star.

Alternatively, it is quite possible that the reason the progenitor
stars of SN~2008S and N300-OT were obscured (and possibly why they had
low luminosities) is because the stars suffered a previous recent
outburst that had not been documented.  LBVs are known to suffer
multiple successive eruptions (see \S 4.2).  The most distinguishing
property of the SN~2008S and N300-OT progenitors was their relatively
low luminosity, implying initial masses lower than 20 $M_{\odot}$.
Although the observed IR luminosities (which are really minimum
luminosities) are consistent with some models for the most extreme
super-AGB stars, studies of the star formation history of N300-OT's
environment favor a more massive progenitor star of 12-25 $M_{\odot}$
(Gogarten et al.\ 2009).

Thus, it is difficult to reliably classify SN~2008S and N300-OT as a
wholely new and separate type of transient (note that they reside in
the most common location for SN impostors in
Figure~\ref{fig:compare}).  Until we actually identify the underlying
physical mechanism of the outbursts (see \S 4.6), this difference is
rather semantic, depending on whether one prefers to see them as an
extension of eruptive phenomena in more massive stars or a different
class of eruptions occurring in stars below 20 $M_{\odot}$.  (In
either case, they are extremey interesting, and may be more common
than SN impostors from more massive progenitors simply because of the
slope of the initial mass function.)  For these reasons, we have
included SN~2008S and N300-OT among the other SN impostors, but
perhaps the debate will continue for decades until the stars recover
from the outbursts and reveal themselves or until they finally explode
as core collapse SNe.

Nevertheless, it appears that some recent transients are pushing the
bottom end of the envelope that encompasses LBVs, the strongest
evidence of which is their relatively low-luminosity progenitors and
stellar environments.  The physical mechanism of all these outbursts
remains elusive.  The following transients share some overlap with
LBVs, but do bear some perceived differences well.  We note them in a
separate section here because previous authors interpreted them as
something other than LBVs.

\begin{figure*}\begin{center}
\includegraphics[width=4.8in]{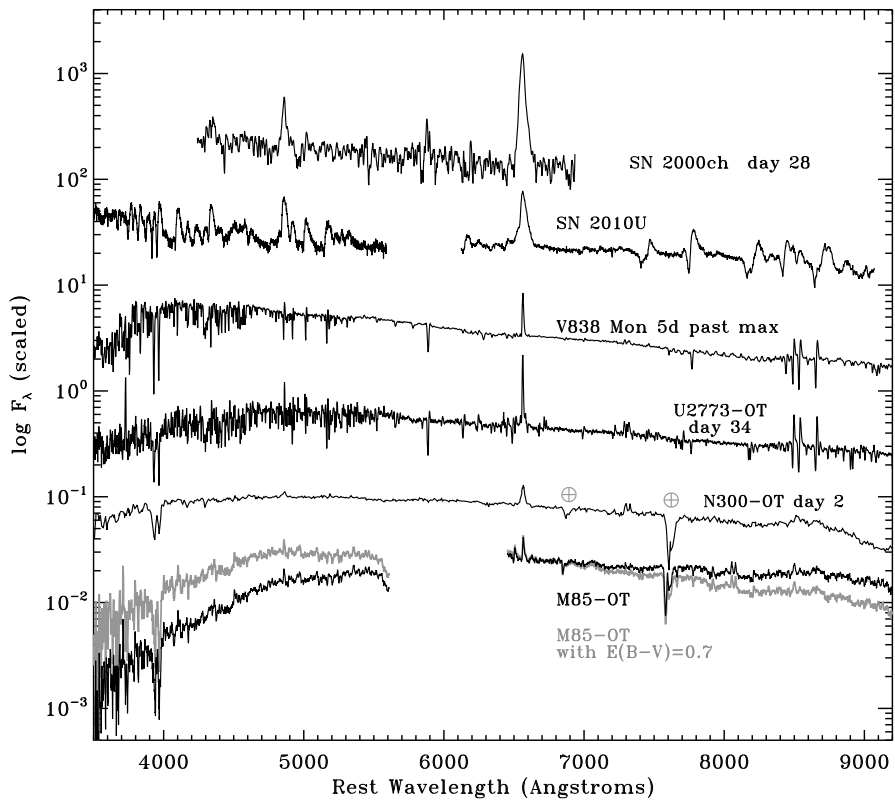}
\end{center}
\caption{Spectra from our database of a few of the
  intermediate-luminosity transients discussed in \S 4.5, compared to
  SN impostors shown earlier in Figure~\ref{fig:spec}.  The spectrum
  of SN~2010U was taken on 2010 Feb.\ 7 with Keck/LRIS, and
  corresponds to day 2 after discovery.  We obtained the spectrum of
  V838~Mon with Lick/KAST on 2002 Feb.\ 11, about 5 days after the
  brightest peak in the $B$ and $V$ light curves (see
  Figure~\ref{fig:novae}).  The observed flux has been corrected for
  $E(B-V)$=0.87 mag, following Munari et al.\ (2005).  The spectrum of
  M85-OT is the Keck/LRIS pectrum from Kulkarni et al.\ (2007); here
  we show it corrected for $E(B-V)=0.14$ mag (black), as in that
  paper, as well as what it would look like corrected for a larger
  value of $E(B-V)$=0.7 mag (gray) for comparison.  The comparison
  spectra of SN~2000ch, UGC~2773-OT, and N300-OT are the same as in
  Figure~\ref{fig:spec}.}
\label{fig:spec2}
\end{figure*}


{\it V838 Mon:} The best studied of this group of unusual transients
is V838~Mon, which erupted at a distance of $\sim$6 kpc in our Galaxy
in 2002, and has since produced a spectacular light echo in its
circumstellar reflection nebula (Sparks et al.\ 2008; Bond et al.\
2003).  The transient had a peak absolute $V$ magnitude of $-$9.8, and
studies of its associated cluster of B-type stars implies an age of
$\la$25 Myr and an initial mass of $\ga$8 $M_{\odot}$ if the transient
is an evolved star (Afsar \& Bond 2007).  The complex, multi-peaked
$BVI$ light curve from Sparks et al.\ (2008) is shown in
Figure~\ref{fig:novae}.  From this, we determine values of $t_{1.5}$
of roughly 12 and 75 days for the main and secondary peaks.

Figure~\ref{fig:spec2} shows a visual spectrum of V838~Mon from our
spectral database, obtained with Kast/Lick on 2002 Feb.\ 11, about 5
days after the main $B$ and $V$-band peak in the light curve.  This
spectrum is representative of the early bright stages of the
transient, whereas the spectrum evolved significantly at late times,
becoming much redder and displaying deep molecular absorption features
(e.g., Bond et al.\ 2003).  We find it quite remarkable that the
spectrum of V838 Mon near maximum light is nearly a carbon-copy of the
visual-wavelength spectrum of U2773-OT, which had a more luminous LBV
progenitor star and a more luminous and longer-lasting eruption than
V838 Mon.  The only substantive difference between the spectra of V838
Mon and U2773-OT is that the narrow absorption lines in U2773-OT are
somewhat stronger than in V838 Mon.

{\it M85-OT:} We discovered M85-OT during the normal course of the
LOSS in January 2006.  Kulkarni et al.\ (2007) presented the first
detailed study of this object, and drew attention to its faint
progenitor star and the transient's apparent differences compared to
novae, SNe, and LBVs.  Alternatively, Pastorello et al.\ (2007a)
argued that it could be a very faint core-collapse SN.  The absolute
$R$-band light curve of M85-OT from Kulkarni et al.\ (2007) is shown
in Figure~\ref{fig:novae}.  Kulkarni et al.\ suggested that the peak
luminosity and decay rate of this transient occupied a ``gap'' between
novae and SNe in luminosity, but faster than LBVs.  As we have seen in
this paper, however, LBV eruptions occupy a larger range of
characteristic fading times than previously recognized, from a day to
a decade, and the light curve of M85-OT with a peak absolute magnitude
of almost $-$13 and a decay time of 80--100 days fits well within the
paramter space occupied by known LBV giant eruptions
(Figure~\ref{fig:compare}).  M85-OT also appeared somewhat redder than
LBVs, suggesting a temperature of roughly 5000~K (Kulkarni et al.\
2007), but this depends on the assumed extinction and reddening.  Note
that these authors estimated an upper limit to the extinction based on
the observed Balmer decrement, assuming that it should follow the
standard Case B low-density recombination value.  However, Balmer line
ratios in dense winds and ejecta do not always follow standard
recombination values, so the reddening could be higher.  Indeed,
Prieto et al.\ (2008b) later showed that M85-OT had a large IR excess,
suggesting a very dense dusty CSM.  This means that the apparent
temperature of M85-OT may have been warmer than 5000~K, and that its
peak magnitude was probably more luminous.  It also means that the
progenitor could have been substantially more luminous than Kulkarni
et al.'s upper limit to the absolute $g$ magnitude of the progenitor
star of $>-$4.1 mag.  While this would still be fainter than the most
luminous LBVs, it approaches the values inferred for N300-OT and
SN~2008S.

In Figure~\ref{fig:spec2}, we show the Keck/LRIS spectrum of M85-OT
from Kulkarni et al.\ (2007).  We show it corrected for the value of
$E(B-V)$=0.14 mag adopted by those authors (black spectrum), as well
as a higher reddening correction of $E(B-V)$=0.7 mag (gray spectrum).
With this higher reddening, the continuum can be approximated by a
temperature around 6500~K, except for wavelegnths below 4500 \AA \
where line blanketing may be important (this is also the case for fits
involving cooler temperatures and lower reddening correction).  The
point of this comparison is that a larger value for the extinction and
reddening is plausible, which could mean that the transient and its
progenitor were more luminous (for a reddening of 0.7 mag, as shown
here, the peak absolute visual magnitude would have been around
$-$14.5, for example).  The spectrum of M85-OT closely resembles that
of N300-OT, also shown in Figure~\ref{fig:spec2} for comparison.  Both
transients have weak emission from the Ca~{\sc ii} IR triplet, weak
narrow H$\alpha$, and fairly strong Ca~{\sc ii} HK absorption,
implying that at the times when the spectra were taken, the emitting
photospheres probably had similar temperatures.  In that case, the
higher reddening correction we have shown here would also bring the
continuum shapes into better agreement.

After considering the distribution of properties of LBV eruptions
described in this paper, as well as a comparison between the spectrum
of M85-OT and N300-OT, it is much less clear based on the properties
of the outburst alone that M85-OT is something altogether different
from other SN impostors, especially if it is shifted upward by 0.5--1
mag in Figure~\ref{fig:compare} due to a higher reddening value.  The
strongest case for a different type of souce comes from its local
environment that implies an initial mass around 7 $M_{\odot}$ or less
(Ofek et al.\ 2008), which is similar to V838~Mon and lower than
SN~2008S and N300-OT.

{\it SN~2010U:} Nakano (2010) discovered SN~2010U in NGC~4212, with an
unfiltered magnitude of 16.0 on 2010 Feb.\ 5.6 UT.  The peak
unfiltered magnitude was 15.9, and the progenitor was undetected at a
limit of 18.0 mag.  At a distance of 3.3 Mpc for NGC~4212, the
corresponding peak absolute magnitude is roughly $-$11.7 mag (not
corrected for reddening).  A spectrum obtained 2 days later by Mario,
Vinko, \& Wheeler (2010) showed a blue continuum with strong narrow
Balmer emission lines with P~Cyg absorption features indicating
outflow speeds of order 900 km s$^{-1}$, similar to many SN impostors.
Humphreys et al. (2010) recently suggested that this source is not an
LBV eruption, but is instead a luminous nova.  The light curve is
shown in Figure~\ref{fig:novae} for comparison.  It does fade faster
than some LBVs like HD~5980 and V12 (shown here), but the rapid fading
is comparable to or even slower than brief events in SN~2009ip (Smith
et al.\ 2010a) and SN~2000ch (Wagner et al.\ 2004; Pastorello et al.\
2010).  Its peak luminosity is closer to SN impostors than to novae
(Figure~\ref{fig:compare}).

We obtained one spectrum of SN~2010U on 2010 Feb.\ 7 using Keck/LRIS,
and the resulting spectrum is shown in Figure~\ref{fig:spec2}.  This
date corresponds to 2 days after discovery and about 1 day after
maximum light (i.e., well before the transient faded significantly).
Among our sample of SN impostors, SN~2010U most closely resembles the
spectrum of SN~2000ch (also shown in Figure~\ref{fig:spec2}).  It is
interesting, then, that the fast decay and peak absolute magnitude of
SN~2010U are also very similar to those of SN~2000ch, which is a
confirmed LBV showing additional multiple eruptions many years later
(Pastorello et al.\ 2010).  It would be very interesting to continue
observing SN~2010U, to see if it follows suit.  Based on the
similarity in both light curves and spectra between SN~2010U and
SN~2000ch, the claim that SN~2010U is a luminous nova and not an
LBV-like eruption becomes less secure; the observed properties of the
transient seem more consistent with LBVs than with novae.  As noted by
Humphreys et al.\ (2010), however, the upper limits for the progenitor
star and its surrounding population seem to argue that it had an
initial mass $\la$5 $M_{\odot}$ in this case.  This provides another
case where eruptions that closely resemble LBVs seem to occur in
lower-mass stars as well.

{\it M31~RV1:} Rich et al.\ (1989) discovered a luminous red variable
star in M31, which rose to a bolometric absolute magnitude of $-$9.8
in September 1988.  In the $I$-band it then faded about 3 mag in 43
days, and had been 5 mag fainter in pre-discovery images.  It did not
fade nearly as much in the $K$ band after discovery, suggesting that
the bolometric luminosity did not necessarily change much, and that
either dust obscuration increased or the object cooled after the
outburst. The nature of this transient is unclear, but it showed the
absorption-line spectrum of an M0 supergiant plus narrow Balmer line
emission.  IR data are not available before discovery, so one cannot
exclude the possibility that the progenitor was heavily obscured.
Kulkarni et al.\ (2007) compared it to M85-OT, although that source
was more luminous and faded more slowly, and M85-OT showed the Ca~{\sc
  ii} IR triplet and other features in emission and did not exhibit
the TiO and other absorption bands characteristic of a cool M
supergiant.  In terms of its absolute magnitude and strong evolution
to the red as it faded, M31~RV1 was more like V838~Mon than M85-OT,
although it is not known to have had a complex multi-peaked light
curve like V838~Mon.  Recently, Shara et al.\ (2010) examined archival
{\it HST} images obtained 10 years after the transient event and
suggested that a detected source is consistent with an old nova rather
than a merger event, but continued study of the remnant object is
needed.


{\it PTF10fqs:} This is an apparently red-colored transient located in
outer parts of a spiral arm of M99, with peak absolute visual
magnitude of about $-$11, expansion speeds of around 800 km s$^{-1}$,
and a spectrum very similar to V838~Mon as well as SN impostors
SN~2008S and N300-OT (Kasliwal et al.\ 2010).  Overall, the outburst
of PTF10fqs appeared extremely similar to other SN impostors on the
faint end of the distribution.  As with M85-OT, visual upper limits to
its progenitor would imply a star less massive than about 8
$M_{\odot}$ if there were no local extinction.  However, based on the
red color of the transient and the presence of [Ca~{\sc ii}] emission
in its spectrum, the progenitor may have been heavily obscured like
SN~2008S and N300-OT, but unfortunately, upper limits in the mid-IR
are well {\it above} the IR luminosities for the progenitors of
SN~2008S and N300-OT, so a more massive star cannot be ruled out for
PTF10fqs.


The initial masses for these objects are uncertain, but at least one
can be confident that they are not among the most luminous stars
known.  Stellar ages for the local environments of both V838~Mon and
M85-OT are consistent with initial masses curiously close to the
dividing line between core-collapse SNe and massive white dwarfs at
around 8 $M_{\odot}$.  For PTF10fqs a more massive star cannot be
ruled out because of uncertainty in the pre-outburst extinction, but
it may be in this range as well.  The progenitor of SN~2010U was
proposed to be somewhat lower at about 5~$M_{\odot}$.  One obvious
possible suggestion for these transients is that they may arise from
electron-capture SNe, expected to occur at initial masses around 8--10
$M_{\odot}$.  This possibility was suggested for SN~2008S and N300-OT
as well (Thompson et al.\ 2009; Botticella et al.\ 2009); mounting
evidence argues against this interpretation for these particular
sources, but it remains a possibility for the other objects.

Other potential explanations include a wide variety of failed
core-collapse SNe (e.g., Fryer et al.\ 2009; Moriya et al.\ 2010), the
explosive birth of a massive white dwarf initiating the planetary
nebula (PN) phase (Thompson et al.\ 2009), or stellar mergers and
other tidal encounters (see below).  The possibility that the PN phase
might be initiated by a sudden explosive or eruptive event is
extremely interesting from the point of view of understanding the late
evolution of intermediate-mass stars and the dynamics of PNe, but well
beyond the scope of our present paper and in need of further
theoretical investigation.  If true, there is potentially a great deal
of synergy between studies of the transient sources and the associated
nebulae in the mass ranges above and below 8 $M_{\odot}$.  Stellar
mergers and tidal encounters represent another attrative explanation
for these transients and other SN impostors, since there is no clear
obstacle to binary encounters above or below $\sim$8 $M_{\odot}$.

In summary, we find that based on criteria such as absolute magnitude,
rate of fading, light curve shape, or even color and spectra, it is
difficult to reliably distinguish LBV eruptions from non-LBVs (if
indeed the objects discussed in this section are a distinct set of
events).  The only reliable way to establish a difference is based on
having good information (and indeed, when one considers the
possibility that they may be obscured at visual wavelengths by CSM
dust, we must also include deep IR data) about their faint progenitor
stars or their progenitor environment.  Such information is rarely
available except for nearby objects.  Without such detailed
information, claims of new types of transients may be unsubstantiated
or highly speculative.  This is a sobering fact to keep in mind as we
embark on an era of more intensive transient studies.

{\it Caveat:} An interesting twist involves binary evolution, which
should not be overlooked.  We have emphasized that estimates of a
progenitor star's initial mass based on studies of the surrounding
stars (e.g., Gogarten et al.\ 2010) provide some of our most important
contraints on the nature of the progenitor stars.  One potential
pitfall, however, is the following: A star with an initial mass below
8 $M_{\odot}$ may accrete a substantial amount of mass from its
companion, raising it to more than 12 $M_{\odot}$, and changing its
evolutionary fate and perhaps leading to the types of eruptions
encountered in initially more massive stars.  Similarly, a secondary
star with initial mass of, say, 12--15 $M_{\odot}$ may gain enough
mass to raise its luminosity and make it behave like a 20--25
$M_{\odot}$ star, and so on.  The point is that even in cases where we
have good constraints on the surrounding stellar population, the
progenitor star may actually have been more massive than we are led to
believe.  This complexity is somewhat unsettling.

\subsection{What is the underlying mechanism?}

After more than a half-century of research since the Hubble-Sandage
variables were identified (Hubble \& Sandage 1953), the underlying
cause and trigger of LBV giant eruptions remains unexplained.  This
makes it very difficult to say whether a given observed non-SN
transient event is or is not an LBV, and this is exacerbated by the
huge diversity in observed LBV properties demonstrated in this paper.
Some LBVs are highly obscured by their own ejected dust shells (some
are even {\it completely} obscured for decades) while others show no
sign of dust whatsoever; some LBV giant eruptions involve
$\ga$10$^{50}$ erg explosions and 10--20 $M_{\odot}$ of ejected mass,
while others have only 10$^{47}$ ergs and 0.01 $M_{\odot}$; and so on.
As described in the previous section, there is considerable overlap
with transients that are purported to be non-LBVs.

As with the case of distinguishing between SNe~Ia and all other types
of SNe, some clue of the physical mechanism is needed before we can
reliably differentiate LBV giant eruptions from other transient events
arising in moderately massive and intermediate-mass stars.
Unfortunately, we do not yet have a clear working hypothesis for the
physical mechanism behind LBV eruptions, so one hopes that
observational clues can help narrow the field.  Among the most
important observational clues are the ejection speed of material
launched from the star, as well as the total mass and energy budget.
As discussed in \S 4.3, the outflow velocities observed in most
sources are a few 10$^2$ up to a little more than 10$^3$ km s$^{-1}$,
and these are suggestive of either strong supergiant winds or CSM
interaction, whereas some sources (e.g., $\eta$ Car and SN~2009ip)
show evidence for a small mass of much faster material moving at
$\sim$5000 km s$^{-1}$ (Smith 2008; Smith et al.\ 2010a; Foley et al.\
2010), probably requiring the presence of a leading blast wave as
well.  Whether or not impulsive or explosive acceleration of the
envelopes is at work in all SN impostors is uncertain, but shock waves
clearly are at work in a few of them, and so any successful theory
must incorporate this.  The total mass and energy budgets are harder
to evaluate.  Nearby examples with resolved nebulae from past
outbursts allow us to measure the mass of ejecta directly, and here we
see a huge range from 0.01 to more than 10 $M_{\odot}$ ejected in a
single event.  Unfortunately, the mass ejected in distant SN impostors
is poorly constrained.

We can, however, estimate the total escaping radiated energy budget
for each, which is given roughly by $E_{\rm rad} = \zeta t_{1.5}
L_{\rm peak}$, where $\zeta$ is factor of order unity that depends on
the exact shape of the light curve.  $L_{\rm peak}$ is the luminosity
corresponding to the inferred peak absolute bolometric magnitude
corrected for extinction.  From Table~\ref{tab:list2}, then, one can
deduce a huge range in values of $E_{\rm rad}$ from
$\sim$2$\times$10$^{49}$ ergs ($\eta$ Car), down to $\la$10$^{46}$
ergs.  {\it Caveat:} we must remember a lesson from nearby examples
such as $\eta$ Car, however, where the kinetic energy budget of more
than 10$^{50}$ erg greatly outweighs the escaping radiative energy
budget of $\sim$10$^{49}$ ergs.

An interesting timescale to consider is the ``buildup'' or
``recovery'' timescale for the radiated energy budget, which is the
time the star would require to supply $E_{\rm rad}$ in its quiescent
state, given by

\begin{displaymath}
t_{\rm rad} = E_{\rm rad} / L_* = t_{1.5} \frac{\zeta L_{\rm peak}}{L_*}
\end{displaymath}

\noindent where $L_*$ is the quiescent pre-outburst bolometric
luminosity of the progenitor star.  This is relevant in a type of
model where the output core luminosity of the star is constant over a
long timescale compared to the event, and where the extra radiated
energy is presumed to be the result of thermal energy being stored in
the star's enveloped and then released suddenly by some mechanism.  It
is also relevant for the time it takes the star to re-establish
thermal and radiative equilibrium after a disruptive event.  Here too
we see a wide range of values, with $t_{\rm rad} \approx $ 40 years
for $\eta$ Car, $\sim$1.1 yr for SN~2009ip, and 32 yr for SN~2008S.
If one can establish that $t_{\rm rad}$ is considerably longer than
any observed timescale of variability in the progenitor, then it is
likely that an additional energy reservoir is required (the need for
extra energy obviously increases if one makes an allowance for kinetic
energy as mentioned above).  It also seems likely that $t_{\rm rad}$
may be related to the amount of mass ejected from the star or the
amount of envelope mass involved in the adjustment of the star,
although this is based only on the vague notion that the
Kelvin-Hemholz timescale for the ejected mass plays a critical
role. These considerations, while not conclusive, may be kept in mind
when thinking about various models mentioned below.

Investigating and evaluating theoretical possibilities is far beyond
the scope of this paper, but here we list some hypothetical
mechanisms, as well as their pros and cons from the perspective of
explaining the observed phenomena associated with SN impostors.


{\it Continuum-driven super-Eddington winds?}  In addition to
$\eta$~Carinae, all of the luminous SN impostors clearly exceed the
classical (i.e., electron-scattering) Eddington limit during the
brightest phases of their outbursts.  In the case of $\eta$~Car, the
star apparently exceeded the classical Eddington limit by a factor of
$\Gamma$=5 for more than a decade.  Other SN impostors that appear to
have lower progenitor masses but similar peak luminosities can achieve
much more extreme values of $\Gamma$=40 to 80.  Regardless of the
origin of this super-Eddington (SE) luminosity, it is unavoidable that
such sustained high luminosities will in fact drive a strong wind from
the star (unless of course the emerging radiation is from an
already-successful hydrodynamic explosion).  A few examples that have
been studied in detail (e.g., V1 in NGC~2366; U2773-OT) are clearly
consistent with wind-like spectra rather than explosions, so models of
SE winds are certainly applicable to at least {\it some} of the SN
impostors.  Much of the work on the properties of SE so far has been
conducted by Owocki and collaborators (Owocki et al.\ 2004; van Marle
et al.\ 2008, 2009; Shaviv 2000).  These studies on continuum-driven
SE winds assume a strong increase in luminosity as a precondition for
the models, concentrating primarily on the physics of driving mass
from the surface of the star in quasi-steady-state.  These models do
not, however, address the deeper question of what triggers the
required increase in bolometric luminosity, or what the ultimate
energy source is.

{\it Runaway Pulsations?}  Following early work on the pulsational
instability of massive stars (Ledoux 1941; Schwarzschild \& H\"arm
1959; Appenzeller 1970), there is an expectation that the outer
envelopes of massive stars should be quite unstable.  Can runaway
pulsational instability give rise to sudden mass ejections and
luminous transients like SN impostors?  Stothers \& Chin (1993)
proposed that an ionization-induced dynamical instability in their
models of very massive stars could lead to violent outbursts such as
that experienced by $\eta$~Carinae, but Glatzel \& Kiriakidis (1998)
criticized this model because the adiabatic approximation is not valid
for the envelopes of these stars, and their non-adiabatic models could
not reproduce the instability except at very low temperatures.
Non-linear growth of non-adiabatic {\it strange mode} pulsations,
however, may occur in the envelopes of luminous stars where the
thermal timescale is short and comparable to the dynamical timescale
(e.g., Glatzel et al.\ 1999; Glatzel \& Kiriakidis 1993; Kiriakidis et
al.\ 1993; Gautschy \& Saio 1995, 1996).  Non-linear growth of
strange-mode pulsations is expected in very luminous stars, but may
also occur in less massive stars such as AGB stars (Gautschy \& Saio
1995, 1996); thus, the full range of initial mass over which these
pulsations are effective at triggering instability is uncertain, but
potentially interesting for SN impostors and related transients.
Strange-mode instabilties depend on the iron opacity bump and occur
primarily in the outer envelope of standard stellar evolution models
(containing less than 1\% of the stellar mass); they therefore lead
only to relatively minor increases in luminosity (a few tenths of a
magnitude) and perhaps somewhat enhanced wind mass loss.  It has
therefore been challenging to explain the major outbursts
characteristic of SN impostors with the strange-mode instability.
(Stange mode pulsations may help trigger the normal S~Doradus
variations of LBVs, however, and can potentially account for their
observed microvariations.)  Further work is needed to determine if
similar instabilities might occur deeper in the star, if they are to
explain the ejection of several $M_{\odot}$ and 10$^{49}$--10$^{50}$
ergs, as in a major outburst like $\eta$~Carinae.  For example, Young
(2005) has described alternative stellar evolution models that include
the effects of wave-driven mixing and rotation in the core evolution,
and find that these stars are more extended, and that they therefore
have the iron opacity bump deeper in the star.  With the critical
opacity bump in deeper layers, more mass and thermal energy are above
the potentially unstable region, and Young (2005) hypothesizes that
this stellar structure might give rise to more energetic and massive
eruptions like SN impostors.

{\it Runaway mass loss and the Geyser model?}  Much of the observed
phenomenology of LBV eruptions is reminiscent of geophysical geysers
or volcanoes (see Humphreys \& Davidson 1994).  As described above,
LBVs sometimes are seen to exhibit growing instability leading up to a
large eruption.  The occurance of multiple shells in some cases
suggests that eruptions may be followed by a more quiescent recovery
time before the instability builds again.  This has led to the
suggestion of a geyser-like model for LBV giant eruptions (Maeder
1992), where very luminous stars reach cool temperatures in their
post-main-sequence evolution, allowing a recombination front (akin to
a boiling front in a geyser) to proceed into the star, thereby
initiating a rise in mass loss because of the change in opacity.  This
increased mass loss continues until the star contracts to warmer
temperatures, when the cycle begins again.  The simplicity of such a
thermal engine is appealing, although more detailed calculations are
needed to study the hydrodynamic response of a star in these
conditions.  It seems unlikely that this mechanism can explain the
extreme amounts of mass ejected in brief energetic events, the sharp
increases in bolometric luminosity, or the explosive property of some
eruptions.  Another potential drawback of this mechanism is that it
will only occur in very luminous stars near the classical Eddington
limit, and so cannot explain the full diversity of SN impostor
eruptions, some of which apparently occur at relatively modest initial
masses below 20 $M_{\odot}$.  It is nevertheless an interesting
possibility for eruptions in the most luminous stars.

{\it Pulsational pair-instability ejections?}  Heger \& Woosley (2002)
have described a type of severe mass-loss event known as pulsational
pair-instability (PPI) ejections, when a very massive star can eject
of order 10 $M_{\odot}$ in an explosive but non-terminal event.  This
is the same pair-formation instability that leads to a
pair-instability SN (PISN; Barkat et al.\ 1967; Rakavy \& Shaviv 1967;
Bond et al.\ 1984; Heger \& Woosley 2002), but it occurs in a mass
range below that of successful PISNe, where the explosive burning is
not enough to completely unbind the star, resulting in a
$\sim$10$^{50}$ erg ejection of the outer envelope only.  Woosley et
al.\ (2007) and Smith et al.\ (2007, 2010b) have mentioned the PPI as
an attractive explanation for the precursor LBV-like mass ejections
that precede some very luminous SNe~IIn like SN~2006gy.  Smith et al.\
(2010b) has emphasized that expectations for PPI ejections match
properties of some giant LBV eruptions in every {\it observable} way
(large mass ejected, H-rich envelopes, total energy of $\sim$10$^{50}$
erg, etc.), but also noted several problems with attributing PPI
events as a general explanation for all LBV giant eruptions.  First,
the PPI occurs during final burning phases and is expected to
transpire in the few years to decades immediately before core
collapse.  However, many LBVs have massive shells with dynamical ages
of 10$^3$ -- 10$^4$ yr, indicating that they have survived for
millenia after the giant eruption that ejected their
shells.\footnote{Heger \& Woosley (2002) did note a rare case where
  the PPI eruption can delay the resumption of nuclear burning,
  leading to intervals of as much as 10$^3$ yr between bursts, but
  this is not generally the case.}  Second, the PPI is only predicted
to occur for the most massive stars with initial masses above $\sim$95
$M_{\odot}$, and usually only at low metallicity,\footnote{This exact
  mass range, however, depends on mass-loss rates assumed in stellar
  evolution models throughout the lifetime of the star.} whereas LBVs
are know to arise from stars with initial masses as low as 20--25
$M_{\odot}$ (Smith et al.\ 2004).  Recent observations of
low-luminosity progenitors may extend this mass range even lower, to
within 10--20 $M_{\odot}$, as described above.  These lower-mass LBVs
will never encounter the PPI, so the PPI can only provide a possible
explanation for the most extreme LBV giant eruptions in the most
luminous stars like $\eta$~Car, not the full range of the observed
LBV-like eruption phenomenon.

{\it Other shell-burning explosions?}  What about other types of
explosive burning events, analogous to the PPI, but not necessarily
restricted to the late-phase O or Si burning?  This may relax the
requirements for the very high core temperatures needed for the PPI,
and hence, may relax the restrictions on initial masses that
experience explosive buring instabilities.  This is an old idea, first
suggested (somewhat ironically) as a possibility for SN~1961V by
Branch \& Greenstein (1971), and revisited several times since then
(Guzik et al.\ 1999, 2005; Smith et al.\ 2003a; Smith 2008; Smith \&
Owocki 2006; Dessart et al.\ 2009). Substantive models for such an
event do not yet exist, but should be pursued.  Hypothetically, one
can imagine that the energy source could be nuclear fusion of a small
amount of material, if oscillations (i.e. non-radial g-modes, unsteady
convection, external perturbations, etc.) in the lower envelope mix
fresh H-rich fuel into deeper and hotter layers of the star,
triggering explosive burning.  Initial simulations suggest that
boundary layers within the star may be susceptible to dynamic
disturbances (Meakin \& Arnett 2007; Guzik et al.\ 1999, 2005).  Even
a few percent of a solar mass of burnt H, for example, or a few tenths
of a solar mass in silicon burning, would be sufficient to provide the
extra energy inferred for giant LBV eruptions.  Different amounts of
energy deposited at different depths within the star could conceivably
account for the wide diversity in observed properties of SN impostors,
over a wide range of masses, as suggested by some recent exploratory
models (Dessart et al.\ 2009).  A relatively large amount of deposited
energy compared to the binding energy would initiate a large
hydrodynamic explosion, whereas a smaller amount of deposited energy
may just temporarily increase the luminosity of the star above the
Eddington limit, at which point the physics of continuum-driven SE
winds becomes relevant, as discussed above.  Observations show
evidence for both phenomena.  Further progress in this direction
requires substantial effort in multidimensional and hydrodynamic
simulations of stellar interiors, plus estimates of the resulting
observables.  The observed radiation from SN impostor events coupled
with the detailed kinematics of nearby circumstellar shells can
provide important constraints on such models.  Dessart et al.\ (2009)
have argued that this type of energy deposition may be particularly
likely in stars with initial masses of 8--12 $M_{\odot}$, with obvious
possible implications for SN~2008S, N300-OT, and some of the sources
discussed in the preceding section.


{\it Failed SNe?}  Models that fail to generate successful
core-collapse SNe may nevertheless produce partial explosions, and
therefore, observable transient sources, as discussed recently by
Fryer et al.\ (2009; see also Moriya et al.\ 2010).  With a wide range
of possible absolute magnitues around $-$14, these certainly may be
applicable to some of the observed SN impostors, especially in cases
with dense CSM discussed by Fryer et al.\ (2009).  Such mechanisms are
unlikely to explain the full diversity of SN impostors, however, since
several examples exist of LBVs that have survived giant eruptions as
relatively stable hot supergiant stars, and there is considerable
evidence that these eruptions can repeat {\it multiple times} on a
variety of timescales up to millenia.  Nevertheless, such failed SNe
remain viable explanations for distant SN impostors unless deep
follow-up observations are available to establish the post-eruption
state of the (surviving?) star.

{\it Electron-capture SNe?}  This idea has been discussed above.  It
does not offer an attractive explanation for the diversity observed in
most SN impostors, because this type of event is only expected for a
narrow range of initial masses around 8 $M_{\odot}$.  It does,
however, provide a potential explanation for either faint SNe~II-P
(not discussed here) or some of the relatively faint transients
discussed in \S 4.5.


{\it Close binary interaction events?}  Through the course of
post-main sequence evolution of a massive star, its luminosity goes up
as the core contracts, its total mass goes down due to mass loss, and
so its proximity to the Eddington limit becomes more precarious.  This
is presumed to lead --- somehow -- to the instability we see as LBVs,
by making the star more susceptible to internal {\it or external}
disturbances.  But in addition, as a massive star evolves off the main
sequence, it migrates to cooler temperatures, and so its radius
increases by a huge factor.  An increasing radius leads to inevitable
dangerous encounters in binary systems with periods less than several
years.  Smith (2011) notes that in the case of $\eta$~Car (5.5 yr
orbital period), even with the {\it quiescent} pre-outburst luminosity
of $\eta$~Car and a likely temperature around 8000~K, that the
companion star would plunge well inside the apparent photosphere of
the primary during periaston passages.  A violent periastron encounter
is therefore inevitable, and may help explain the brief brightening
episodes that occurred in 1838 and 1843; these two events are, in
fact, closely associated with times of periastron (Smith \& Frew
2010).  Exactly how this works in unclear, and explaining the
energetics is not trivial.  Similarly, in the case of the eclipsing
binary HD~5980, Koenigsberger and collaborators (see Koenigsberger
2004) have proposed that tidal interactions in the close binary may
have triggered the eruption observed in the 1990s.\footnote{Soker and
  collaborators (e.g., Soker 2001) have envisioned a much more
  complicated model for $\eta$~Carinae, where the main-sequence
  secondary star accretes from the primary wind during close passages
  and blows a pair of collimated jets, as an attempt to explain the
  bipolar shape and kinematics of Homunculus; in their model, however,
  an eruption from the primary star was an assumed and necessary
  precondition.}

Interacting binary events are attractive in the sense that the
seemingly endless free parameters in binary models (mass ratios,
stellar radii, orbital period, eccentricity, conservative
vs. non-conservative mass transfer/mass loss, posible instability of
either star, etc.) may provide a natural origin for the wide diversity
observed in SN impostor outburst properties.  Furthermore, close
binary interactions could conceivably operate over a wide range of
initial masses, even in stars that are not dangerously close to the
Eddington limit on their own.  Specifically, these enounters may occur
for initial masses both above and below 8$M_{\odot}$, regardless of
differences in core evolution, providing a possible link between LBV
eruptions and very similar transients from lower-mass stars (see \S
4.5).  The detailed way in which binary encounters could account for
the energetics of SN impostor events looms on the horizon as a major
open question, however.  A fruitful possibility for explaining SN
impostors is that such a model would require two suitable conditions:
1) an evolved primary star that approaches instability anyway, and 2)
a rather sudden increase in stellar radius so as to initiate a
catastrophic encounter.  Whether such binary interactions are the key
to causing LBV eruptions or whether they simply modify the temporal
behavior by triggering an instability that would have occurred anyway,
is obviously a key question for future theoretical research.  In the
case of $\eta$~Carinae, though, it seems clear that a simple mechanism
such as kinetic heating of the primary star's envelope by the invading
secondary is insufficient, since the gravitational binding energy of
the binary orbit is substantially less than the kinetic energy of the
expanding Homunculus nebula (Smith et al.\ 2003a).  However, it must
also be noted that binary interactions such as this are unlikely to
explain {\it all} LBV eruptions; P~Cygni, for example, has shown no
evidence of binarity despite decades of detailed study.  There must be
some mechanism than can lead to eruptions of single massive stars as
well.


A main emphasis of this paper has been to demonstrate the wide
diversity in observed properties of SN impostors and their
progenitors, but a fair question is whether the group is {\it too
  diverse}.  In other words, can this group be explained by a single
mechanism operating over a wide range of energy and mass, or must it
be a collection of different mechanisms operating in different stars
that are susceptible to perturbations?  Can these different mechanisms
give rise to transient sources that overlap in
Figure~\ref{fig:compare} and have similar spectra?  Is it {\it
  required} that multiple mechanisms work together to initiate an
eruption?  Since several potential mechanisms listed above seem at
least plausible, there may be more than one cause of SN impostors.
Inventing ways to connect observations to theory and to distinguish
between these will be a major task for future work.

\subsection{Summary and Future directions}

While SN impostors are intrinsically fainter than SNe, and are
therefore discovered less easily, their numbers are growing.  They
will continue to be discovered in increasing frequency in upcoming
surveys, and so a better framework to understand them is needed.  In
this paper, we have attempted to compile some of the basic observables
of SN impostors and related transients known to date, including nearby
historical examples and more recent events discovered in modern
SN/transient searches.  We also presented new spectra and light curves
for a number of SN impostors.

Examining the full distribution of observed properties --- including
peak absolute magnitude, characteristic fading timescale, outflow
velocity, spectral morphology, and progenitor properties --- the most
striking result is that SN impostors are extremely diverse, filling
essentially all the available parameter space between SNe and novae.
We find no clear correlations between spectral morphology, luminosity,
or fading timescale, as exhibited by other transients like SNe and
novae.  Moreover, the diversity exhibited by well-studied cases where
the progenitor is known to be an LBV fully encompases the range of
parameter space occupied by transients that are supposedly not LBVs.
In some cases, therefore, previous claims of new types of transients
based on observed properties of the eruption appear to have been too
strong.  On the other hand, the mechanism behind these eruptions is
still unknown, and so multiple different types of outburst phenomena
may overlap in parameter space, so we are not arguing that all these
sources are necessarily LBV giant eruptions.  Indeed, LBVs may be a
subset of a larger group of nonterminal eruptive phenomena.  A great
deal of theoretical work is needed before confident conclusions can be
drawn.

Nevertheless, even though the distribution of SN impostor properties
is very diverse, we did find one extreme outlier among the sample,
which stood out in every measurable way: the supposedly prototypical
impostor SN~1961V.  We find that SN~1961V is more naturally explained
as a true core-collapse SN of Type~IIn, similar to SN~2005ip and
SN~1988Z, but with delayed CSM interaction.  We propose that the
strange light curve shape of SN~1961V can be explained by a relatively
normal SN~II, followed by a late turn-on of CSM interaction luminosity
that causes its rise to its peak luminosity after $\sim$100 days.
That late peak luminosity was the same as SN~2005ip, and comparable
late turn-on of CSM interaction has been documented in previous
SNe~IIn.  This requires that the CSM shell had an interior cavity, and
reasonable velocities would imply that the shell would have been
ejected within a few years before core collapse.  Indeed, the
progenitor of SN~1961V was observed at an absolute magnitude of
roughly $-$14 about a year before its main brightening, and we suspect
that this was the {\it direct} detection of a precursor LBV-like
outburst.  This erradicates the notion that the progenitor of SN~1961V
must have been an astoundingly massive star, and instead, suggests
that it had an initial mass and luminosity comparable to $\eta$
Carinae.


There is considerable room for improvement in our understanding of LBV
eruptions and SN impostors.  The most glaring deficiency is in our
theoretical understanding.  A theory for these eruptions should strive
to identify a physical mechanism that can account for a range of
ejected mass (0.01--10 $M_{\odot}$) and kinetic energy (10$^{46}$ --
10$^{50}$ ergs), total radiated energy (10$^{46}$--10$^{49.3}$ ergs),
peak luminosity ($-$10 to $-$15 mag), outflow speeds (100--1000 km
s$^{-1}$), and different spectral properties through all luminosities
(relatively cool and hot, varying emission line strengths, etc.).
This is admittedly a tall order.  Although more realistic models for
the structure of post-main-sequence massive stars are needed to assess
the susceptibility and outcomes of various instabilities, it is also
likely that simple toy models can be useful to investigate the
hydrodynamics of envelope ejection and the star's dynamical and
thermal response.  Detailed radiative transfer calculations for these
ejections are needed in order to connect observable spectra and
luminosities to derived properties (see, e.g., Dessart et al.\ 2009).
Finally, dynamical models of close binary interactions and the
transients they might produce are sorely needed.

On the observational front, our understanding of SN impostor
statistics will improve in the near future, since these kinds of
transients will be a major emphasis of upcoming photometric surveys.
In this paper we have only examined about 2-dozen SN impostors and a
few additional cases whose nature is debated.  While this has been
sufficient to demonstrate the diversity in observed properties, it is
not sufficient to examine their intrinsic statistical distribution.  A
prohibitive weakness is that this sample is not drawn from a uniform
survey with understood systematics, so we have been careful not to
draw conclusions about how common SN impostors with various peak
luminosities are, for example, or how common they are compared to
core-collapse SNe.  Understanding the intrinsic rates of SN impostors
is key, as has been done for SNe (e.g., Li et al.\ 2010), but it has
been difficult to address for SN impostor statistics because they are
so faint.  The Large Synoptic Survey Telescope (LSST) will provide a
critical advance in this area, allowing estimates of control times and
completeness of a large sample.

It may seem discouraging that the diversity of SN impostors is so
large, because it follows that there is limited utility from spotty
observations of a transient's spectrum or a monochromatic light curve
of only the time around peak luminosity.  Follow-up spectroscopy and
photometry are extremely useful, however, when combined with good
coverage at late times or with cases where detections of a progenitor
star are available.  In particular, followup observations that may
(eventually) detect a second outburst or multiple eruptions can be
extremely useful for understanding the phenomenon, although this may
take several years of monitoring.  We should keep a watchful eye on
all nearby and historical examples, in case they erupt again or
explode as real SNe. Late-time data and upper limits can potentially
help us understand the recovery of a star after a disruptive event,
which is a problem that has received little attention so far.

Lastly, these transient sources are associated with substantial mass
ejection.  The resulting circumstellar shells are potentially
observable for a much longer time than the outburst itself.  Thus,
continued detailed study of nearby examples of resolved circumstellar
shells around all types of stars is needed, as it offers our only way
(in the absence of good models for the outbursts) to evaluate the
amount of ejected mass.  Comparing the statistics of circumstellar
shells to the properties of SN impostors and other transients may
prove enlightening when a statistical sample is available.  If nearby
examples are any guide, then the mass ejection of SN impostors is
probably not spherical.  It is therefore likely that considerations
associated with asymmetry (spectrapolarimetry, detailed line profiles,
rotation, asymmetric explosions and winds, binary encounters) will
figure more prominently in upcoming studies.

\smallskip\smallskip\smallskip\smallskip
\noindent {\bf ACKNOWLEDGMENTS}
\smallskip
\scriptsize

We thank A.J.\ Barth, S.B.\ Cenko, R.\ Chornock, A.L.\ Coil, R.J.\
Foley, C.V.\ Griffith, M.T.\ Kandrashoff, I.K.W.\ Kleiser, M.\ Modjaz,
J.\ Bloom, A.\ Miller, D.\ Perley, and D.\ Poznanski for their
assistance with some of the observations and data reduction.  We thank
S.\ Kulkarni and E.\ Ofek for providing us with their spectrum of
M85-OT.  A.V.F.'s group is supported by NSF grants AST-0607485 and
AST-0908886, the TABASGO Foundation, US Department of Energy SciDAC
grant DE-FC02-06ER41453, and US Department of Energy grant
DE-FG02-08ER41653. KAIT and its ongoing operation were made possible
by donations from Sun Microsystems, Inc., the Hewlett-Packard Company,
AutoScope Corporation, Lick Observatory, the NSF, the University of
California, the Sylvia \& Jim Katzman Foundation, and the TABASGO
Foundation.


\end{document}